\documentclass[fleqn,usenatbib]{mnras}

\usepackage{newtxtext,newtxmath}
\usepackage{bm}

\usepackage[T1]{fontenc}

\DeclareRobustCommand{\VAN}[3]{#2}
\let\VANthebibliography\thebibliography
\def\thebibliography{\DeclareRobustCommand{\VAN}[3]{##3}\VANthebibliography}

\usepackage{graphicx}	
\usepackage{amsmath}	
\usepackage{placeins}
\usepackage{mathtools}
\usepackage{cuted}

\makeatletter
\newcommand*{\hyperlinkcite}[1]{\hyper@link{cite}{cite.#1}}
\newcommand{\OmK}{\Omega_\mathrm{K}}
\newcommand{\re}{\operatorname{Re}}

\makeatother

\title[Instabilities in proto-planetary discs]{Instabilities in dusty non-isothermal proto-planetary discs}

\author[M. Lehmann $\&$ M.-K.~Lin]{
Marius Lehmann $^{1}$\thanks{E-mail: mlehmann@asiaa.sinica.edu.tw}
Min-Kai Lin,$^{1,2}$
\\
$^{1}$Institute of Astronomy and Astrophysics, Academia Sinica, Taipei 10617, Taiwan\\
$^{2}$Physics Division, National Center for Theoretical Sciences, Taipei 10617, Taiwan\\
}

\date{Accepted XXX. Received YYY; in original form ZZZ}

\pubyear{2021}

\begin{document}
\label{firstpage}
\pagerange{\pageref{firstpage}--\pageref{lastpage}}
\maketitle

\begin{abstract}
Protoplanetary discs (PPDs) can host a number of instabilities that may partake directly or indirectly in the process of planetesimal formation. These include the Vertical Shear Instability (VSI), Convective Overstability (COS), Streaming Instability (SI), and Dust Settling Instability (DSI), to name a few. Notably, the VSI and COS have mostly been studied in purely gaseous discs, while the SI and DSI have only been analyzed in isothermal discs. How these instabilities operate under more general conditions is therefore unclear. To this end, we devise a local model of a PPD describing a non-isothermal gas interacting with a single species of dust via drag forces. Using this, we find that dust drag sets minimum length scales below which the VSI and COS are suppressed. Similarly, we find that the SI can be suppressed on sufficiently small scales by the gas' radial buoyancy if it cools on roughly a dynamical timescale. 
We show that the DSI can be effectively stabilized by vertical buoyancy, except at special radial and vertical length scales. We also find novel instabilities unique to a dusty, non-isothermal gas. These result in a dusty analog of the COS that operates in slowly cooled discs, and a dusty version of the VSI that is strongly enhanced by dust settling. We briefly discuss the possible implications of our results on planetesimal formation.
\end{abstract}




\section{Introduction}\label{sec:intro}


Dust plays an important role in various processes taking place in PPDs.
Planetesimals, the building blocks of planets, are believed to consist to a large extent of originally micron-sized dust grains, which first coagulated with one another to form $mm$-sized particles, and subsequently clumped together and underwent self-gravitational collapse to form $km$-sized planetesimals \citep{chiang2010,johansen2014,lesur2022,drazkowska2022}.  

For almost two decades now, the most viable mechanism for producing large enough dust-to-gas ratios in young PPDs, which can result in self-gravitational collapse, is considered to be the Streaming Instability [SI: \citep{youdin2005,johansen2007,youdin2007,johansen2009a}].
For the SI to be sufficiently efficient, however, it already requires concentrations of dust, essentially in the disc midplane, which are larger than what is expected in young, newly formed PPDs. The local dust-to-gas ratio can be raised, for example by dust settling \citep{nakagawa1986,dubrulle1995,takeuchi2002} or trapping by pressure bumps \citep{haghighipour2003a,haghighipour2003b,taki2016,onishi2017,huang2020}.  

One of the main processes opposing dust concentration in the midplane of PPDs is assumed to be turbulence. As PPDs are poorly ionized, the magneto-rotational-instability [MRI: \citep{balbus1991}] is likely suppressed in most regions \citep{lesur2020}. This circumstance has led to the discovery of a number of purely hydrodynamic instabilities, capable of driving weak to moderate levels of turbulence \citep{lesur2022}. While the corresponding turbulent angular momentum transport of these instabilities is unlikely sufficient to explain observed accretion rates, it can play an important role in the process of dust concentration.

Among the many possible instabilities that may in principle exist in PPDs, three have been found to be potentially relevant in driving hydrodynamic turbulence. 
These are the Vertical Shear Instability (VSI) \citep{urpin1998,urpin2003,nelson2013,barker2015,lin2015}], which requires vertical differential rotation, as well as rapid cooling of the gas; the Convective Overstability [COS: \citep{klahr2014,lyra2014,latter2016}], which requires an unstable radial entropy gradient coupled with a cooling time comparable to the orbital time scale; and the Zombie Vortex Instability \citep{marcus2015,lesur2016,barranco2018}, which requires slow cooling and a strong stable vertical entropy gradient. For the process of planetesimal formation, these instabilities are particularly interesting due to their ability to form large-scale zonal flows, as well as large-scale vortices. While the former is a possible explanation for the presence of dust rings, the latter is a possible explanation for non-axisymmetric dust concentrations in PPDs \citep{vdmarel2021,bae2022,lesur2022}. 

As stated above, which of the aforementioned hydrodynamic instabilities operate depends critically on the gas' cooling timescale. This, in turn, is largely controlled by the distribution of small dust grains tightly coupled to the gas 
\citep{malygin2017,barranco2018,pfeil2019,fukuhara2021}. Calculations of the gas cooling time scale across the radial and vertical extent of a PPD by these authors revealed that different hydrodynamic instabilities may be active in different regions, and also different epochs. However, the results are largely model-dependent and details differ between the studies. Thus, for the time being, the cooling time scale of gas in PPDs remains uncertain.  Nevertheless, these estimates suggest that one or more of the aforementioned instabilities are likely active in planet-forming regions of the disc.

Indeed, radiative hydrodynamic simulations have shown that the VSI operates in the outer tens of au in PPDs \citep{stoll2014, stoll2016, flock2017,flock2020}. These authors included to some extent the role of dust in gas thermodynamics, by solving for the balance between stellar irradiation and radiation transport including dust opacities. They also modeled large grains as passive tracers, which can be effectively stirred up by VSI-turbulence. This `puff up' has direct observational implications \citep{dullemond2022}.

Furthermore, dust also has dynamic feedback onto the gas. For example, the VSI is weakened by dust-induced buoyancy \citep{lin2017}, which allows dust grains --- if sufficiently abundant --- to settle against VSI stirring \citep{lin2019,lehmann2022}. This is consistent with analyses of the VSI with dust grains perfectly coupled to the gas \citep{lin2015,lin2017}.  
The COS has been simulated with dust feedback \citep{raettig2015,lyra2018,raettig2021}, but these simulations focus on dust-trapping by large-scale vortices that are formed by the nonlinear evolution of the COS. How dust dynamically affects the COS mechanism itself, remains unexplored.


When dust and gas are imperfectly coupled, an entire class of drag instabilities becomes possible. The prime example here is the SI mentioned above. The classical SI of \citeauthor{youdin2005} is powered by the relative radial drift between dust and gas, itself due to the disc's radial pressure gradient. Recently, there have also been several generalisations of the SI: the Dust Settling Instability [DSI: \citet{squire2018,krapp2020}], which is powered by the dust settling; the Vertically Shearing SI \citep{lin2021}, powered by the vertical gradient in the midplane dust layer's rotation velocity; and the Azimuthal-drift SI \citep{lh2022}, powered by the azimuthal velocity difference between dust and gas when the latter is torqued within the disc plane. However, these drag instabilities have only been studied assuming an isothermal gas, which can be considered to cool instantaneously such that buoyancy effects are absent. Thus, the effect of finite cooling timescales and buoyancy forces on known dusty-gas instabilities is largely unknown.


Motivated by the above considerations, in this study we formulate a local hydrodynamic model for a non-isothermal gas interacting with dust. We apply a two-fluid formulation that self-consistently describes the dynamical impact of dust onto the gas and vice versa. The gas is subject to optically thin cooling, which is adopted as it provides a simple means to shift between the isothermal and adiabatic regimes. The cooling time scale is an input parameter in our model, thus ignoring the effect of dust on the gas thermodynamics. This is expected to be a minor shortcoming though, as it is the small grains that largely control the gas cooling, whereas it is the larger grains (as considered here) that impose dynamical effects onto the gas. 

We utilise our model to perform linear stability analyses of a number of local instabilities as described above. Specifically, we will study the effect of dust on the COS and the VSI. Moreover, we will investigate the impact of cooling and buoyancy on the SI and the DSI. While we find the classical COS to be damped by dust, we discover a dust-induced variant of the COS which operates at large cooling times, and which co-exists with the SI, possibly exceeding the SI's growth rates. 

Furthermore, while the SI turns out to be unaffected by radial buoyancy for most values of the gas cooling time, we find that it can be substantially suppressed on sufficiently short length scales if the cooling time is comparable to the orbital time scale. 

Similarly, while the VSI turns out to be damped by dust sufficiently close to the disc midplane, farther away, we discover a new instability that combines vertical shear and dust-settling to become a combination of the VSI and the DSI, with growth rates possibly exceeding those of the latter instabilities. Finally, we find the DSI to be progressively suppressed with increasing gas cooling times on all scales apart from a narrow band of `resonant' modes in wavenumber space. 
The paper is structured as follows.
In \S \ref{sec:hydromodel} we introduce our hydrodynamic model of dust and gas. Here we first discuss briefly important properties of the global disc structure, before we focus on a local description and present the set of linearised equations that will be used in our analyses. Subsequently we derive values for all the parameters that define our local linear model and we briefly introduce diagnostic techniques applied in our analyses. In \S \ref{sec:prelim_instab} we provide a short introduction of the linear instabilities that we study in this paper with connections to previous works. In \S \ref{sec:instab_midp} we present our analysis of the COS and the SI
and in \S \ref{sec:instab_z} our analysis of the DSI and the VSI. Finally, in \S \ref{sec:summary} we discuss the potential relevance of our findings to planetesimal formation PPDs, and we summarise in \S\ref{sec:final_summary}.


\section{Hydrodynamic Model}\label{sec:hydromodel}

We consider a three-dimensional (3D) PPD comprised of gas and dust orbiting a central star of mass $M_*$. We adopt cylindrical coordinates $(r,\phi, z)$, centered on the star. Disc self-gravity, viscosity, and magnetic fields are neglected. We will eventually adopt a local description of the disc defined by a number of parameters related to the global disc. To anchor our local model, and by way of introducing notation, we first describe the global gas disc in the absence of dust. In this limit, the disc admits exact, steady state, axisymmetric equilibria with density, pressure, and velocity fields $\rho_g(r,z)$, $P(r,z)$, and $\vec{\mathbf{v}}_g=(0,r\Omega_g,0)$, respectively, where $\Omega_g(r,z)$ is the gas rotation profile. These are set by centrifugal balance and vertical hydrostatic equilibrium. Explicit profiles are given in \S \ref{sec:pars}. 

We define
\begin{align}\label{eq:eta}
\eta \equiv -\frac{1}{2r\OmK^2\rho_g}\frac{\partial P}{\partial r}
\end{align}
as a dimensionless measure of the global radial pressure gradient, where 
\begin{equation}
\OmK\equiv\sqrt{\frac{G M_{*}}{r^3}}
\end{equation}
is the local Keplerian frequency and  $G$ is the gravitational constant. The global radial pressure gradient drives the SI and $\eta$ features prominently in studies of the SI. Typically, $\eta \sim h^2$, where the disc aspect ratio $h\equiv H/r\ll 1$ and $H$ is the disc scale-height, formally defined via
\begin{equation}\label{eq:h0}
   \frac{1}{H} \equiv \frac{1}{\rho_{g}} \left| \frac{\partial \rho_{g}}{\partial z} \right|_{z=H}.
\end{equation}

PPDs are generally baroclinic with $\nabla \rho_g\times\nabla P\neq0$. 
In these cases, the disc exhibits vertical shear with $\partial_z \Omega_g \neq0$. We therefore define 
\begin{align}\label{eq:qz}
q_{z} \equiv \frac{r}{\Omega_g}\frac{\partial \Omega_g}{\partial z}
\end{align}
as a dimensionless measure of the vertical shear rate. The parameter $q_z$ appears in local models of the VSI. In PPDs, vertical shear is typically weak with $|q_z|\lesssim h$ within the disc bulk. 

We shall also be concerned with buoyancy effects, which are measured by the radial and vertical buoyancy frequencies $N_r$ and $N_z$ defined via
\begin{align}
  N_{r}^2 & \equiv -\frac{1}{\gamma\rho_{g}}\frac{\partial P}{\partial r}\frac{\partial S}{\partial r},\label{eq:nr2} \\
  N_{z}^2 & \equiv -\frac{1}{\gamma\rho_{g}}\frac{\partial P}{\partial z}\frac{\partial S}{\partial z},\label{eq:nz2}
\end{align}
where $\gamma$ is the gas' adiabatic index and 
\begin{align}\label{eq:ent}
    S \equiv \ln{\frac{P}{\rho_g^\gamma}}
\end{align}
is the gas' (dimensionless) entropy. In typical disc regions  $|N_{r}^2|\sim h^2\OmK^2$. The value of $N_{z}^2$ depends on the distance $z$ away from the disc midplane. At $|z|\sim H$ one finds $N_{z}^2 \sim \OmK^2$. 

It is also convenient to define the characteristic entropy length scales $H_r$ and $H_z$ via \citep{lp2017}
\begin{align}
  \frac{1}{H_{r}} & \equiv \frac{1}{\gamma}\frac{\partial S}{\partial r}\label{eq:dSdr},\\
        \frac{1}{H_{z}} &  \equiv \frac{1}{\gamma}\frac{\partial S}{\partial z} \label{eq:dSdz}.
\end{align}
Then $H_{r,z}N_{r,z}^2 = -\rho_g^{-1}\partial_{r,z}P$. Typically $|H_{r}| \sim r$ and $|H_{z}| \sim H \ll r $ in radially-smooth, vertically thin discs. 
However, in general, the length scales (\ref{eq:dSdr}) and (\ref{eq:dSdz}) can by definition take positive and negative values.
In this paper, we consider only discs that are convectively stable in the vertical direction, i.e. we assume $N_{z}^2\geq 0$. This implies $H_{z}<0$ below the midplane (since $\partial_z P>0$ there). On the other hand, $N_{r}^2$ can be positive or negative. Specifically, if $\partial_r P <0$, which is typical for PPDs, then a negative radial entropy gradient ($H_r<0$) can yield $N_r^2<0$ and drive the COS (see \S \ref{sec:prelim_cos}). More on these quantities follows in \S \ref{sec:pars}.

\subsection{Local description}\label{sec:local}

We focus on local dynamics with characteristic length scales $\lambda \ll r$ and adopt the `shearing box' framework to model a small patch of the disc \citep{goldreich1965}. The box is centered on a fiducial point $(r_0,\phi_0 - \Omega_0 t, -z_0)$, that rotates around the star at approximately the local Keplerian frequency $\Omega_0\approx \OmK(r_0)$. At the disc midplane $\Omega_0$ co-incides exactly with $\OmK(r_0)$, whereas for $z_{0}>0$ small deviations on the order $(z_{0}/r_{0})^2$ occur due to vertical stellar gravity. In our convention, the shearing box is placed below the midplane, such that $z_{0}>0$, without loss of generality. The parameters defined above are assumed to be constant in the box and equal to their equilibrium, dust-free values at $r=r_0$, $z=-z_0$ in the global disc. We introduce local Cartesian coordinates $(x,y,z)$ with unit vectors $\vec{\mathbf{e}}_x,\, \vec{\mathbf{e}}_y, \, \vec{\mathbf{e}}_z$ in the box corresponding to the radial, azimuthal, and vertical directions in the global disc. 


\subsubsection{Governing equations: gas}


In this paper, we consider instabilities that are incompressible in the gas' motions. However, they may involve perturbations in the gas entropy or temperature in order to operate, which translates to weak density perturbations in accordance with the ideal gas law. Therefore we adopt the Boussinesq shearing box model of \citet{lp2017} to describe the gas. 
The governing equations read
\begin{align}
  \vec{\nabla} \cdot \vec{\mathbf{v}}_{g} & = 0, \label{eq:contrhog}\\
\begin{split}
 \left(\frac{\partial}{\partial t} + \vec{\mathbf{v}}_{g}\cdot\vec{\nabla}\right) \, \vec{\mathbf{v}}_{g}    & =  - 2 \Omega_{0} \vec{\mathbf{e}}_{z}\times \left[\vec{\mathbf{v}}_{g}-\left(-\frac{3}{2}\Omega_{0} x + q_{z} \Omega_{0} z\right)\vec{\mathbf{e}}_{y} \right]\\
 & \quad +\left( 2 \eta \Omega_{0}^2 r_{0}  -H_{r}N_{r}^{2} \frac{\delta \rho_{g}}{\rho_{g0}} \right) \vec{\mathbf{e}}_{x}\\
 & \quad -\left( \epsilon_{0} z_{0} \Omega_{0}^2 + H_{z}N_{z}^{2}\frac{\delta \rho_{g}}{\rho_{g0}}  \right)\vec{\mathbf{e}}_{z} \\
 & \quad - \frac{\epsilon}{t_{s}} \left(\vec{\mathbf{v}}_{g} -\vec{\mathbf{v}}_d \right)- \frac{1}{\rho_{g0}} \vec{\nabla} \delta P\label{eq:contvg},  
 \end{split}\\
\begin{split}
 \left(\frac{\partial}{\partial t} + \vec{\mathbf{v}}_{g}\cdot\vec{\nabla}\right) \, \delta \rho_{g}    & = \rho_{g0} \left( \frac{1}{H_{r}} \delta u_{g} + \frac{1}{H_{z}}\delta w_{g}\right) + \Lambda_{c}.  \label{eq:conteg}  
 \end{split}
\end{align}
The first two equations are the incompressibility condition and the momentum equation, respectively. 
The gas velocity field is $\vec{\mathbf{v}}_g=(u_g, v_g, w_g)$. The gas has a constant background density $\rho_{g0}$ with density perturbations $\delta\rho_g$ from the background. Similarly, $\delta P$ and $\delta\vec{\mathbf{v}}_g$ denote pressure and velocity perturbations from equilibrium, respectively.


Eq.~(\ref{eq:conteg}) corresponds to the gas entropy equation in the Boussinesq approximation, such that entropy perturbations are directly related to density perturbations, i.e. $\delta S = -\left(\delta\rho_g/\rho_{g0}\right)$, and  the gas temperature perturbation $\delta T = -\left(\delta\rho_g/\rho_{g0}\right)T_0$, where $T_0$ is the equilibrium temperature. 
Furthermore, the gas is subject to optically thin cooling on a time scale $t_{c}$, described through
\begin{equation}\label{eq:beta}
    \Lambda_{c} = -\frac{1}{t_{c}}\delta \rho_{g}
\end{equation}
in the entropy equation, where we absorbed $\gamma$ into the cooling time for notational convenience. 
 In what follows we will work with the dimensionless cooling time
\begin{equation}
\beta =t_{c} \Omega_{0}.
\end{equation}


Our local model incorporates several effects from the global disc. The gas' offset from pure Keplerian rotation, due to the global radial pressure gradient, is (as in numerous other studies) modeled via the constant radial forcing term $\propto\eta$. Furthermore, vertical shear appears as the forcing term $\propto q_z z$, which leads to a linear shear flow in $z$ (see \S \ref{sec:equilibrium}). The terms $\propto N_r^2$ and $\propto N_z^2$ in Eq. (\ref{eq:contvg}) represent radial and vertical buoyancy, respectively, which require gas density perturbations to take effect. This, in turn, requires a non-isothermal gas with $t_c>0$.  

In passing we note that in the case of a disc with entropy stratifiction only in the radial \emph{or} the vertical direction, it is common practice to replace the gas density perturbation $\delta\rho_{g}$ with a `buoyancy variable' $\propto H_{r} \delta\rho_{g}$ or $\propto H_{z} \delta \rho_{g}$, respectively. In that case the stratification length no longer appears in the equations. However, since we will consider stratifications in single, as well as both directions in our analyses below, we will --- for clarity --- not follow this practice.

Finally, the terms $\propto \epsilon$ and $\propto\epsilon_0$ in Eq. (\ref{eq:contvg}) are associated with feedback from the dust, described next.


\subsubsection{Governing equations: dust}\label{sec:dust_local}

We model a single species of dust grains coupled to the gas via a drag force characterized by a constant particle stopping time $t_{s}$, which corresponds to a Stokes number 
\begin{equation}\label{eq:tau}
    \tau \equiv t_{s} \Omega_{0}.
\end{equation}
We assume small dust grains with $\tau\ll 1$, such that the dust is tightly coupled to the gas and may be described as a second, pressureless fluid \citep{jacquet2011}.

The dust component is governed by the set of fluid equations
\begin{align} 
\left(\frac{\partial}{\partial t} + \vec{\mathbf{v}}_{d}\cdot\vec{\nabla}\right) \, \rho_{d} &=  - \rho_{d} \left( \vec{\nabla} \cdot \vec{\mathbf{v}}_d \right) \label{eq:contrhod},\\
\begin{split}
\left(\frac{\partial}{\partial t} + \vec{\mathbf{v}}_{d}\cdot\vec{\nabla}\right) \, \vec{\mathbf{v}}_{d}  &= - 2 \Omega_{0} \vec{\mathbf{e}}_{z}\times \left[\vec{\mathbf{v}}_{d}-\left(-\frac{3}{2}\Omega_{0} x + q_{z} z \Omega_{0}\right)\vec{\mathbf{e}}_{y} \right] \\
& \quad  +z_{0} \Omega_{0}^2 \vec{\mathbf{e}}_{z}  + \frac{1}{t_{s}} \left(\vec{\mathbf{v}}_{g} - \vec{\mathbf{v}}_{d} \right), \label{eq:contvd} 
\end{split}
\end{align}
 where $\rho_d$ and $\vec{\mathbf{v}}_d=(u_d, v_d, w_d)$ are the dust density and velocity, respectively. Unlike gas, dust is compressible and it is assumed that dust-related pressure can be neglected. Therefore, the dust fluid does not have an energy equation. However, the dust indirectly (via drag) experiences the gas' radial pressure gradient and buoyancy forces. We define
\begin{align}\label{eq:eps}
\epsilon \equiv \frac{\rho_d}{\rho_{g0}}
\end{align}
as the local dust-to-gas ratio, which for a Boussinesq gas is equivalent to the (scaled) dust density. The equilibrium dust-to-gas ratio is $\epsilon_0$.


At general heights $z_0\neq0$, dust is expected to settle towards the disc midplane. Following \citet{krapp2020}, we model this background settling by applying a constant vertical gravitational acceleration $z_{0}\Omega_{0}^2$ to the dust. Since, as outlined above, we assume 
 that the shearing box is placed at a distance $z_0>0$ below the disc midplane, dust 
experiences a vertically upwards acceleration. 
A corresponding downwards vertical acceleration,$-\epsilon_{0} z_{0} \Omega_{0}^2$, is applied to the gas (Eq. \ref{eq:contvg}) by hand so that a steady equilibrium state can be defined.



Note that we use the same parameter $q_z$ to impose the same vertical shear rate for the dust as for the gas. This is because small grains tightly coupled to the gas are expected to possess the same rotation profile to a good approximation \citep{lin2021}.  

Furthermore, our local model neglects vertical dust stratification, which naturally occurs away from the midplane in PPDs as dust settles to the midplane. The related vertical gradient in the dust density is known to produce a vertical buoyancy, and this dusty buoyancy has been found to stabilise vertical motions \citep{lin2017} and is expected to dominate gas-related buoyancy sufficiently close to the midplane $z_{0} \ll H_{0}$.

In addition to our two-fluid description outlined above, we will also make use of a one-fluid description of dust and gas \citep{laibe2014,lin2017,lovascio2019}. The corresponding equations are presented in Appendix \ref{app:onefl}.

\subsubsection{Remarks}

What distinguishes the `Boussinesq' shearing box model applied here from a general compressible local model are the assumptions that the length scale of the phenomena under investigation $\lambda \ll H_{z},H_{r}$ (rather than only $\lambda \ll r_{0}$), that the flow is strongly subsonic ($\delta \vec{\mathbf{v}}_{g}$ much smaller than the sound speed), and that the fractional perturbations of the gas density $\delta \rho_{g}/\rho_{g0}\ll 1$ and the gas pressure $\delta P/P \ll 1$, but with $\delta \rho_{g}/\rho_{g0} \gg \delta P/P $. The latter assumption allows for gas buoyancy effects to enter the model (see \citealt{lp2017} for details). 
It should be noted that the background pressure gradient $\eta$ formally vanishes in the expansion performed in \citet{lp2017} (their Section 3.2) to obtain (\ref{eq:contvg}). That is, for the gas in isolation it is negligible. However, in combination with dust it can result in the SI and the DSI, as studied here.
As we will see below, our model successfully reproduces previous results on the pure gas instabilities COS \citep{lyra2014,latter2016} and VSI \citep{latter2018}, as well as the dust-gas drag instabilities SI \citep{youdin2005} and DSI \citep{squire2018,krapp2020}.

\subsection{Equilibrium state}\label{sec:equilibrium}


Eqs. (\ref{eq:contrhog})-(\ref{eq:contvd}) admit equilibrium solutions with constant $\rho_d$ and $\delta P = \delta\rho_g=0$. The equilibrium velocity fields can be written as
\begin{align}\label{eq:shearflow}
\vec{\mathbf{v}}_{g0} = \left(-\frac{3}{2}x + q_z z\right)\Omega_0 \vec{\mathbf{e}}_y + \vec{\mathbf{v}}_{g0}^\prime,
\end{align}
and similarly for $\vec{\mathbf{v}}_{d0}$, with
\begin{align}
u_{g0}^\prime &= \frac{2 \epsilon_{0} \tau (\eta r_{0} - \left(1 + \epsilon_{0}\right) q_{z}  z_{0}) }{\left(1 +  \epsilon_{0}\right)^2 + \tau^2} \Omega_{0}\label{eq:ug0},\\ 
v_{g0}^\prime &= \frac{-\eta r_{0}\left(1 + \epsilon_{0} + \tau^2\right) + \epsilon_{0}  \tau^2 q_{z} z_{0} }{\left(1 +  \epsilon_{0}\right)^2 + \tau^2} \Omega_{0}\label{eq:vg0},\\
w_{g0}^\prime &= 0\label{eq:wg0},
\end{align}
and
\begin{align}
u_{d0}^\prime & = -\frac{2  \tau (\eta r_{0} +  \left(\epsilon_{0} + \epsilon_{0}^2 + \tau^2\right)  q_{z} z_{0})}{\left(1 +  \epsilon_{0}\right)^2 + \tau^2} \Omega_{0}\label{eq:ud0},\\
  v_{d0}^\prime & = -\frac{\left(1 + \epsilon_{0}\right) \eta r_{0} +  \tau^2 q_{z} z_{0}}{\left(1 +  \epsilon_{0}\right)^2 +\tau^2} \Omega_{0}\label{eq:vd0},\\
w_{d0}^\prime &= \tau  z_{0}\Omega_{0}\label{eq:wd0}, 
\end{align}
which are constants. In the limit of $q_{z}, z_0\to 0$, these  coincide with the standard solutions as derived in \citet{nakagawa1986}. For clarity, hereafter we drop the primes with the understanding that the equilibrium velocities are relative to the linear shear flow in $x$ and that in $z$ as given by (\ref{eq:shearflow}).

It is worth noting that in addition to the radial pressure gradient $\eta$, also vertical shear contributes to the radial and azimuthal velocities for dust and gas, and causes the center of mass velocity [Eq. (\ref{eq:cms})] to attain a finite value. However, it should be kept in mind that while in our local model $q_{z}$ and $\eta$ are freely adjustable parameters, in a global disc model both parameters depend on the radial temperature profile of the disc (see \S \ref{sec:pars} for details).

\subsection{Conservation properties}
Following \citet{youdin2007}, from (\ref{eq:contrhog})---(\ref{eq:conteg}) and (\ref{eq:contrhod})---(\ref{eq:contvd}) we can derive conservation equations for the angular momentum and the energy of the dust-gas mixture.
For simplicity we assume axisymmetry of the flow, and entropy stratification only in vertical direction, i.e. $N_{r}^2=0$, $1/H_{r}^2=0$, $N_{z}^2 >0$ and $H_{z}<0$.

From the $y$-component of (\ref{eq:contvg}) and (\ref{eq:contvd}) we obtain the conservation law for the angular momentum density $\mathcal{L} \equiv \rho_{g0} v_{g} + \rho_{d} v_{d}$:
\begin{equation}\label{eq:lflux}
\frac{\partial }{\partial t} \mathcal{L} + \vec{\nabla}\cdot \vec{\mathcal{F}}_{\mathcal{L}} = -\frac{\Omega_{0}}{2} \mathcal{F}_{\rho,x}-q_{z}\Omega_{0} \mathcal{F}_{\rho,z},
\end{equation}
where
\begin{align}
    \vec{\mathcal{F}}_{\mathcal{L}} & \equiv \rho_{g0} v_{g}\vec{\mathbf{v}}_g + \rho_{d} v_{d} \vec{\mathbf{v}}_d,\\
     \vec{\mathcal{F}}_{\rho} & \equiv \rho_{g0} \vec{\mathbf{v}}_g + \rho_{d}  \vec{\mathbf{v}}_d ,  
\end{align}
are the angular momentum flux and mass flux, respectively.

For the equilibrium described by (\ref{eq:ug0})---(\ref{eq:wd0})
we find that the right hand side of (\ref{eq:lflux}) vanishes, as well as
\begin{align}
    \begin{split}
\mathcal{F}_{\mathcal{L},x} & = -\frac{2 \eta^2 v_{K}^2 \,\rho_{d}\tau^3}{\Delta^2},\\
\quad & + \frac{2 \eta v_{K}^2  \,\rho_{d} \tau q_{z} \frac{z_{0}}{r_{0}}(1+\epsilon) (\Delta^2 + 2 \tau^2)}{\Delta^2} + \mathcal{O}\left( \tau^5 q_{z}^2\right),
    \end{split}\\
  \mathcal{F}_{\mathcal{L},z} & =-\frac{v_{K}^2 \,\rho_{d} \tau \frac{z_{0}}{r_{0}}\left((1+\epsilon)\eta  + \tau^2 q_{z} \frac{z_{0}}{r_{0}}\right)}{\Delta},\\
  \mathcal{F}_{\rho,x} & = -2 v_{K} \, \rho_{d} \tau q_{z}\frac{z_{0}}{r_{0}},\\
   \mathcal{F}_{\rho,z} & = v_{K} \,\rho_{d} \tau \frac{z_{0}}{r_{0}} ,
\end{align}
where $v_{K}\equiv \Omega_{0} r_{0}$ and $\Delta \equiv (1+\epsilon_{0})^2 + \tau^2$.
Compared to the model applied in \citet{youdin2007}, our model incorporates vertical settling and vertical shear. 
For $z_{0}=0$ we recover Eq. (18b) of \citet{youdin2007} for $\mathcal{F}_{\mathcal{L},x}$, while the other flux components vanish identically.
For $z_{0}>0$ (i.e. below the midplane) vertical settling and vertical shear result in a flow of angular momentum radially outward, and away from the midplane, and a related mass flow directed radially inward and toward the midplane. Using $\eta\sim h_{0}^2$ and $q_{z}\sim h_{0} z_{0}/H_{0}$ (see \S \ref{sec:pars} below), we find that the radial angular momentum flux changes sign for $z_{0}\gtrsim \tau H_{0}$.

Furthermore, we can derive the conservation equation for energy, which reads
\begin{equation}
\begin{split}\label{eq:econserv}
    \frac{\partial}{\partial t} \mathcal{E} + \vec{\nabla}\cdot \vec{\mathcal{F}}_{\mathcal{E}} & = \dot{\mathcal{E}}_{\text{drag}}  -\epsilon_{0}\rho_{g0}z_{0}\Omega_{0}^2 w_{g} +\rho_{d} z_{0} \Omega_{0}^2 w_{d} \\
    \quad & - \vec{\mathbf{v}}_g\cdot \nabla P  + \frac{3}{2}\Omega_{0} \mathcal{F}_{\mathcal{L},x} - q_{z} \Omega_{0} \mathcal{F}_{\mathcal{L},z} \\
    \quad & -\rho_{g0}\frac{N_{z}^2 \theta^2 }{t_{c}},
    \end{split}
\end{equation}
where \citep{youdin2007,lp2017}
\begin{align}
\mathcal{E} & = \frac{1}{2}\rho_{g0} |\vec{\mathbf{v}}_g |^2 + \frac{1}{2}\rho_{d} |\vec{\mathbf{v}}_d |^2 +\frac{1}{2}\rho_{g0} N_{z}^2 \theta^2,\\
\vec{\mathcal{F}}_{\mathcal{E}} & = \frac{1}{2}\rho_{g0} |\vec{\mathbf{v}}_g |^2  \vec{\mathbf{v}}_g +
\frac{1}{2}\rho_{d} |\vec{\mathbf{v}}_d |^2  \vec{\mathbf{v}}_d + \frac{1}{2}\rho_{g0} N_{z}^2 \theta^2  \vec{\mathbf{v}}_g,\\
\dot{\mathcal{E}}_{\text{drag}} & = -\rho_{d} |\vec{\mathbf{v}}_g-\vec{\mathbf{v}}_d|^2/t_{s},\\
\theta & = \frac{H_{z}}{\rho_{g0}}\delta \rho_{g}.
\end{align}
The interpretation of most terms is as described in \citet{youdin2007}. Since the gas considered here is non-isothermal, we include in addition the thermal energy $\propto \theta^2$ \citep{lp2017}, which prompts the thermal cooling term $\propto 1/t_{c}$ on the right hand side of (\ref{eq:econserv}). 
At equilibrium, the right hand side of (\ref{eq:econserv}) vanishes. Specifically, the energy released by vertical dust settling cancels with the vertical contribution to the drag dissipation term $\dot{\mathcal{E}}_{\text{drag}}$.

As was done in \citet{lp2017}, one can in addition consider the conservation equation of vorticity. However, rather than doing so, here we merely point out that drag forces and the radial pressure gradient $\eta$ do not enter the corresponding equation for the total vorticity $(\rho_{g0} \vec{\nabla}\times \vec{\mathbf{v}}_g + \rho_{d} \vec{\nabla}\times \vec{\mathbf{v}}_d)$ of dust and gas.

\subsection{Linear perturbation equations}\label{sec:linpert}

To the equilibrium described in \S \ref{sec:equilibrium}, we add wave-like, axisymmetric perturbations of the form
\begin{align}
 &\left\{\delta \rho_{g},\delta \vec{\bm{u}}_{g},\delta P,\delta \rho_{d},\delta \vec{\bm{u}}_{d} \right\} \notag \\
 &= \re\left[\left\{\delta \rho_{g1},\delta \vec{\bm{u}}_{g1},\delta P_1,\delta \rho_{d1},\delta \vec{\bm{u}}_{d1} \right\} \cdot e^{i \left(k_{x} x + k_{z} z\right) + \sigma t} \right]
 \label{eq:perturbations}
\end{align}
("Re" denoting the real part) with positive real radial and vertical wavenumbers $k_{x}$ and $k_{z}$, respectively, and generally complex frequency 
\begin{equation*}
\sigma \equiv \sigma_{R} + i \sigma_{I},
\end{equation*}
where $\sigma_R$ is the real growth rate and $\sigma_I$ is the real oscillation frequency. The subscript 1 denotes complex perturbation amplitudes, but will henceforth be dropped for clarity. 
%
In our analysis presented below we will work with dimensionless quantities. Length will be scaled with $r_{0}$, time will be scaled with $1/\Omega_{0}$ and densities will be scaled with $\rho_{g0}$.

Linearising Eqs. (\ref{eq:contrhog})---(\ref{eq:conteg}) and Eqs. (\ref{eq:contrhod})---(\ref{eq:contvd}) with respect to the above equilibrium and assuming Eq.~(\ref{eq:perturbations}) yields the dimensionless perturbation equations for the gas:
\begin{align}
        \sigma \delta \rho_{g}  = &-\left(  i k_{x} u_{g0}+ \frac{1}{\beta} \right) \delta \rho_{g} +\left(\frac{1}{H_{r}}  -\frac{k_{x}}{k_{z}} \frac{1}{H_{z}}\right) \delta u_{g}\label{eq:linrhog}, \\
    \begin{split}
        \sigma \delta u_{g}  =& -i k_{x} u_{g0} \delta u_{g} +2 \delta v_{g} - i k_{x} \delta P  - H_{r}N_{r}^2 \delta \rho_{g} \\
        \quad & +\frac{\epsilon_{0}}{\tau}\left(\delta u_{d}-\delta u_{g} \right) +\delta \rho_{d} \frac{1}{\tau}\left(u_{d0}-u_{g0}\right)\label{eq:linug} , 
    \end{split}\\
    \begin{split}
        \sigma \delta v_{g}  =& -i k_{x} u_{g0} \, \delta v_{g} - \left(\frac{1}{2} -\frac{k_{x}}{k_{z}}q_{z}\right)  \delta u_{g}\\
        \quad & +\frac{\epsilon_{0}}{\tau}\left(\delta v_{d}-\delta v_{g} \right)
        +\delta \rho_{d} \frac{1}{\tau}\left(v_{d0}-v_{g0}\right)\label{eq:linvg}  ,
    \end{split}\\
    \begin{split}
        \sigma \frac{k_{x}}{k_{z}}\delta u_{g}  =&  -i \frac{k_{x}^2}{k_{z}} u_{g0} \delta u_{g} + i k_{z} \delta P +H_{z}N_{z}^2 \delta \rho_{g} \\
        \quad & -\frac{\epsilon_{0}}{\tau}\left(\delta w_{d}+\frac{k_{x}}{k_{z}}\delta u_{g} \right) -\delta \rho_{d} \frac{1}{\tau} w_{d0}\label{eq:linwg} ,
    \end{split}
\end{align}
and for the dust:
\begin{align}
        \sigma \delta \rho_{d}  = & - i \left(k_{x} u_{d0} +k_{z} w_{d0}\right)\delta \rho_{d}  - i \epsilon_{0} \left(k_{x} \delta u_{d}  + k_{z} \delta w_{d} \right)  ,\label{eq:linrhod}\\  
        \sigma \delta u_{d}  = &   -   i \left(  k_{x} u_{d0} +  k_{z} w_{d0}\right)\delta u_{d}+2 \delta v_{d} + \frac{1}{\tau}\left(\delta u_{g} -\delta u_{d} \right) \label{eq:linud} , \\
    \begin{split}
        \sigma \delta v_{d}  = & -   i\left(  k_{x} u_{d0} +  k_{z} w_{d0}\right) \delta v_{d}  -\frac{1}{2} \delta u_{d} + \frac{1}{\tau}\left(\delta v_{g}-\delta v_{d} \right)  \\
    \quad & - q_{z} \delta w_{d}\label{eq:linvd}, 
    \end{split}\\
        \sigma \delta w_{d}  = & -  i\left(  k_{x} u_{d0} +  k_{z} w_{d0}\right) \delta w_{d} 
        +\frac{1}{\tau} \left( -\frac{k_{x}}{k_{z}}\delta u_{g}-\delta w_{d}\right) \label{eq:linwd} .
\end{align}
where we used the linearised version of the incompressibility condition (\ref{eq:contrhog}): 
\begin{equation}\label{eq:dincomp}
    \delta w_{g} = -\frac{k_{x}}{k_{z}} \delta u_{g},
\end{equation}
to eliminate the vertical gas velocity. Eqs. (\ref{eq:linug}) and (\ref{eq:linwg}) can be combined to obtain the pressure perturbation
\begin{equation}\label{eq:dpress}
\begin{split}
  \delta P & = \frac{i}{k^2} \bigg[ \left(H_{r}N_{r}^2 k_{x} + H_{z}N_{z}^2 k_{z}\right) \delta \rho_{g} -2 k_{x} \delta v_{g} \\
  \quad &   -\frac{k_{x} \left(u_{d0}-u_{g0}\right) + k_{z} w_{d0}}{\tau}\delta \rho_{d} - \frac{\epsilon_{0}k_{x}}{\tau} \delta u_{d}  
  - \frac{\epsilon_{0} k_{z}}{\tau} \delta w_{d}  \bigg],
  \end{split}
\end{equation}
which contains contributions from buoyancy (the first bracket), radial and vertical dust-gas drift (the third term), as well as dust-gas drag (the last two terms). Eq.~(\ref{eq:dpress}) is plugged into Eq.~(\ref{eq:linug}) such that the final set of perturbation Eqs. (\ref{eq:linrhog})---(\ref{eq:linvg}) and (\ref{eq:linrhod})---(\ref{eq:linwd}) are seven Eqs. for seven unknowns. 

Thus, the gas in isolation is characterized by the parameters $\beta$, $H_{r}$, $H_{z}$, $N_{r}^2$, $N_{z}^2$ and $q_{z}$. On the other hand, the description of gas and dust in co-existence requires in addition the parameters $\epsilon$, $\tau$ and $\eta$. Table \ref{tab:anysymbols} summarises the most important parameters, scales, and definitions that appear throughout the paper.
Possible inter-dependencies of these parameters and their connection to the global disc model are described in \S \ref{sec:pars}. It should be noted that the aspect ratio $h_{0}$ is not a parameter of our model, but is used to set the values of several model parameters.

The perturbation equations for gas and dust can be written as 
\begin{align}\label{eq:eigenproblem}
    M\vec{b} = \sigma \vec{b},
\end{align}
where $M$ is a $7\times7$ complex matrix and $\vec{b}=\left\{\delta \rho_{g}, \delta u_{g}, \delta v_g, \delta \rho_{d}, \delta \vec{\mathbf{u}}_d \right\}^T$ is a corresponding eigenvector. In the general case, we solve this eigenvalue problem with standard packages in IDL.

\begin{table}
 \caption{Important symbols and definitions. $(^{*})$: "equ."=equilibrium, "vel." =velocities, "pert."=perturbation, "cms"=center of mass.}
 \label{tab:anysymbols}
 \begin{tabular*}{\columnwidth}{@{}l@{\hspace*{50pt}}l@{\hspace*{50pt}}l@{}}
  \hline
  Symbols  & Definitions\\
  \hline
  $\sigma$  & complex eigenvalue of pert.$^*$\\[2pt]
   $k_{x/z}$  & radial/vertical wavenumber of pert.$^{*}$\\[2pt]
  $k=\sqrt{k_{x}^2+k_{z}^2}$  & total wavenumber of pert.$^{*}$\\[2pt]
  $\mu_{x/z}=\frac{k_{x/z}}{k}$  & scaled radial/vertical wavenumber of pert.$^*$\\[2pt]
  $\beta$  & dimensionless gas cooling time\\[2pt]
  $\gamma$  & gas' adiabatic index\\[2pt]
          $p$ & density slope parameter of global disc \\[2pt]
 $q$ & temperature slope parameter of global disc \\[2pt]
  $\eta$  &  dimensionless radial pressure gradient\\[2pt]
  $\Omega_{0}$  & orbital reference frequency\\[2pt]
  $r_{0}$  & reference disc radius\\[2pt]
  $\epsilon_{0}$  & equilibrium dust-to-gas density ratio\\[2pt]
  $\tau$  & particle Stokes number\\[2pt]
  $N_{r/z}^2$  & squared radial/vertical buoyancy frequency\\[2pt]
  $H_{r/z}$  &  radial/vertical entropy stratification length\\[2pt]
  $H_{0}$  &  vertical disc scale height\\[2pt]
  $z_{0}$  & distance away from the disc midplane\\[2pt]
   $h_{0}=\frac{H_{0}}{r_{0}}$  & disc aspect ratio \\[2pt]
  $q_{z}$  & vertical shear parameter\\[2pt]
  $\widetilde{q}_{z}=q_{z}\frac{k_{x}}{k_{z}}-\frac{1}{2}$  & VSI threshold parameter \\[2pt]
   $\delta u_{d}$, $\delta v_{d}$, $\delta w_{d}$  & linear dust pert.$^{*}$ vel.$^{*}$\\[2pt]
    $\delta u_{g}$, $\delta v_{g}$, $\delta w_{g}$  & linear gas pert.$^{*}$ vel.$^{*}$\\[2pt]
  $\delta \rho_{d},\delta \rho_{g}$  & linear dust and gas pert.$^{*}$ densities \\[2pt]
  $\delta u, \delta v, \delta w$  & linear one-fluid pert.$^{*}$ vel.$^{*}$ \\[2pt]
   $a_{d,x}=k_{x} u_{d0}$  &  radial dust advection parameter\\[2pt]
  $a_{d,z}=k_{z} w_{d0}$  & vertical dust advection parameter\\[2pt]
  $a_{g,x}=k_{x} u_{g0}$  &  radial gas advection parameter\\[2pt]
  $u_{d0},v_{d0},w_{d0}$  & dust equ.$^{*}$ vel.$^{*}$\\[2pt]
  $u_{g0},v_{g0},w_{g0}$  & gas equ.$^{*}$ vel.$^{*}$\\[2pt]
  \hline
 \end{tabular*}
\end{table}


\subsection{Pseudo-energy decomposition}\label{app:pseudo}

Following \citet{ishitsu09}, we perform an energy decomposition of the linearised two-fluid Eqs. (\ref{eq:linrhog})-(\ref{eq:linwd}). This technique has been proven to be a useful tool to better understand linear instabilities in terms of their driving forces and damping agents \citep[see for instance][]{lin2021,lh2022}. We therefore multiply Eqs. (\ref{eq:linug}) and (\ref{eq:linvg}) with $\delta u_{g}^{*}$ and $4 \delta v_{g}^{*}$, respectively, and Eqs. (\ref{eq:linud}), (\ref{eq:linvd}) and (\ref{eq:linwd}) with $\delta u_{d}^{*}$, $4 \delta v_{d}^{*}$ and $\delta w_{d}^{*}$, respectively. We then add the results and take the real part to obtain 
the total `pseudo'-energy\footnote{The factor 4 has been chosen as in \citet{ishitsu09} to eliminate the contributions due to the epicyclic terms.}
\begin{equation}
\begin{split}
    E_{\text{pseudo}} & \equiv |\delta u_{g}|^2 + 4 |\delta v_{g}|^2 + |\delta w_{g}|^2 + \epsilon_{0} \left(|\delta u_{d}|^2 + 4 |\delta v_{d}|^2 + |\delta w_{d}|^2\right)\\
    \quad & \equiv \sum\limits_{i=1}^{4} E_{i},
    \end{split}
\end{equation}
where we used (\ref{eq:dincomp}), and find
\begin{equation}
    E_{1}  = -H_{r}N_{r}^2 \,\text{Re}\left[\delta u_{g} \delta \rho_{g}^{*}\right] - H_{z}N_{z}^2\, \text{Re}\left[\delta w_{g} \delta \rho_{g}^{*}\right],\\
 \end{equation}
 \begin{equation}
    E_{2}  = -4 q_{z} \left(\text{Re}\left[\delta w_{g} \delta v_{g}^{*}\right] + \epsilon_{0} \,\text{Re}\left[\delta w_{d} \delta v_{d}^{*}\right] \right), \\
 \end{equation}
 \begin{equation}
 \begin{split}
    E_{3} & = \frac{1}{\tau}\bigg[ \left(u_{d0}-u_{g0}\right) \text{Re}\left[\delta u_{g} \delta \rho_{d}^{*}\right] \\
    \quad & + 4 \left(v_{d0}-v_{g0}\right) \text{Re}\left[\delta v_{g} \delta \rho_{d}^{*}\right]  + w_{d0}\, \text{Re}\left[\delta w_{g} \delta \rho_{d}^{*}\right]\bigg],\\
  \end{split}
 \end{equation}
 \begin{equation}
    E_{4}  = -\frac{\epsilon_{0}}{\tau}\left( |\delta u_{d} -\delta u_{g}|^2 + 4 |\delta v_{d}- \delta v_{g}|^2 + |\delta w_{d}- \delta w_{g}|^2 \right),
 \end{equation}
 where "$^*$" indicates the complex conjugate of a quantity.
In turn, $E_{1}$-$E_{4}$ are contributions due to buoyancy, vertical shear, dust-gas drift, as well as dust-gas drag, respectively. 
Correspondingly, in our analyses presented in \S \ref{sec:gas_midplane} and \S \ref{sec:gas_z} we will refer to terms $\propto \frac{\epsilon_{0}}{\tau}\left( \delta\vec{\mathbf{v}}_{d} -\delta\vec{\mathbf{v}}_{g}\right)$ as `drag force terms', and to terms $\propto \delta \rho_{d} \left(\vec{\mathbf{v}}_{d0} -\vec{\mathbf{v}}_{g0} \right)$ as `dust-gas drift terms'.

\subsection{Dimensionless parameters}\label{sec:pars}


Our local hydrodynamic model defined above is specified by a number of (dimensionless) parameters.
Here we briefly motivate the choice of all parameter values used in our numerical calculations below. We do this by evaluating these parameters at a fiducial location ($r=r_0, z=-z_0$) in a given global disc model. In this section, we use a `hat' notation to indicate the dimensionless versions of specific \emph{dimensional} quantities defined above by adopting the scaling as described in \S \ref{sec:linpert}. For clarity, the `hats' will, as also done in \S \ref{sec:linpert}, be dropped in the remainder of the paper.

We consider locally isothermal equilibria with  
\begin{equation}\label{eq:pres}
    P  =  c_{s}^2 \rho_{g},
\end{equation}
where $c_s$ is a prescribed sound-speed profile. We assume radial power law dependencies for sound-speed and the midplane gas density profiles, while the vertical gas profile is Gaussian. That is, 
\begin{align}
    c_{s}^2\left(r\right) & \propto \left(\frac{r}{r_{0}}\right)^q,\label{eq:plaw_cs}\\
        \rho_{g}\left(r,z\right) & \propto \left(\frac{r}{r_{0}}\right)^p \exp \left(-\frac{z^2}{2 H^2}\right)\label{eq:plaw_dens},
\end{align}
where the temperature slope parameter $q$ is not to be confused with the vertical shear parameter $q_{z}$. Typically, $p$ and $q$ are negative and of order unity. Note that the vertical hydrostatic equilibrium of pure gas, thin discs yields $H=c_s/\OmK$.

\subsubsection{Global radial pressure gradient and vertical shear rates}

Evaluating the radial pressure gradient parameter [Eq. (\ref{eq:eta})] using the above profiles gives:
\begin{align*}
\eta = -\frac{h_0^2}{2}\left(\overline{p}+q\right),
\end{align*}
where 
\begin{align*}
\overline{p} \equiv p + \frac{z_0^2}{H_{0}^2}\left.\frac{\partial \ln{H}}{\partial \ln{r}}\right|_0
\end{align*}
is the height-dependent, local radial gas density gradient. For radially-smooth discs, $\eta$ is of order $h_0^2$. For simplicity, in this study, we set $\eta = h_{0}^2$ when considering a non-zero radial pressure gradient. All calculations in \S \ref{sec:instab_midp} are performed assuming a disc aspect ratio $h_{0}=0.05$. On the other hand, in \S \ref{sec:instab_z} we use $h_{0}=\sqrt{10^{-3}}$, to facilitate a better comparison with the results on the linear DSI presented in \citet{krapp2020}. 

%

The vertical shear rate for a locally isothermal, thin gas disc is approximately
\begin{align}
q_z \simeq -\frac{q h_0}{2}\frac{z_0}{H_{0}}.
\end{align}
This is obtained from the definition of $q_z$ [Eq. (\ref{eq:qz})] and the exact expression $r\partial_z\Omega_g^2 ~= ~q(z/r)\OmK^2\left(1+z^2/r^2\right)^{-3/2}$  
\citep[see ][ Eq. 20]{lin2015}.
Thus $|q_z| \lesssim h_0$ and is typically (for $q<0$) positive below the disc midplane. In practice, if vertical shear is considered, we adopt a positive value $q_{z}=0.05 z_{0}/H_{0}$. 

\subsubsection{Buoyancy frequencies, entropy stratification lengths, and cooling}
We find from (\ref{eq:nr2}) using (\ref{eq:ent}) and (\ref{eq:pres})---(\ref{eq:plaw_dens}) the dimensionless radial buoyancy frequency 
\begin{equation}\label{eq:nr2_hat}
   \widehat{N}_{r}^2  = h_{0}^2 q^2 f(\zeta),
\end{equation}
%
%
where
\begin{equation}\label{fx}
f(\zeta)  = \zeta\left(\zeta+1\right) - \frac{\left(\zeta+1\right)^2}{\gamma}, \hspace{0.5cm} \text{with} \hspace{0.1cm} \zeta  =\frac{\overline{p}}{q}.
\end{equation}
The quantity $-\widehat{N}_{r}^2$ is also known as the Richardson number \citep{latter2016}. 
Since $f(\zeta)\sim 
\mathcal{O}\left(1\right)$, for smooth discs  $|\widehat{N}_{r}^2|\sim h_0^2 $. Depending on the values of $\overline{p}$ and $q$,  $\widehat{N}_{r}^2$ can be positive or negative. At the disc midplane ($\overline{p}=p$), a rather flat radial density profile is needed to make $\widehat{N}_r^2<0$, which is required for the COS.  However, away from the midplane, $|\overline{p}|$ can be reduced on account of the disc flaring.  In all cases with a negative value we use $\widehat{N}_{r}^2=-0.01$. In some calculations in \S \ref{sec:si_cool} we use a positive value $\widehat{N}_{r}^2=0.05$. 

The radial entropy stratification length (\ref{eq:dSdr}) is computed via
\begin{equation}
    \widehat{H}_{r} = \frac{\widehat{H}_{r}\widehat{N}_{r}^2}{\widehat{N}_{r}^2} = 2\frac{ \eta}{\widehat{N}_{r}^2},
\end{equation}
so that $\,|\widehat{H}_{r}| \sim 2\frac{ \eta}{h_{0}^2}$ is $\mathcal{O}\left(1\right)$. 


 Furthermore, we find from (\ref{eq:nz2}) using (\ref{eq:h0}), (\ref{eq:ent}) and (\ref{eq:pres})---(\ref{eq:plaw_dens}) the dimensionless vertical buoyancy frequency 
\begin{equation}\label{eq:nz_eval}
    \widehat{N}_{z}^2 = \left(1-\frac{1}{\gamma}\right)\frac{z_0^2}{H_{0}^2}. 
\end{equation}
For $\gamma=1.4$ and at $z_0=H_{0}$, we find $\widehat{N}_{z}^2\simeq 0.28$. In calculations that include vertical buoyancy, we use this value unless otherwise stated.

Furthermore, using again (\ref{eq:h0}), (\ref{eq:ent}) and (\ref{eq:pres})---(\ref{eq:plaw_dens}) one finds
\begin{equation}
\widehat{H}_{z}=-\frac{\gamma h_0}{\gamma-1}\left(\frac{H_{0}}{z_0}\right).
\end{equation}
Thus, in smooth disc regions away from the disc midplane, one typically has $|H_{z}|\ll \left|H_{r}\right|$ and $|N_z^2|\gg |N_r^2|$. However, near special locations such as pressure bumps, ice lines, or dead zone edges $|N_{r}^2|$ may take larger values, whereas $|H_{r}|$ is expected to take smaller values.

Finally, we use the dimensionless cooling time $\beta$ (Eq. \ref{eq:beta}) as a convenient parameter to explore different thermodynamic responses of the gas. The isothermal limit corresponds to $\beta\to0$, while the adiabatic limit corresponds to $\beta\to\infty$. However, in practice values of $\beta \lesssim 10^{-2}$ are identical to the isothermal limit, whereas values $\beta\gtrsim 10^3$ reproduce  adiabatic results.

\subsubsection{Dust size and abundance}

The implementation of dust in our model introduces two additional parameters compared to the pure gas model. These are the particle Stokes number $\tau$ [Eq.~(\ref{eq:tau})] and the equilibrium dust-to-gas ratio $\epsilon_{0}$ [cf. Eq.~(\ref{eq:eps})].

 In this paper, we assume tightly coupled dust and gas with small Stokes numbers $\tau \lesssim 10^{-2}$, although specific illustratory calculations are performed with larger values. In a Minimum Mass Solar Nebula disc, $\tau \sim 10^{-3}-10^{-2}$ corresponds to particle sizes from $\sim 280 \, \mathrm{\mu m}$ at 10 au to $\sim 9\, \mathrm{\mu m}$ at 100 au for $\tau =10^{-3}$, or between $\sim 3\,\mathrm{mm}$ and $90 \,\mathrm{\mu m}$ for $\tau =10^{-2}$ \citep{chen2020}.

 In the midplane of PPDs, dust is expected to settle to a relatively dense layer with $\epsilon_{0}\sim 1$ if the disc is laminar.  However, this value may be significantly smaller in the presence of external turbulence. We therefore consider a range of values $10^{-2}\lesssim \epsilon_{0}\lesssim 1$ in the disc midplane in \S \ref{sec:instab_midp}. On the other hand, for our calculations in \S \ref{sec:instab_z} away from the midplane at $z=H_{0}$ we assume, unless otherwise stated, $\epsilon_{0}=10^{-3}$, as in \citet{krapp2020}.

\subsubsection{Parameter inter-dependencies}

While some of the above parameters can be adjusted independently in our local model, they can be connected in the global disc. This is for instance true for the vertical shear $q_{z}$ and the radial pressure gradient, both of which depend on the radial temperature profile of the global disc. Furthermore, in the presence of a finite gas cooling time $\beta>0$ and radial buoyancy $N_{r}^2\neq 0$, one necessarily has $\eta \neq 0$.
That being said, it can still be useful to study the effect of a finite $N_{r}^2$ while setting $\eta=0$, since the two parameters result in physically different effects. The same applies to vertical shear $q_{z}$ and vertical buoyancy $N_{z}^2$. Of course, in the isothermal limit $\beta \to 0$ the effect of buoyancy vanishes, effectively setting $N_{r}^2\to 0$ and $N_{z}^2\to 0$, while $\eta$ and $q_{z}$ can still be non-zero.

\section{Preliminaries on the studied instabilities}\label{sec:prelim_instab}

The set of linearised equations (\ref{eq:linrhog})---(\ref{eq:linwd}) encapsulates a number of known gas and dust-gas instabilities, namely the COS, DSI, SI, and VSI, in the appropriate limits. In this paper, we generalise these instabilities within the aforementioned local model of dusty, non-isothermal gas. We shall also find new instabilities unique to various combinations of physical effects (vertical shear, dust settling or drift, buoyancy, etc.). To guide our investigation in a more systematic manner, we here provide a summary of the "base" instabilities above.

The COS, DSI, SI, and VSI are all destabilized inertial waves (IW), which are therefore central to our study. IWs are restored by the Coriolis forces and thus only exist in rotating flows. They are most easily extracted from the linearized equations by considering an isothermal, pure gas disc without vertical shear ($\beta=\epsilon_0=q_z=0$). We then find $\sigma = i\sigma_{\text{IW}}$, with
\begin{align}
\sigma_{\text{IW}} \equiv \pm\frac{k_z}{\sqrt{k_x^2+k_z^2}},
\end{align}
in dimensionless units. IWs, therefore, require $k_z\neq 0$, as do the  base instabilities. In the limit $|k_x/k_z|\ll 1$, IWs attain frequencies of $\pm 1$, corresponding to the local Keplerian frequency. In the discussion below, for convenience, we define $\mu_z\equiv k_z/k$, where $k=\sqrt{k_x^2+k_z^2}$.



\subsection{Pure gas instabilities}\label{sec:prelim_gas}
%
\subsubsection{COS}\label{sec:prelim_cos}

As mentioned earlier, the COS requires the square of the radial buoyancy frequency to be negative, $N_{r}^2<0$, and its growth rate is maximal for gas cooling times $\beta \sim 1$. Under typical conditions where $\partial P/\partial r<0$, the former criterion amounts to a negative radial entropy gradient, $\partial S / \partial r<0$. The COS is an overstability (oscillatory instability) of inertial waves which feeds off of the disc's unstable radial entropy gradient: Consider a gas parcel that is displaced radially outward from its original reference radius. Due to disc rotation it will undergo epicyclic oscillation. If $\partial S/\partial r<0$, during the first half of its epicycle the parcel will get in contact with `colder' material such that it will deposit some of its excess entropy. When it has returned to its original radius it is `colder' (and hence denser) than its surroundings and will therefore experience an inward buoyancy force, which accelerates the inward motion. The reverse process occurs during outward motion, which is hence amplified as well, and the process runs away. For an illustrative explanation of this mechanism, the reader is referred to \citet{latter2016}. 

 The linear, gaseous COS was first studied by \citet{klahr2014}, \citet{lyra2014}, and \citet{latter2016}. Its potential role in planetesimal formation has been demonstrated via the collection of dust in vortices in local shearing box hydrodynamics simulations \citep{lyra2018,raettig2021}. In \S \ref{sec:cos_nodp}-\ref{sec:cos_onef} we study in some detail the effect of dust on the COS in the disc midplane. In addition, in \S \ref{sec:cos_general} we briefly consider the effect of dust settling on the COS away from the midplane. 

\subsubsection{VSI}\label{sec:prelim_vsi}
The \emph{local} VSI studied here is a non-oscillatory instability of inertial waves, tapping into the vertical gradient of the disc's equilibrium orbital velocity $\partial \Omega_{g}/\partial z$. This contrasts to the vertically-global VSI modes found in semi-global analyses \citep{barker2015,lin2015} and numerical simulations \citep{nelson2013, stoll2014}, which are overstable. Nevertheless, a local approach is much more analytically tractable, while still capturing the essence of the instability.

\begin{figure}
\centering 
\includegraphics[width = 0.25\textwidth]{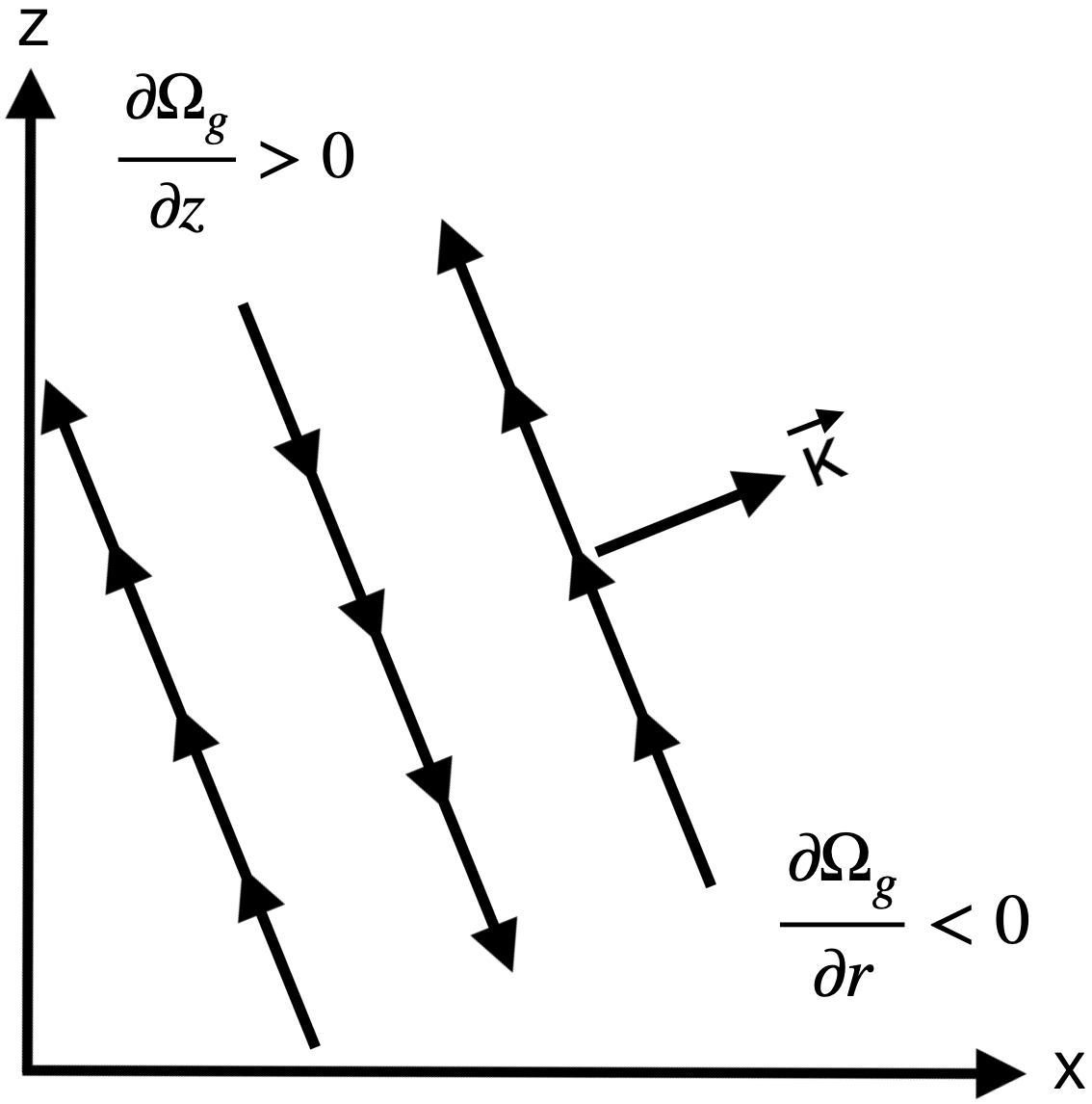}
\caption{Illustration of gas motions in an inertial wave with wavenumber $\vec{\mathbf{k}} = k_{x} \vec{\mathbf{e}}_{x}+k_{z} \vec{\mathbf{e}}_{z} $ with $k_{x}\gg k_{z}>0$, which is ought to be subject to the VSI: The destabilizing vertical gradient of angular velocity $\partial \Omega/\partial z$ dominates the stabilizing radial gradient $\partial \Omega/\partial r$, as explained in the text.}
\label{fig:vsi_cartoon}
\end{figure}

In the local picture, we imagine an inertial wave with wave vector $\vec{\mathbf{k}} = k_{x} \vec{\mathbf{e}}_{x} + k_{x} \vec{\mathbf{e}}_{z}$ with $k_{x} \gg k_{z}$ in an incompressible gas, as depicted in Figure \ref{fig:vsi_cartoon}. On account of incompressibility, gas motions occur on straight `lines' inclined 
 to the midplane. In the presence of vertical shear $q_{z}>0$ [Eq.~(\ref{eq:qz})], a gas parcel will possess a larger specific angular momentum than its surroundings upon downward motion (as indicated by the arrows). This will force it to move radially outward, which amplifies the downward motion. Similarly, upward motion is amplified as well, and instability ensues. For this mechanism to work, the ratio $k_{x}/k_{z}$ has to be sufficiently large and positive. Otherwise, motions are stabilized by the disc's radial angular momentum gradient. This also implies that the instability is non-oscillatory, as the restoring inertial forces are over-compensated by the destabilizing vertical angular momentum gradient. In the case $q_{z}<0$ the same argument applies to waves with $k_{x}/k_{z} \ll -1$.

The VSI has largely been studied in pure gas discs. In the limit of perfectly coupled dust, the VSI can be dampened by dust-induced buoyancy as it "weighs down" the gas \citep{lin2017,lin2019}. However, this only occurs in a stratified dust layer and is thus absent in our vertically local models. Instead, 
in \S \ref{sec:gas_z} we will study the effect of partially coupled dust on local VSI modes, as well as the interaction of the VSI and the DSI. 


\subsection{Drag instabilities}\label{sec:res_drag_instab}

The SI and the DSI apply to dusty gas in which there exists an equilibrium drift between dust and gas. For small dust-to-gas ratios, these instabilities belong to a general class of 
"resonant drag instabilities" \citep[RDI: ][]{squire2018}. According to RDI theory, a dusty fluid (here a mixture of gas and a single species of dust) can be subject to a linear `drag instability' if the equilibrium drift velocity between dust and gas matches the phase velocity of \emph{neutral} 
waves that exist in the pure gas limit. This condition reads
\begin{equation}\label{eq:krdi} 
\vec{\bm{k}} \cdot \Delta \vec{\mathbf{v}}_{0} = \sigma_{I}(\vec{\bm{k}}),      
\end{equation}
where $\sigma_{I}$ is the (real-valued) oscillation frequency of an underlying gas wave, and $\Delta \vec{\mathbf{v}}_0=\vec{\mathbf{v}}_{d0} - \vec{\mathbf{v}}_{g0}$ is the drift velocity at equilibrium. Eq.~(\ref{eq:krdi}) relates the radial and vertical wavenumbers that yield maximum growth (i.e. those which are resonant). 

In the case of the standard SI and the DSI, instability results from a resonance between IWs and the background drift. Then $\sigma_{I}=\sigma_{\text{IW}}$, while the radial component of the dust-gas drift, $\Delta u_0$, can be found by subtracting Eqs. (\ref{eq:ud0}) and (\ref{eq:ug0}) and the vertical component is the dust settling velocity directly given by (\ref{eq:wd0}). 

We note that the original RDI recipe was formulated for neutral, purely oscillitory gas waves with $\sigma= i \sigma_{I}$. However, in our analyses below, the gas is generally subject to buoyancy and cooling. This implies (as shown in \S \ref{sec:gas_midplane}) that generally, $\sigma=\sigma_{R} + i \sigma_{I}$, with $\sigma_{R} \neq 0$, i.e the corresponding gas waves are not neutral. Nevertheless, when analyzing the dusty gas problems below,  we will apply the RDI condition (\ref{eq:krdi}) to the imaginary part $\sigma_{I}$ of the complex pure gas frequency. Our results in \S \ref{sec:si_cool} and \S \ref{sec:dsi} will confirm that this approach is appropriate.

\subsubsection{SI}\label{sec:prelim_SI}

The classical SI \citep{youdin2005} was derived for an incompressible, isothermal dusty gas (i.e. without buoyancy and cooling/heating effects) in the disc midplane with $w_{d0}=0$ and no vertical shear $q_{z}=0$. Applying the RDI condition, Eq. (\ref{eq:krdi}), to IWs and a background radial dust-gas drift, we find
\begin{equation}\label{eq:res_si}
    k_{x} \Delta u_{0} = \pm \mu_{z},
\end{equation}
which gives the resonant vertical wavenumber as
\begin{equation}\label{eq:kres_iso}
k_{z,\text{res}} = \pm \frac{k_{x}^2 \Delta u_{0}}{\sqrt{1-\Delta u_{0}^2 k_{x}^2}}.
\end{equation}
Although this equation might suggest that there are two different curves $k_{z,\text{res}}(k_{x})$, in a linear analysis (as considered in this paper) only one of the two solutions needs to be considered, since we only consider equilibria with $\Delta u_{0}<0$. 


Previous theoretical studies of the SI have been restricted to the isothermal limit. In this paper, we explore the possible influence of buoyancy effects, the degree of which is controlled by the gas cooling time, on the SI (\S \ref{sec:si_cool}). We will find that the SI can be modified because cooling affects the underlying IWs or the RDI condition. 

\subsubsection{DSI}\label{sec:prelim_DSI}

The DSI is an extension of the SI that emerges when vertical dust-gas drift with the dust settling velocity $w_{d0}$ occurs in addition to radial drift $\Delta u_{0}$. The resonant criteria for the standard, isothermal DSI is thus  
\begin{equation}\label{eq:krdi_dsi} 
    k_{x} \Delta u_{0} + k_{z} w_{d0} = \pm\mu_z. 
\end{equation}
Both components of the drift velocity drive the DSI, with the vertical component being the dominant one if $|w_{d0}| \gtrsim |\Delta u_{0}| $. This typically occurs at sufficiently large (dimensional) heights $z_{0}\gtrsim h_{0} H_{0}$ from the disc midplane. In contrast to the resonance condition (\ref{eq:kres_iso}) for the SI, Eq. (\ref{eq:krdi_dsi}) for the DSI generally yields two distinct solutions, which in the general case need to be computed numerically.

In this paper, we investigate the effects of buoyancy and also vertical shear on the DSI (\S\ref{sec:dsi}). An interesting result we find is that, for non-isothermal gas, there is a third resonance with "cooling modes", which have $\sigma_I=0$. 
For such a mode, the resonance condition reads
\begin{equation}\label{eq:rdi_zero}
    k_{z,\text{res}} = -\frac{\Delta u_{0}}{w_{d0}} k_{x}.
\end{equation}
Although the original RDI recipe applies to neutral gas waves, while the cooling modes decay in the pure gas limit, we nevertheless find them to resonate with dust-gas drift and cause instability (see \S \ref{sec:dsi}).

\begin{table*}\label{tab:pars1}
\centering
 \caption{Overview of parameter-values used in the various analyses presented in \S \ref{sec:instab_midp}. The upper row indicates the instabilities and the corresponding section in the paper. A value of $H_{r/z}=\infty$ indicates that no entropy stratification exists in $r/z$-direction.}
 \begin{tabular*}{0.85\textwidth}{@{}l@{\hspace*{25pt}}l@{\hspace*{35pt}}l@{\hspace*{35pt}}l@{\hspace*{35pt}}l@{\hspace*{35pt}}l@{\hspace*{35pt}}l@{\hspace*{35pt}}l@{\hspace*{35pt}}l@{\hspace*{35pt}}l@{\hspace*{35pt}}l@{\hspace*{35pt}}l@{\hspace*{35pt}}l@{\hspace*{35pt}}l@{\hspace*{35pt}}l@{\hspace*{35pt}}l@{\hspace*{35pt}}l@{}}
  \hline
      & COS (gas) & COS (gas) & COS(gas+dust) & DCOS+SI &  SI  &  COS+SI \\
     & \S\ref{sec:cos_novbuo} & \S\ref{sec:cos_vbuo} & \S\ref{sec:cos_nodp} & \S\ref{sec:cos_dp} & \S\ref{sec:si_cool}  & \S\ref{sec:cos_vs_si} \\
      \hline
  $\beta$ & $>0$ & $>0$ & $>0$ & $\gg 1$ & $>0$ &  $>0$      \\[2pt]  
  $N_{r}^2$    & $<0$ & $<0$ & $<0$ & $<0$  & $>0$ & $<0$ \\[2pt]
  $H_{r}$  & $<0$ & $<0$ & $<0$ & $<0$ & $>0$  & $<0$ \\[2pt]
  $N_{z}^2$   & $0$ & $>0$ & $0$ & $0$ & $0$  & $0$    \\[2pt] 
  $H_{z}$     & $\infty$ & $<0$ & $\infty$ & $\infty$ & $\infty$  & $\infty$     \\[2pt]
  $q_{z}$  & $0$ & $0$ & $0$ & $0$ & $0$  & $0$      \\[2pt] \hline
  $\epsilon_{0}$   & $0$ & $0$ & $>0$ & $>0$ & $>0$  & $>0$  \\[2pt]
  $\tau$    & $0$ & $0$ & $>0$ & $>0$ & $>0$  & $>0$    \\[2pt]
  $\eta \,[\times 10^{-3}]$   & $0$ & $0$ & $0$ & $2.5$ & $2.5$ &  $2.5$  \\[2pt]
  \hline
 \end{tabular*}
\end{table*}

\section{Analysis: The COS and the SI}\label{sec:instab_midp}


In this section, we consider linear instabilities that do not involve vertical shear or dust settling by setting $q_z=w_{d0}=0$. This certainly applies to the disc midplane ($z_0=0$), but our analysis below is not limited to this location. For example, for a barotropic equilibrium, vertical shear vanishes identically; and dust settling is irrelevant in pure gas discs. We generally allow $N_z^2>0$ and finite $H_z$, but it should be kept in mind that these quantities formally vanish and diverge, respectively, if one applies our results to the disc midplane. Note that the quantity $H_zN_z^2\propto \partial_zP \to 0$ as $z_0\to 0$.

Neglecting vertical shear and dust settling simplifies the problem by eliminating the VSI and DSI, respectively, allowing us to first focus on the COS and SI. Compared to previous studies, we will study here for the first time the linear COS in the presence of dust, as well as the effect of buoyancy on the linear SI. 
Table \ref{tab:pars1} summarises the analyses considered in this section and lists the relevant parameters involved in each analysis.


\subsection{Pure gas, the COS}\label{sec:gas_midplane}


We begin by examining the COS in the pure gas limit. To this end, we further set $\epsilon_0=\delta\rho_d=0$ and discard the dust equations (\ref{eq:linrhod})---(\ref{eq:linvd}). The setup here generalises previous COS analyses \citep{klahr2014,lyra2014} by incorporating the effect of vertical buoyancy $N_z^2\neq0$, which only exists off of the midplane. However, as discussed above, the fact that we neglect vertical shear means that we are implicitly considering a barotropic equilibrium or that we are considering the disc midplane.

In this limit, we obtain the cubic dispersion relation
\begin{equation}\label{eq:deteqz}
   0 = \beta \sigma^3 +\sigma^2  + \xi \beta \sigma +\mu_{z}^2,
\end{equation}
where 
\begin{equation}\label{eq:xi}
    \xi =   \mu_{z}^2 \left(1+N_{r}^2\right) +\mu_{x}^2 N_{z}^2 -\mu_{x}\mu_{z} \left(\frac{H_{r}}{H_{z}}N_{r}^2 + \frac{H_{z}}{H_{r}} N_{z}^2 \right),
\end{equation}
and recall $\mu_z=k_z/k$ and
%
%
%
$\sigma=\sigma_{R} + i \sigma_{I}$.  We now follow \citet{lyra2014} and consider the real and imaginary parts of Eq.~(\ref{eq:deteqz}):
\begin{align}
0 & = \beta \sigma_{R}^3 +  \left(\sigma_{R}^2 -\sigma_{I}^2 +\mu_{z}^2 \right) + \beta \left[\xi -3 \sigma_{I}^2 \right] \sigma_{R} \label{eq:rp},   \\
0 & = \left(\beta \xi  -\beta \sigma_{I}^2 + 3 \beta \sigma_{R}^2 + 2 \sigma_{R}\right) \sigma_{I}\label{eq:ip}.
\end{align} 
Assuming $\sigma_I\neq0$, first we solve (\ref{eq:ip}) for $\sigma_{I}$ to obtain
\begin{equation}\label{eq:freq}
    \sigma_{I} = \pm \sqrt{ \xi + 3  \sigma_{R}^2 + \frac{2}\beta \sigma_{R}}.
\end{equation}
Plugging this relation into (\ref{eq:rp}) and assuming $|\sigma_{R}|\ll 1/\beta$ (which turns out to be fulfilled) 
we find
\begin{equation}\label{eq:grate}
\sigma_{R} = -\frac{\left(\xi-\mu_{z}^2\right)\beta}{2\left(1+\beta^2 \xi\right)}.
\end{equation}

There is no simple instability criterion on the disc structure in the general case, but instructive results may be obtained in the limiting cases examined below. 
Furthermore, there is one subtlety involved in the analytical solutions (\ref{eq:freq}) and (\ref{eq:grate}), which for sufficiently small $\mu_{x}$ describe inertial waves affected by buoyancy. These have been derived under the assumption that $\sigma_{I}\neq 0$, since otherwise Eq. (\ref{eq:ip}) would be trivially fulfilled. For example, the full problem includes the cooling mode, which possesses a purely real eigenvalue, and which is discarded by this assumption. This is unimportant for the discussion here as we focus on the COS. 

However, in the case that $N_{z}^2 \neq 0$, the full numerical solution of the cubic (\ref{eq:deteqz}) turns out to be more complex than what is described by (\ref{eq:freq})-(\ref{eq:grate}).
That is, the full numerical solution, which provides three eigenvalues for all parameter values, reveals that the growth rates of the inertial waves undergo a `bifurcation' for a given $k_{x}\equiv k_{x,\text{bif}}$, the latter depending on the values of $k_{z}$,  $N_{z}^2$ and $\beta$. At this bifurcation the wave frequencies drop to zero. We explore this in Appendix \ref{app:iw_bifurc}.  

%

Our analysis of the COS in this section is not affected by this bifurcation. The reason is that in the presence of vertical buoyancy, we find below that the COS modes rapidly damp for $k_{x}$-values much smaller than those for which the bifurcation occurs. For these values $k_{x}\ll k_{x,\text{bif}}$ the analytical expressions (\ref{eq:freq})-(\ref{eq:grate}) correctly apply.




\subsubsection{No vertical buoyancy}\label{sec:cos_novbuo}
We can recover the standard COS in an unstratified disc by neglecting vertical buoyancy, $N_{z}^2\to 0$, and setting $H_{z}\to \infty$ (such that $H_{z} N_{z}^2\to 0$). Then $\xi \to \mu_{z}^2 \left(1+N_{r}^2\right)$, which gives
\begin{equation}\label{eq:grate_cos}
\sigma_{R} = -\frac{\beta \mu_{z}^2 N_{r}^2}{2 \left(1 + \beta^2 \mu_{z}^2 \left[1+N_{r}^2 \right]\right)}.
\end{equation}
Typically $|N_r^2|\ll 1$ in PPDs (cf. \S \ref{sec:pars}). Thus Eq.~(\ref{eq:grate_cos}) signifies instability when $N_r^2<0$ (an unstable radial stratification). 

\begin{figure*}
\centering 
\includegraphics[width = 0.7\textwidth]{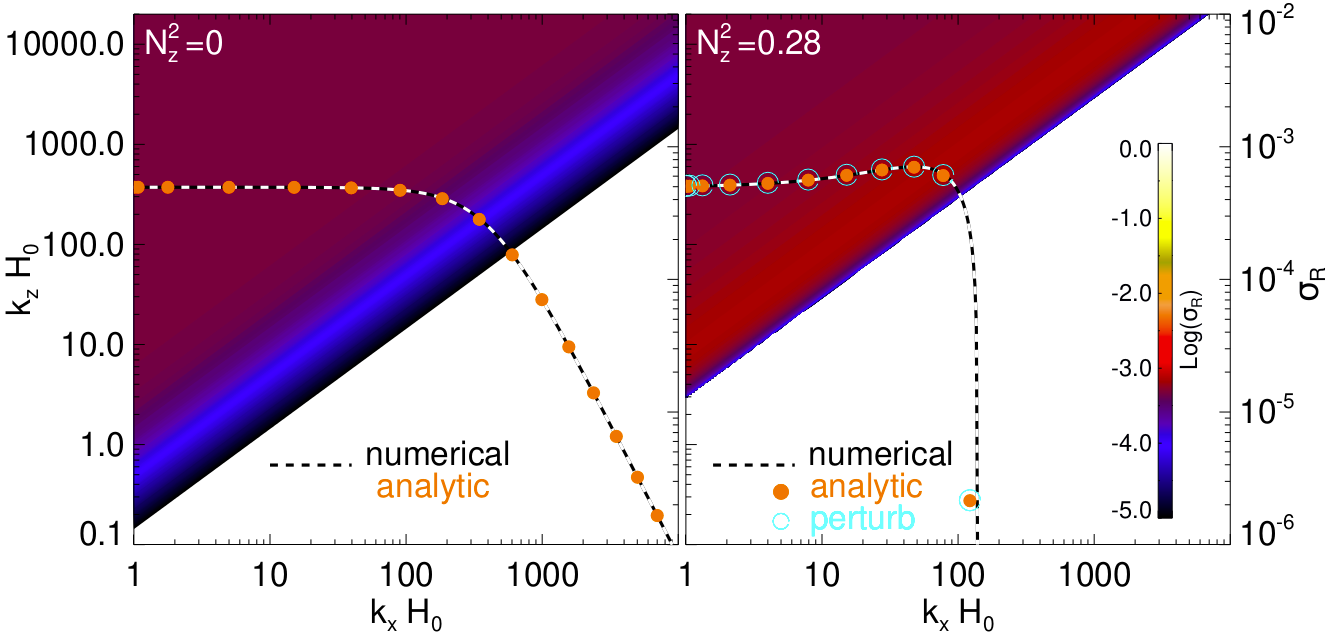}
\caption{Contours of linear growth rates of the COS in a pure gas with  $N_{r}^2=-0.01$, $H_{r}=-0.5$ and $\beta=0.1$, obtained from the numerical solution of (\ref{eq:deteqz}) for different radial and vertical wavenumbers $k_{x}$ and $k_{z}$, respectively.
\emph{Left:} calculation in absence of vertical buoyancy ($N_{z}^2=0$ and $1/H_{z}=0$, such that $H_{z} N_{z}^2=0$), with $\xi = \mu_{z}^2\left(1+N_{r}^2\right)$. The dashed curve represents the growth rate of modes with fixed $k_{z} H_{0}=400$ and varying $k_{x}$ (values indicated on the right axis). The circles are the same growth rate computed using the analytical expression (\ref{eq:grate_cos}).
\emph{Right:} calculation including vertical buoyancy $N_{z}^2=0.28$ with $H_{z}=-0.175$ [cf. (\ref{eq:nz_eval})]. The filled circles now correspond to (\ref{eq:grate}) with $\xi$ given by (\ref{eq:xi}). The open circles represent the perturbation expression (\ref{eq:gr_cos_z}). Compared to the case with $N_{z}^2=0$ we find that the unstable region shrinks to smaller wavenumbers. Furthermore, COS modes now experience a cutoff at a radial wavenumber given by (\ref{eq:kx_cut_nz}). All of the displayed growth rate curves in each panel, respectively, are in excellent agreement.}
\label{fig:cos_gas}
\end{figure*}

By taking $\partial\sigma_{R}/\partial \beta =0$, we recover the maximum growth rate of the COS \emph{over cooling times}
\begin{equation}\label{eq:cos_unstrat_growth}
    \sigma_{R,\text{max}}=-\frac{1}{4}\left|\frac{k_{z}}{k} \right| \frac{N_{r}^2}{\sqrt{1+N_{r}^2}},
\end{equation}
which occurs for $\beta=1/(\mu_{z}\sqrt{1+N_{r}^2})\approx 1/\mu_{z}$ such that the optimum cooling time depends on the wavenumbers \citep{lyra2014,latter2016}. In the limit of long radial wavelengths $k_{x}\to 0$ this yields
\begin{equation}
 \sigma_{R,\text{max}}\approx -\frac{1}{4}N_{r}^2   
\end{equation}
for $\beta\approx 1$.
Figure \ref{fig:cos_gas} (left panel) shows contours of linear growth rates for $\beta=0.1$ and $N_{r}^2=-0.01$, obtained from the numerical solution of (\ref{eq:deteqz}) in the unstratified limit (see above). The dashed curve represents the growth rate for $k_{z} H_{0} \sim 400 $ and is compared to the analytical expression (\ref{eq:grate_cos}) represented by asterisks, in excellent agreement. 

\subsubsection{Effect of vertical buoyancy}\label{sec:cos_vbuo}
We now consider the effect of vertical buoyancy on the COS. 
Figure \ref{fig:cos_gas} (right panel) shows numerically computed growth rates from the full dispersion relation (Eq. \ref{eq:deteqz}) including vertical buoyancy $N_{z}^2=0.28$ and a comparison with the analytical expression (Eq. \ref{eq:grate}) for fixed $k_z H_{0}\sim 400$.
We see that the region of instability (in blue) has shrunken towards larger wavenumbers compared to the unstratified case (the left panel). Furthermore, vertical buoyancy introduces a maximum $k_x$ beyond which growth rates drop to zero; whereas in the unstratified case growth rates decay smoothly as $k_x^{-2}$ [Eq.~(\ref{eq:grate_cos})].

To explain the above differences, we consider the cooling time $\beta$ as a small perturbation parameter\footnote{The effect of vertical (and also radial) buoyancy is eventually controlled by the value of the cooling time $\beta$.} and adopt the perturbation series 
\begin{equation*}
    \sigma = \sigma_{0} + \sigma_{1} \beta + \sigma_{2} \beta^2 + \ldots
\end{equation*}
for the complex eigenvalue, where the $\sigma_i$ are complex. Inserting this in (\ref{eq:deteqz}) and expanding the latter in orders of $\beta$ prompts the equation
\begin{align}
0 = \mu_{z}^2 + \sigma_{0}^2 + \sigma_0\left(\xi  + \sigma_{0}^2 + 2\sigma_{1}\right)\beta + \mathcal{O}\left(\beta^2\right),\label{eq:cos_exp_z}
\end{align}
 with $\xi$ given by Eq.~(\ref{eq:xi}). 
Solving (\ref{eq:cos_exp_z}) at each order of $\beta$ separately yields $\sigma_0^2=-\mu_z^2$ and $\sigma_1=\left(\mu_z^2-\xi\right)/2$, which gives the complete frequency as 
\begin{equation}\label{eq:gr_cos_z}
\begin{split}
    \sigma & = \pm i \mu_{z} 
     - \frac{1}{2}\mu_{z}^2N_{r}^2 \beta -\frac{1}{2}\mu_{x}^2 N_{z}^2\beta + \frac{1}{2} \mu_{x} \mu_{z} \beta \left(\frac{H_{z}}{H_{r}}N_{z}^2 + \frac{H_{r}}{H_{z}}N_{r}^2 \right) ,
    \end{split}
\end{equation}
to first order in $\beta$. The second term agrees with Eq.~(\ref{eq:grate_cos}) in the limit $\beta \ll 1$ and can result in the COS. The third term results in a damping effect for sufficiently large $k_{x}$ due to vertical buoyancy for $N_z^2>0$. On the other hand, the last term can either be damping or amplifying, depending on the values of the involved quantities. It is noteworthy that this term can in principle be \emph{destabilizing} for specific combinations of radial and vertical wavenumbers above and below the disc midplane, respectively, even if the disc is convectively \emph{stable} in both radial and vertical directions. 
Also, one sees that this amplification is strongest if $|k_{x}|\sim |k_{z}|$ \citep[see][ for a detailed analysis]{volponi2016}. For the fiducial parameter values adopted in this paper (\S \ref{sec:pars}) this term is indeed slightly amplifying, but small compared to the other terms, explaining the mild rise of the growth rates with increasing $k_{x}$ before these eventually drop to negative values (Figure \ref{fig:cos_gas}, the right panel).

 The cutoff radial wavenumber beyond which growth rates drop to negative values, which is induced by vertical buoyancy, is found by setting $\xi =\mu_{z}^2$ [cf. Eq. (\ref{eq:grate})], or equivalently $\sigma_1=0$, yielding the quadratic equation for $k_x$
\begin{equation}
    N_{z}^2 k_{x}^2 -\left(\frac{H_{r}N_{r}^2}{H_{z}} + \frac{H_{z}N_{z}^2}{H_{r}}\right)k_{x}k_{z} + N_{r}^2 k_{z}=0.
\end{equation}
Using $H_{r}<0$, $N_{r}^2<0$, $H_{z}<0$ and $N_{z}^2>0$, we find
\begin{equation}\label{eq:kx_cut_nz}
     k_{x,\text{cutoff}} = k_{z}\frac{H_{z}}{H_{r}}.
 \end{equation}
This yields $k_{x} H_{0} \sim 140$ for the parameters used in Figure \ref{fig:cos_gas}, which is consistent with the numerical results. We note that it can easily be shown that (for the parameter regimes considered in this paper) the values of $k_{x,\text{cutoff}}$ are much smaller than those for which the eigenvalues of inertial waves undergo a bifurcation, such that their frequencies vanish (Appendix \ref{app:iw_bifurc}). Therefore the linear COS modes studied here are not affected by this bifurcation.

\subsection{COS in the presence of dust}\label{sec:cos_dust}

In this section, we consider the COS in the presence of dust, so that $\epsilon_0>0$ and $\delta\rho_d\neq 0$. First, we will ignore the effect of a radial background pressure gradient in order to suppress the SI. This allows us to better understand the effect of dust on linear COS modes (\S \ref{sec:cos_nodp}). After this is settled, we will include a radial pressure gradient, which leads to the co-existence of the COS and the SI, but also to a new instability which turns out to be a dust-induced version of the COS in the limit of long cooling times (\S \ref{sec:cos_dp}).
Throughout this section, vertical buoyancy is neglected.

\subsubsection{Excluding a background radial dust-gas drift: dusty damping of the COS}\label{sec:cos_nodp}
We here consider the simplified case without a background radial dust-gas drift, $\vec{\mathbf{v}}_{g0}=\vec{\mathbf{v}}_{d0}=0$. This corresponds to  $\eta=0$, but since $\eta\propto\partial_rP$, this would actually imply $N_{r}^2=0$ and eliminate the COS. Thus, our setup below with $\eta=0$ but $N_r^2\neq0$ is not fully self-consistent, but is useful for understanding how dust-gas drag in the perturbed state affects the COS without complications from the SI.

\begin{figure}
\centering 
\includegraphics[width = 0.45\textwidth]{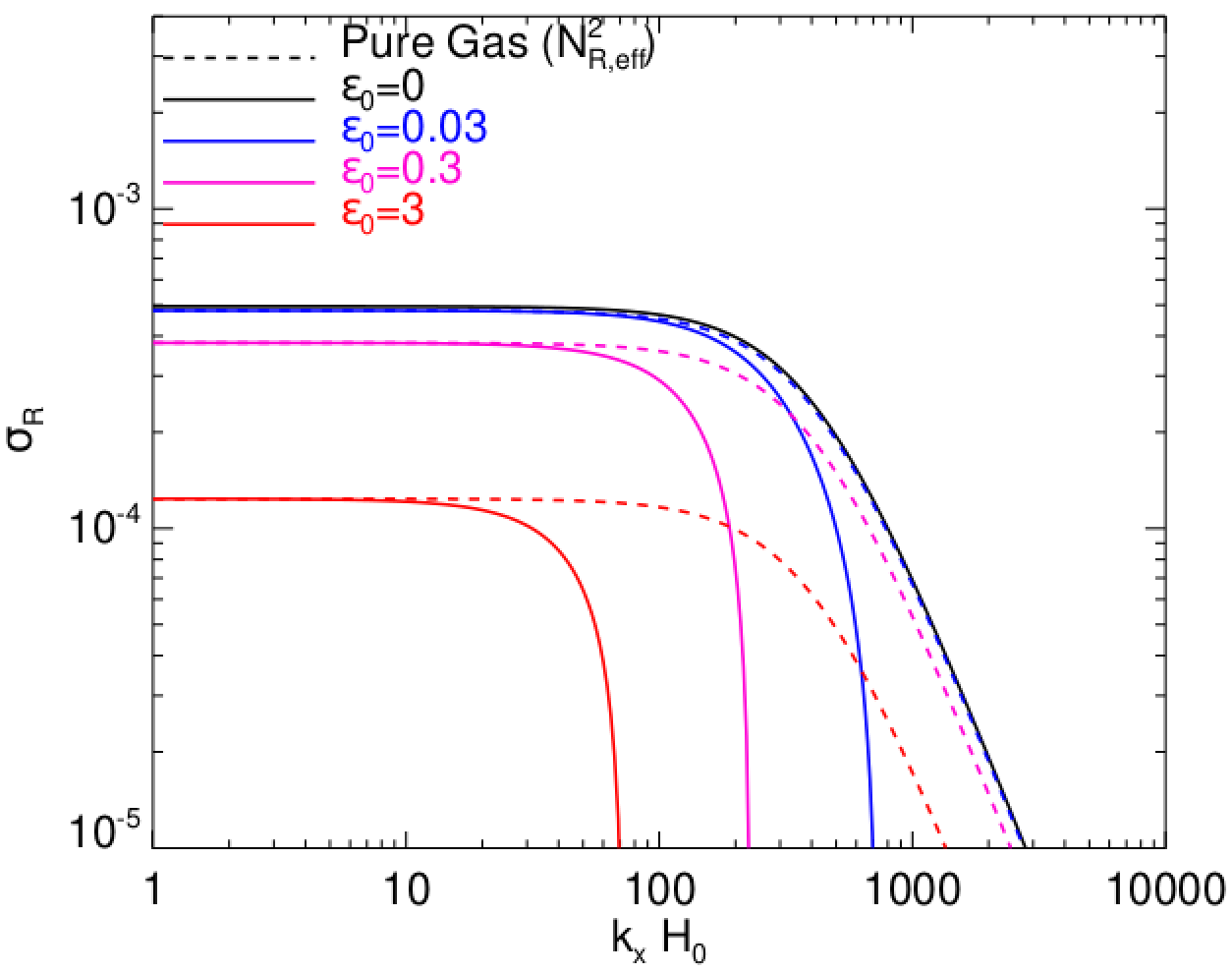}
\caption{Illustration of the effect of dust loading on linear COS modes with $k_{z} H_{0} =400$. The solid curves are linear growth rates of the COS in a dusty gas obtained from numerical solution of (\ref{eq:eigenproblem}) using $\beta=0.1$, $N_{r}^2=-0.01$, $H_{r}=-0.5$, $\tau=10^{-2}$ and various values of $\epsilon_{0}$ as indicated. Furthermore, as explained in the text, we used $\eta=0$. The solid curve with $\epsilon_{0}=0$ is the pure gas growth rate for the same parameters obtained from numerical solution of (\ref{eq:deteqz}) with $\xi=\mu_{z}^2\left(1+N_{r}^2\right)$. On the other hand, the dashed curves are also pure gas growth rates of the COS, obtained from numerical solution of (\ref{eq:deteqz}), but with $N_{r}^2$ replaced the effective buoyancy frequency (\ref{eq:nreff}). Note that the blue dashed curve corresponding to $\epsilon_{0}=0.03$ almost indistinguishable from the black solid curve corresponding to $\epsilon_{0}=0$.}
\label{fig:cos_dustdamp}
\end{figure}

In the absence of a radial background dust-gas drift, we find the effect of dust on linear COS modes to be always damping. This damping can be described in terms of two distinct features. Compared to the dust-free limit, dust lowers the overall level of the growth rates, depending on the dust-to-gas ratio $\epsilon_{0}$. In addition, the presence of dust leads to a sharp cutoff at a critical radial wavenumber, such that modes with larger wave numbers decay, similar to the effect of vertical buoyancy [cf. Eq.~(\ref{eq:kx_cut_nz})].
This is illustrated in Figure \ref{fig:cos_dustdamp}, where the solid lines describe  
growth rates of COS modes with $k_{z} H_{0} =400$, $\beta=0.1$, $\tau=10^{-2}$ and different values of $\epsilon_{0}$. As $\epsilon_0$ increases, growth rates decrease and the instability is limited to longer radial scales. The dashed curves will be explained below. 

\begin{figure}
\centering 
\includegraphics[width = 0.4\textwidth]{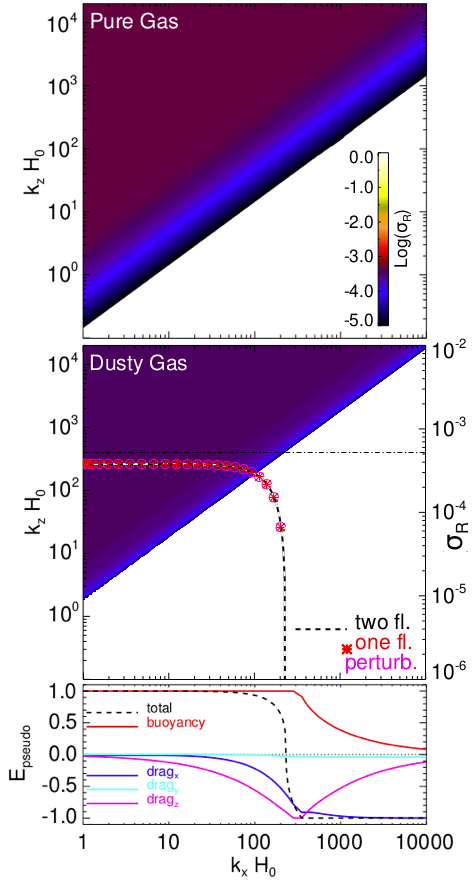}
\caption{\emph{Upper panels:} Contours of linear growth rates of the COS in a pure gas (same as Figure \ref{fig:cos_gas}) and a dusty gas, the latter being obtained from numerical solution of (\ref{eq:eigenproblem}) with $\beta=0.1$, $N_{r}^2=-0.01$, $H_{r}=-0.5$, $\epsilon_{0}=0.3$, $\tau=10^{-2}$, without the effect of a radial pressure gradient $\eta=0$. The dashed curve represents growth rates with fixed $k_{z} H_{0}=400$ (values are indicated by the right hand axis). The red asterisks and magenta circles are the corresponding growth rates obtained from the one-fluid Eqs. (\S \ref{app:onefl}) and the perturbation expression (\ref{eq:cos_exp}), respectively, both in excellent agreement with the numerical two-fluid result. \emph{Lower panel:} Pseudo-energy decomposition of the linear modes with $k_{z} H_{0}=400$ (Appendix \ref{app:pseudo}), showing that these are powered by radial buoyancy (as expected for the COS) and are damped mainly by radial and vertical dust-gas drag.}
\label{fig:cos_dust}
\end{figure}

Within our two-fluid model (\S \ref{sec:hydromodel}), the damping effects just described are necessarily related to the drag-force terms $\propto \left(\delta \vec{\mathbf{v}}_{g} -\delta \vec{\mathbf{v}}_{d}\right)$ in Eqs. (\ref{eq:linug})-(\ref{eq:linvg}). This is demonstrated in Figure \ref{fig:cos_dust}, where we plot the contours of growth rates obtained from numerical solution of (\ref{eq:eigenproblem}). The over-plotted curve represents the growth rates for $k_{z} H_{0} \sim 400$ which are compared to growth rates resulting from the one-fluid model (Appendix \ref{app:onefl}, see also \S \ref{sec:cos_onef} below), as well as an analytical expression obtained below. All of the compared growth rates are practically identical in this case. The lower panel shows the relevant energies involved in the excitation and damping of COS modes (Appendix \ref{app:pseudo}) for the same vertical wavenumber. We find that it is essentially the radial and vertical drag forces that damp the modes, whereas azimuthal drag forces appear to be negligible.
%

Next, we wish to verify the above results by means of a simple perturbation analysis.
We start with the dispersion relation of the system of Eqs. (\ref{eq:linrhog})-(\ref{eq:linwd}). The full expression is lengthy and will not be displayed here.
Next we consider the Stokes number $\tau \ll 1$ as a perturbation parameter and adopt the perturbation series
\begin{equation*}\label{eq:sig_exp}
    \sigma = \sigma_{0} + \sigma_{1} \tau + \sigma_{2} \tau^2 + \ldots
\end{equation*}
for the eigenvalue.
Inserting this in the full dispersion relation and expanding the latter in orders of $\tau$ yields
\begin{equation}
\begin{split}\label{eq:cos_exp}
& c_{0}(\epsilon_{0},k_{x},k_{z},\beta,N_{r}^2,\sigma_{i})\\
\quad & + c_{1}(\epsilon_{0},k_{x},k_{z},\beta,N_{r}^2,\sigma_{i})\,  \tau\\
\quad & + c_{2}(\epsilon_{0},k_{x},k_{z},\beta,N_{r}^2,\sigma_{i})\, \tau^2 + \mathcal{O}\left(\tau^3\right) =0,
\end{split}
\end{equation}
where $i=1,2,\ldots$ and
where the coefficient terms $c_{0}-c_{2}$ generally contain a vast number of terms.
We further assume that the cooling time is small such that $\beta \ll 1$ and the gas is nearly (but not exactly) isothermal.
Solving (\ref{eq:cos_exp}) order by order in $\tau$ gives
\begin{align}
\sigma_{\text{cool}} & = -\frac{1}{\beta} + \mu_{z}^2\frac{\beta N_{r}^2}{1+\epsilon_{0}} -\mu_{z}^2\frac{\epsilon_{0}\tau N_{r}^2}{\left(1+\epsilon_{0}\right)^2} +\mathcal{O}\left(\tau^2\right), \label{eq:cool_tau}\\
\begin{split}
\sigma_{\text{COS}} & =    \pm i\mu_{z} -\frac{1}{2}\mu_{z}^2\frac{\beta  N_{r}^2}{1+\epsilon_{0}} - \mu_{x}^2\frac{\epsilon_{0}\tau}{2 \left(1+\epsilon_{0}\right) } \\
\quad & - i \frac{\epsilon_{0} \tau N_{r}^2}{4 \beta \left(1+\epsilon_{0}\right)^2} \mu_{z}\mu_{x}^2+\mathcal{O}\left(\tau^2\right)\label{eq:cos_tau},
\end{split}
\end{align}
representing the cooling mode and the two COS modes\footnote{The remaining modes, which are not presented, are generally decaying on account of dust and are therefore not of interest here.}, affected by tightly coupled dust, respectively. 
Since we assume $\beta \ll 1$, we expanded\footnote{We note that the cooling mode (\ref{eq:cool_tau}) is expressed here as truncated Laurent series to the next leading order. Therefore also negative powers of $\beta$ occur, in contrast to a Taylor series, which is not applicable in this particular case.} all terms to the leading order in $\beta$. 
%
The terms $\propto \tau$ have in addition been expanded to leading order in $N_{r}^2$ to simplify, but not qualitatively affect the result. Since $|N_{r}^2|\ll 1$ the first term in the cooling mode (\ref{eq:cool_tau}) always dominates over the remaining terms such that stability is assured. The frequency and the growth rate of the COS modes (\ref{eq:cos_tau}) agree with (\ref{eq:freq}) and (\ref{eq:grate}), respectively, in the limit of small $\beta$ and $\epsilon_{0}\to 0$.

In the above expressions, the radial buoyancy frequency appears only in form of a product with $1/(1+\epsilon_{0})$, which is the equilibrium mass fraction of gas [cf. (\ref{eq:fg})]. This suggests the definition of an \emph{effective} radial buoyancy frequency
\begin{equation}\label{eq:nreff}
    N_{r,\text{eff}}^2 \equiv \frac{N_{r}^2}{1+\epsilon_{0}}
\end{equation}
of the dusty gas.
Thus, we find here that the COS in a dusty gas is excited by a buoyancy frequency (\ref{eq:nreff}) that is effectively reduced by dust loading, leading to an overall reduction of the growth rates. Eq.~(\ref{eq:nreff}) implies that $\epsilon_{0}$-values of order unity are required to efficiently damp the COS on larger wavelengths.

Returning to Figure \ref{fig:cos_dustdamp}, the dashed curves are growth rates of a pure gas with accordingly varying effective buoyancy frequency, i.e. Eq.~(\ref{eq:cos_unstrat_growth}) where we replaced $N_r^2\to N_{r,\text{eff}}^2$ given by (\ref{eq:nreff}). For small $k_{x}$ the growth rates agree very well with the dusty cases. 
In the one-fluid model this effect of dust-loading is more clearly manifested (see \S \ref{sec:cos_onef_nreff}). 

Furthermore, the real-valued term $\propto \tau$ in (\ref{eq:cos_tau}) results in the growth rates' cutoff at larger $k_{x}$.
The cutoff wavelength can readily be obtained by setting the real part of $\sigma_{\text{COS}}$ equal to zero:
\begin{equation}\label{eq:kxcut}
    k_{x,\text{cutoff}}=\sqrt{-\frac{\beta N_{r}^2}{\epsilon_{0}\tau}} k_{z}.
\end{equation}
As expected, the value of the cutoff wavenumber decreases with increasing $\epsilon_{0}\tau$ (as also seen in Figure \ref{fig:cos_dustdamp}) and (of course) only exists for $N_{r}^2<0$. 
%
%

Since the third term in (\ref{eq:cos_tau}), which leads to the cutoff, is independent of $N_{r}^2$, it represents a general effect of dust on inertial waves. In Appendix \ref{app:dusty_iw}, we develop a reduced model to derive the same dusty damping term of isothermal (i.e. not subject to buoyancy) inertial waves. Furthermore, in Section (\ref{sec:cos_onef_cutoff}) we provide an explanation for this effect using the one-fluid formalism (Appendix \ref{app:onefl}).

Alternatively, we can also assume that the \emph{inverse} cooling time $\widetilde{\beta}\equiv \beta^{-1} \ll1$ is small, such that the gas is nearly adiabatic.
By performing again the series expansion in $\tau$ as described above we again find the cooling mode in addition to the pair of inertial waves:
\begin{align}
    \sigma_{\text{cool}} & = -\frac{1+\epsilon_{0}}{1+\epsilon_{0}+N_{r}^2}\frac{1}{\beta} -\frac{\epsilon_{0}\tau N_{r}^2}{\left(1+\epsilon_{0}\right)^2}  + \mathcal{O}\left(\tau^2\right)\label{eq:cool_tau_2},\\
    \begin{split}
    \sigma_{\text{COS}} & = \pm i \mu_{z}\sqrt{\frac{1+\epsilon_{0}+N_{r}^2}{1+\epsilon_{0}}} - \frac{N_{r}^2}{2 \beta \left(1+\epsilon_{0} +N_{r}^2\right)} \\ \quad & -\mu_{x}^2 \frac{\epsilon_{0}\tau }{2\left(1+\epsilon_{0}\right)} + i \frac{3 \epsilon_{0}\tau N_{r}^2}{4 \beta \left(1+\epsilon_{0} \right)^2} \frac{\mu_{x}^2}{\mu_{z}}+ \mathcal{O}\left(\tau^2\right)\label{eq:cos_tau_2}.
    \end{split}
\end{align}
Here we expanded all terms to leading order in $\widetilde{\beta}$ and in addition, the terms $\propto \tau$ to leading order in $N_{r}^2$, prior to replacing $\widetilde{\beta} \to 1/\beta$. 
We now find that if $\epsilon_{0}\to 0$, the real and imaginary parts of Eq.~(\ref{eq:cos_tau_2}) agree with (\ref{eq:freq}) and (\ref{eq:grate}) in the limit $\beta \gg 1$ and $\epsilon_{0}\to 0$, as expected. For $\epsilon_{0} >0$, the leading growth rates again correspond to an effectively reduced radial buoyancy frequency (\ref{eq:nreff}). 

Although the focus of this section lies on the COS, we briefly discuss the behavior of the cooling mode (\ref{eq:cool_tau_2}).
A finite coupling time between dust and gas can in the limit of long cooling times ($\beta\gg 1$) result in a quasi-thermal instability by rendering the cooling mode unstable. This instability feeds off the perturbed relative drift velocity between dust and gas induced by radial buoyancy, the latter acting only on the gas, similar to pressure.
Now, if $N_{r}^2<0$ and $\beta$ is sufficiently large, i.e. if 
\begin{equation}
\beta > -\frac{\left(1+\epsilon_{0}\right)^2}{\epsilon_{0}\tau N_{r}^2},    
\end{equation}
then the second term in (\ref{eq:cool_tau_2}) dominates, effectively rendering the system thermally unstable. The resulting growth rates are small though, and are maximal for $\epsilon_0=1$, such that $\sigma_{R}\sim |N_{r}^2| \tau \sim h_{0}^2 \tau$. On the other hand, if $N_{r}^2>0$ the system is thermally stable and the cooling rate of the dusty gas is enhanced compared to that of a pure gas. Similar to the effect of dust on  buoyancy described above, also the effect of dust on cooling is more clearly manifested in the one-fluid model (see \S \ref{sec:cos_onef_cool})

\subsubsection{Including a background radial dust-gas drift: the DCOS}\label{sec:cos_dp}

\begin{figure}
\centering 
\includegraphics[width = 0.38\textwidth]{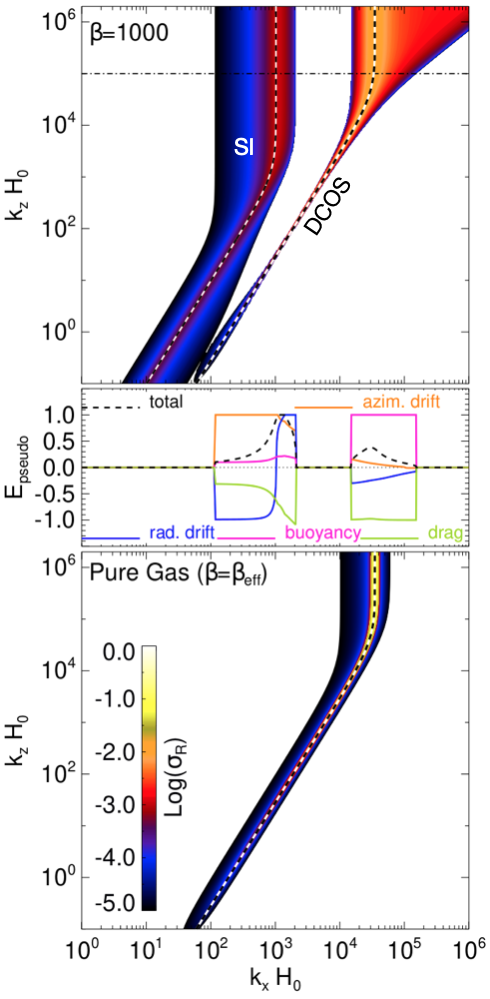}
\caption{The upper panel shows contours of linear growth rates of a dusty gas with $\tau=10^{-2}$, $\epsilon_{0}=0.03$, $\eta=0.0025$, $N_{r}^2=-0.01$, $H_{r}=-0.5$ and $\beta=1000$. In addition to the SI, the DCOS appears, which is powered by radial buoyancy and the radial pressure gradient $\eta$, as explained in the text. This is confirmed by the pseudo-energy decomposition, displayed in the middle panel for fixed $k_{z} H_{0} = 1000$. We note that in principle there is a small amplifying effect of radial buoyancy on the SI (cf. \S \ref{sec:si_cool}) as well. However, this effect is much smaller than suggested by the plot since we display cube roots of the pseudo energies to increase the visibility of small contributions. 
 For reference, the left dashed curve in the upper panel is the resonant wavenumber (\ref{eq:kres_adia}), which is practically identical to (\ref{eq:kres_iso}) as explained in \S \ref{sec:si_cool}. The right dashed curve will be explained in \S \ref{sec:cos_onef_dcos}. The bottom panel shows linear growth rates of a pure gas subject to the effective cooling time (\ref{eq:beteff}). This plot confirms the principle amplification mechanism by dust on the COS resulting in the DCOS, which is explained in detail in Appendix \ref{app:dcos}.}
\label{fig:adiab_instab}
\end{figure}

\begin{figure}
\centering 
\includegraphics[width = 0.35\textwidth]{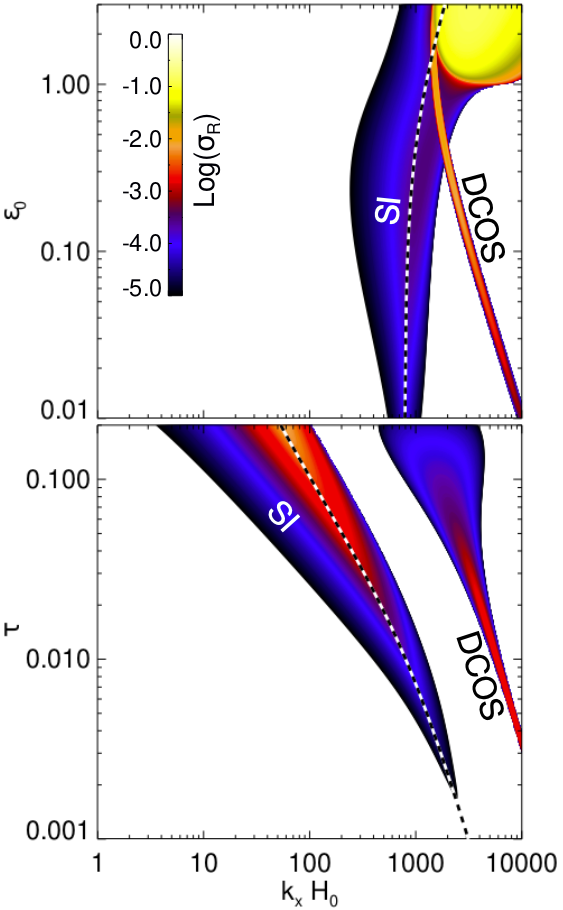}
\caption{Contours of linear growth rates of a dusty gas subject to a radial buoyancy frequency $N_{r}^2=-0.01$, $H_{r}=-0.5$, a cooling time $\beta=1000$ and a radial pressure gradient $\eta=0.05^2$. The prevalent instabilities for these parameters are the SI and the DCOS (as explained in the text), as labeled in the figure. Note that the classical COS is negligible with dimensionless growth rates $<10^{-5}$. The growth rates of the DCOS and the SI are displayed for varying $\epsilon_{0}$ (fixed $\tau=0.01$) in the top panel and varying $\tau$ (fixed $\epsilon=0.03$) in the bottom panel. The dashed curves are the resonant SI wave numbers resulting from (\ref{eq:krdi}), and solved for the corresponding quantity on the vertical axis in both panels.}
\label{fig:adiab_instab_2}
\end{figure}

We now include the effect of a background radial pressure gradient ($\eta\neq0$), which is self-consistent with $N_r^2\neq0$, as required for the COS. However, as a background radial pressure gradient induces a radial dust-gas drift, the system is also susceptible to the SI. The co-existence of the two instabilities will be discussed in \S \ref{sec:cos_vs_si}. 

In this section we focus on long cooling times $\beta \gg 1$ such that the classical (pure gas) COS is negligible. Instead, we will show here that in the presence of dust and a radial background pressure gradient, the COS is modified such that in principle it can attain considerable growth rates for arbitrarily large cooling times $\beta \gg 1$, including the adiabatic limit, with maximum growth rates (across all wavenumbers) actually increasing with increasing $\beta$. 
We, therefore, refer to the corresponding instability as the Dusty Convective Over-Stability (DCOS), which is illustrated in Figure \ref{fig:adiab_instab} for $\beta=1000$ (upper panel), where the DCOS (the right `branch') co-exists with the SI (the left `branch'). 

First of all, we note that a cooling time $\beta=1000$ implies that the classical COS is practically negligible. That is, from (\ref{eq:grate_cos}) we find $\sigma_{R}\sim 10^{-5}$ for the parameters used in Figure \ref{fig:adiab_instab}. This estimate does not include the damping effect of dust described by (\ref{eq:cos_tau_2}), such that the actual growth rates are expected to be even smaller.
On the other hand, the DCOS requires $N_r^2<0$ and $\eta\neq0$, as well as dust. The first two criteria are prerequisites for the COS and the SI, respectively. 
 Similar to the COS, the radial buoyancy term in the momentum equation (\ref{eq:contvg}) eventually drives the DCOS. As such, it is, as its naming suggests, more closely related to the COS than to the SI. 
Indeed, when gradually increasing the cooling time $\beta$ we find that one of the two COS modes is amplified within an increasingly narrow wavenumber range  for decreasing frequency $|\sigma_{I}|<1$, whereas all other wavenumbers (outside of this range), as well as the entire other COS mode gradually damp, as expected anyways for increasing $\beta$. For $\beta=1000$ the band of unstable wavenumbers is rather narrow for smaller wavenumbers, as seen in Figure \ref{fig:adiab_instab}. The principle mechanism driving the DCOS can be well explained using the one-fluid model and will be discussed in \S \ref{sec:cos_onef_dcos}.

Returning to Figure \ref{fig:adiab_instab}, the middle panel displays the pseudo-energy decomposition (\S \ref{app:pseudo}), corresponding to modes with fixed $k_{z} H_{0}\sim 1000$. This plot shows that radial buoyancy is the driving force of the DCOS, which confirms our interpretation as it being a modified (by dust) version of the COS. Since here $\epsilon_{0}\ll 1$, the SI modes, represented by the left `branch', are mainly powered by azimuthal drift\footnote{Only for larger values $\epsilon_{0}\gtrsim 1$ the mode energies are dominated by radial drift.} \citep[as first noted by][]{lh2022}. The bottom panel will be explained in \S \ref{sec:cos_onef_dcos} below.

In addition, Figure \ref{fig:adiab_instab_2} shows growth rates of the DCOS and the SI for varying $\epsilon_{0}$ (top panel) and $\tau$ (bottom panel) at fixed $k_{z} H_{0}=1000$. We find that as long as $\epsilon_{0}\lesssim1$ the growth rates of the DCOS exceed those of the SI. Furthermore, with increasing $\epsilon_{0}$ the DCOS shifts toward the SI in wavenumber space, such that the instabilities increasingly overlap.
On the other hand, with increasing Stokes number $\tau$ both instabilities shift to smaller $k_{x}$-values. While the SI growth rates increase with increasing $\tau$ (at least within the displayed range), those of the DCOS eventually vanish due to increased damping by drag forces (cf. Appendix \ref{app:dusty_iw}). This can be confirmed by considering the modes' pseudo-energies (not shown).

 \subsection{One-fluid description of the COS in a dusty gas}\label{sec:cos_onef}

Many of the results presented in \S \ref{sec:cos_nodp} and \S \ref{sec:cos_dp} can be understood by adopting the one-fluid formalism of a dusty gas, as outlined in Appendix \ref{app:onefl}. Below we will provide explanations for the damping of COS modes in absence of a radial pressure gradient $\eta=0$ (cf. \S \ref{sec:cos_nodp}) from the perspective of the single-fluid approximation. Similarly, the effect of dust on the cooling mode as well as the mechanism driving the DCOS will be discussed (cf. \S \ref{sec:cos_dp}).

\subsubsection{Effective buoyancy of a dusty gas}\label{sec:cos_onef_nreff}
By invoking the one-fluid momentum equation (\ref{eq:contv1f}), the overall reduction of COS growth rates due to dust, as seen in Figure \ref{fig:cos_dustdamp}, can be readily understood as being the result of an effectively reduced radial buoyancy frequency (\ref{eq:nreff}),
the magnitude of which is the driving force of the COS, as it couples to the epicyclic fluid motion in Eq.~(\ref{eq:contv1f}).
This is because the radial buoyancy term in the single fluid approximation (\ref{eq:contv1f}) reads
\begin{equation}
\begin{split}
     F_{N_{r}^2} & \equiv -H_{r}N_{r}^2\frac{\delta \rho_{g}}{\rho}\\
    \quad & = -H_{r}\frac{N_r^2}{(1+\epsilon)}\frac{\delta \rho_{g}}{\rho_{g0}},
    \end{split}
\end{equation}
where $\rho=\rho_{g0}(1+\epsilon)$ is the total density and $\epsilon=\rho_d/\rho_{g0}$. This term is identical to the buoyancy term in the pure gas momentum equation (\ref{eq:contvg}) with $N_{r}^2$ replaced by $N_r^2/(1+\epsilon)$ given by (\ref{eq:nreff}). Note that an equivalent reduction of the vertical buoyancy frequency $N_{z}^2$ occurs. The reduced buoyancy frequency is a result of dust loading and is not related to the (finite) coupling time $\tau$ between dust and gas. In fact, this reduction should be most severe for perfectly coupled dust ($\tau\to0$). 

\subsubsection{Effective compressibility of a dusty gas}\label{sec:cos_onef_cutoff}

On the other hand, the sharp cutoff at larger wavenumbers $k_{x}$ of COS growth rates, which is also seen in Figure \ref{fig:cos_dustdamp} (and Figure \ref{fig:cos_dust}), is a consequence of a finite coupling time between dust and gas. That is, for $\tau\to 0$, the cutoff vanishes (for both one-fluid and two-fluid growth rates) and the growth rates asymptotically approach those of the pure gas for large $k_{x}$. Within the one-fluid model, we identify the term 
in the linearised radial and vertical momentum Eqs. (\ref{eq:dcontu1f}) and (\ref{eq:dcontw1f}):
\begin{equation}\label{eq:fbulk}
\delta\vec{\bm{F}}_{\nu_{b}} = i \frac{\vec{\bm{k}}}{k^2}\frac{1+\epsilon_{0}}{\epsilon_{0}\tau }  i \left(k_{x} \delta u + k_{z} \delta w\right) \equiv  \frac{1}{1+\epsilon_{0}}\vec{\nabla} \left[ \nu_{b} \left(\vec{\nabla}\cdot \delta\vec{\mathbf{v}}\right)\right]
\end{equation} 
with
\begin{equation}\label{eq:nub}
    \nu_{b} \equiv \frac{1}{k^2}\frac{\left(1+\epsilon_{0}\right)^2}{\epsilon_{0}\tau }
\end{equation}
 as the origin of the cutoff. This term stems from the second term in Eq.~(\ref{eq:dpres1f}) when using it to eliminate $\delta P$ in the momentum equations. It is the only term that contains the Stokes number $\tau$ in the one-fluid momentum  equations. 

The second equality in Eq.~(\ref{eq:fbulk}) shows that $\delta\vec{\bm{F}}_{\nu_{b}}$ can be loosely 
interpreted as a `bulk viscous' stress with `bulk viscosity' $\nu_b$. In standard fluid dynamics, bulk viscosity is associated with energy losses due to finite compressibility. 
We, therefore, infer that the compressibility of a dust-gas mixture gives rise to an effective `bulk viscosity' of the single fluid, which damps the COS modes at large $k_{x}$. It should be noted that the above equations are not valid in the limit of vanishing $\epsilon_0\tau$, which would otherwise result in a divergence of $\nu_{b}$ (see Appendix \ref{app:onefl}).

\begin{figure}
\centering 
\includegraphics[width = 0.41\textwidth]{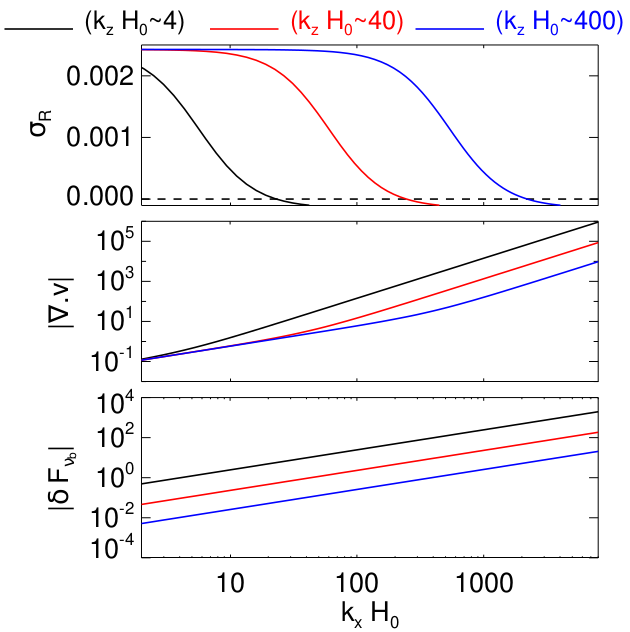}
\caption{Illustration of the relevant terms in the one-fluid model (Appendix \ref{app:onefl}) that lead to a cutoff of the growth rates of linear COS modes at the critical radial wavenumber (\ref{eq:kxcut}). The upper panel displays growth rates for three different vertical wavenumbers $k_{z}$ as indicated, with $\tau=10^{-2}$, $\epsilon_{0}=0.03$, $\beta=1$, and $N_{r}^2=-0.01$ with $H_{r}=-0.5$. The middle and bottom panels describe, respectively, the corresponding divergence of the center of mass velocity (\ref{eq:cms}) and the forcing term (\ref{eq:fbulk}), representing the damping due to an effective `bulk viscosity' (\ref{eq:nub}), resulting from the non-vanishing compressibility of a dusty gas. In particular, while the effective bulk viscosity (\ref{eq:nub}) decreases with increasing radial wave number $k_{x}$, the velocity divergence $\nabla\cdot \vec{\mathbf{v}}$ increases sufficiently fast to overcompensate the former decrease and results in a damping effect.}
\label{fig:cos_tau}
\end{figure}

For illustration, we plot in Figure \ref{fig:cos_tau} COS growth rates, resulting from numerical solution of the linearised one-fluid Eqs. (\ref{eq:dpres1f}), (\ref{eq:dconteps})---(\ref{eq:dcontrhog1f}), with $\tau=10^{-2}$, $\epsilon_{0}=0.03$ and $\beta=1$, for three different values of $k_{z}$ as indicated. As already indicated in Figure \ref{fig:cos_dust}, the one-fluid growth rates of the COS are practically identical to the corresponding two-fluid growth rates for the parameter values considered here. The remaining panels show the magnitudes of the velocity divergence $|\vec{\nabla} \cdot \delta\vec{\mathbf{v}}|$ and the `viscous' term $|\delta \vec{\bm{F}}_{\nu_{b}}|$, respectively, for the same wavenumbers. 
Thus, although $\nu_{b}$ decreases with increasing $k_{x}$, the increase in the magnitude of $\vec{\nabla} \cdot \vec{\mathbf{v}}$ is over-compensating such that the resulting `viscous' term increases with $k_{x}$.
For the parameters used in this figure the radial cutoff wavenumber (\ref{eq:kxcut}) yields $k_{x,\text{cutoff}}\approx 5.8 k_{z}$, in good agreement with the wavenumbers for which the drawn growth rates drop to negative values.

\subsubsection{Effective cooling time of a dusty gas}\label{sec:cos_onef_cool}
Complementary to our discussion at the end of \S \ref{sec:cos_nodp}, and also in preparing our discussion of the DCOS in \S \ref{sec:cos_onef_dcos}, we briefly consider the dynamical effect of dust on gas cooling.
When considering the one-fluid model, the presence of partially coupled dust (i.e. $\epsilon_{0}\tau>0$) gives rise to an \emph{effective} cooling time (\ref{eq:beta_eff_full}) of the dusty gas. The full expression is fairly involved and depends on the disc background structure through $\eta$, $N_{r}^2$, etc., and which enters the entropy equation in the one-fluid model (\ref{eq:dcontrhog1f}). 

Adopting the same approximations as in \S \ref{sec:cos_nodp} (i.e. $\beta \gg 1$ and $\eta=0$, $N_{z}^2=0$, $H_{z}^{-1}=0$, while $N_{r}^2 \neq 0$), the effective cooling time is given by
\begin{equation}\label{eq:beta_eff_a}
\beta_{\text{eff}} = \left[\frac{1}{\beta} +\mu_{z}^2 \frac{\epsilon_{0} \tau N_{r}^2}{(1+\epsilon_{0})^2}  \right]^{-1}.
\end{equation}
Notice that for $k_x\to0$ and $|N_r^2|\ll 1$, the cooling mode growth rate in the two-fluid model, as given by Eq.~(\ref{eq:cool_tau_2}), is $\sigma_\mathrm{cool}\simeq -1/\beta_\text{eff}$, i.e. the standard cooling rate of a gas with a cooling time $\beta_{\text{eff}}$. For general wavenumbers,  exact agreement is not to be expected since Eq.~(\ref{eq:dcontrhog1f}) describes the entropy evolution of the dusty fluid with velocity $\vec{\mathbf{v}}$ given by (\ref{eq:cms}), rather than only the gas with velocity $\vec{\mathbf{v}}_{g}$.

\subsubsection{The DCOS}\label{sec:cos_onef_dcos}

The occurrence of the DCOS --- a dust-induced version of the COS discovered here and described in \S \ref{sec:cos_dp} as well as Appendix \ref{app:dcos} --- can be explained within the one-fluid model as follows.
We first determine that within the one-fluid model, the terms in the entropy equation (\ref{eq:contrhog1f}) that are necessary for its onset are the second ($\propto 1/H_{r}$) and fourth ($\propto \eta$) terms on the right hand side. The former term describes the advection of entropy resulting from the perturbation velocities and is also necessary for the existence of the COS. The latter term results from the equilibrium radial drift between dust and gas.  

By again considering the effective cooling time within the one-fluid model (\ref{eq:beta_eff_full}) we now have
\begin{equation}\label{eq:beteff}
    \beta_{\text{eff}} = \left[ \frac{1}{\beta} + i \frac{2 \eta \tau \epsilon_{0} k_{x}}{(1+\epsilon_{0})^2} \right]^{-1},
\end{equation}
which is complex. The background dust-induced gas drift transports gas density or entropy perturbations, which contributes to an effective cooling. Here we assume that $k_x \gg 1$ (as also seen in Figure \ref{fig:adiab_instab}), so that we can neglect the buoyant contribution to $ \beta_{\text{eff}}$ [cf. Eq.~(\ref{eq:beta_eff_a})]. 

In Appendix \ref{app:dcos}, we provide a detailed explanation of the  mechanism underlying the DCOS. In short, the introduction of dust results in a phase lag between the gas density perturbation and the gas cooling term, induced by the imaginary part of (\ref{eq:beteff}), which requires finite values of $\eta$, $\tau$, $\epsilon_{0}$ and $k_{x}$. This phase lag enhances the maximum growth rates of the COS, and at the same time shrinks the unstable region in wavenumber space, resulting in the DCOS. The phase lag increases with increasing values of $\beta$, $\tau$ and $k_{x}$, and takes largest values if $\epsilon_{0}=1$.

Returning to Figure \ref{fig:adiab_instab}, the bottom panel shows the maximum growth rates in a pure gas subject to the cooling time (\ref{eq:beteff}), which indeed largely reproduces the DCOS as seen in the top panel of the same Figure. The dashed curve delineating the DCOS (top panel) can be obtained as follows. As outlined in Appendix \ref{app:dcos}, the optimal cooling time of the COS (and the DCOS) is given by $|\beta_{\text{eff}}| \propto 1/\sigma_{I}$, where $\sigma_{I}$ denotes the frequency of pure gas inertial waves. In the adiabatic limit $\beta\to \infty$ we have $\sigma_{I} \approx \mu_{z}$ (\S \ref{sec:cos_novbuo}). Using (\ref{eq:beteff}) we then find
\begin{equation}
   k_{z,\text{DCOS}} = \frac{k_{x}^2}{\sqrt{\frac{\left(1+\epsilon_{0}\right)^4}{\left(2 \eta \tau \epsilon_{0}\right)^2}-k_{x}^2}}.
\end{equation}
Although the DCOS clearly does not fit into the concept of RDIs (cf. \S \ref{sec:res_drag_instab}), the close resemblance to the resonant SI wavenumber (\ref{eq:kres_iso}), which is also evident from Figure \ref{fig:adiab_instab}, is intriguing. This resemblance is a consequence of the similar criteria from which these wavemumbers have been derived. That is, in both cases quantities which are directly related to dust-gas drift are required to match the frequency of inertial waves.
Moreover, since $\tau \ll 1$ we have  $\delta u_{0} \sim -2 \eta \tau /(1+\epsilon_{0})$, such that for $\epsilon_{0} \gg 1$ $k_{z,\text{DCOS}}$ becomes equal to (\ref{eq:kres_iso}).


\subsection{SI in the presence of a finite cooling time}\label{sec:si_cool}
In this section, we turn our attention to the SI and examine how it is modified by gas buoyancy. Since the original SI of \cite{youdin2005} is incompressible in the gas (i.e. $\delta\rho_g=0$), one may expect buoyancy to have a negligible effect. However, we shall find that this need not be the case.
We neglect vertical buoyancy $N_z^2=0$ with $H_z^{-1}=0$. This is assumed to be the most relevant situation because the SI is expected to operate in the disc midplane where the dust settles, and where this condition applies. In general, however, radial buoyancy may be present: $N_{r}^2 \neq 0$. 


Since the SI is powered by the radial and azimuthal equilibrium drift between gas and dust (cf. Figure \ref{fig:adiab_instab})
 one might expect the energy equation to play little role on this instability.
%
Indeed, in the limit of small and large cooling times $\beta \ll 1$ and $\beta \gg 1$, respectively, we find no influence of radial buoyancy on linear SI modes.
However, not so at intermediate cooling times $\beta \sim 1$, where we find a damping effect. 
Here we consider the case $N_{r}^2>0$. If $N_{r}^2<0$ 
the SI and the COS can co-exist, which will be discussed in the next section.

\begin{figure*}
\centering 
\includegraphics[width = \textwidth]{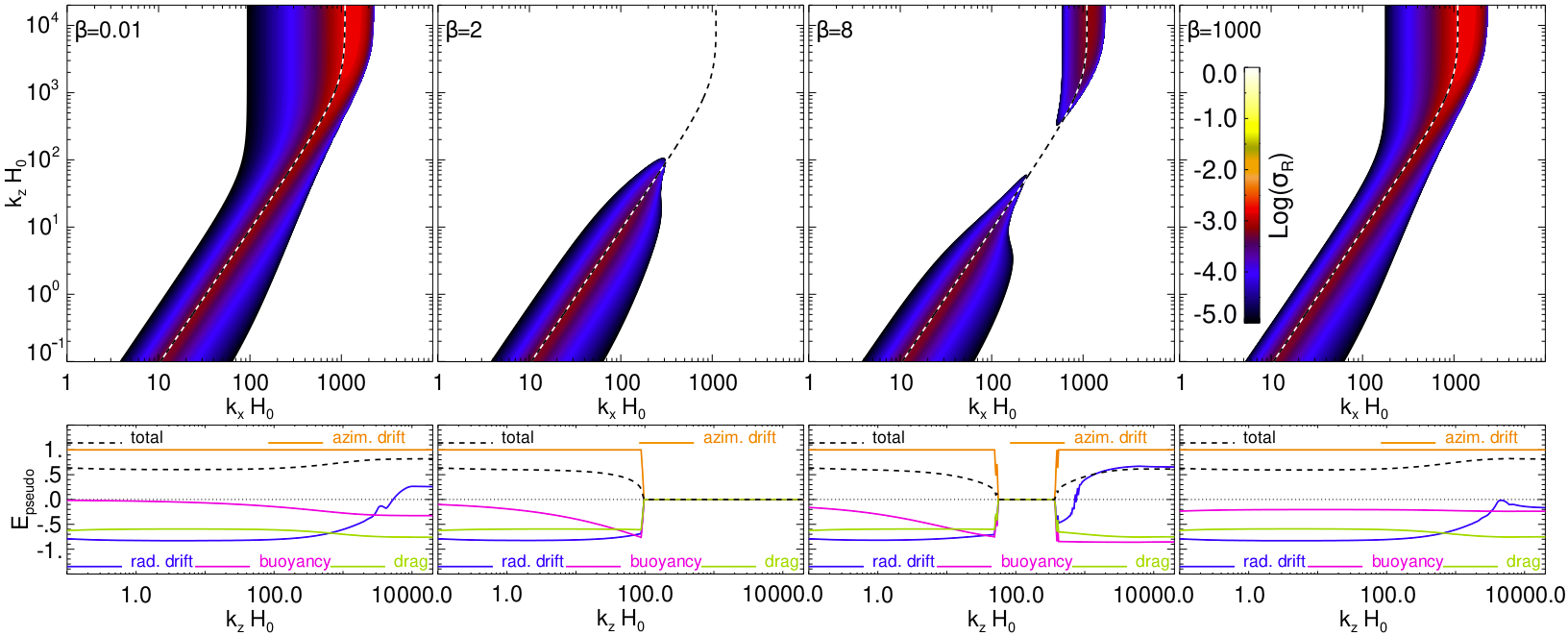}
\caption{Illustration of the effect of radial buoyancy on the linear SI with $\epsilon_{0}=0.1$, $\tau=10^{-2}$, $\eta=0.05$, $N_{r}^2=0.03$, and $H_{r}=0.16$, such that the disc is convectively stable. From left to right panels the cooling time $\beta$ increases - as indicated - from the nearly isothermal limit $\beta=0.01$ to the nearly adiabatic limit $\beta=1000$. The strongest effect of radial buoyancy occurs at intermediate cooling times $\beta \sim 1$, where the damping of the underlying (pure gas) inertial waves is strongest, as described by (\ref{eq:grate}). This is also clearly seen in the pseudo-energy decomposition, which is computed for the resonant wavenumbers (\ref{eq:krdi}) and displayed as function of $k_{z}$ in the lower panels. On the other hand, SI growth rates in the adiabatic limit are nearly identical to those in the isothermal limit.}
\label{fig:si_cool}
\end{figure*}

In Figure \ref{fig:si_cool} we compare SI growth rates in a disc with $N_{r}^2=0.03$ for $\epsilon_{0}=0.1$ and $\tau=10^{-2}$ and for different values of $\beta$ . These parameter values will be motivated later in \S \ref{sec:disc_si}.
The plots show that the SI can be stabilized (i.e. its growth rates are reduced to negligible values) on an appreciable range of wavenumbers, depending on the value of the cooling time $\beta$. For the parameters adopted here, the fastest growing SI modes are significantly suppressed if $\beta\sim 2$.  

%
The dashed curves represent the resonant wavenumber of the SI given by (\ref{eq:krdi}) with $\sigma_{I}$ given by (\ref{eq:freq}), using the parameters specified above.
In the isothermal limit, $\beta\to 0$, the resonant wavenumber is given by (\ref{eq:kres_iso}).
On the other hand, in the adiabatic limit $\beta\to \infty$ we have $\sigma_{I}=\pm \sqrt{\xi}$, where in the current situation $\xi=\mu_{z}^2 \left(1+N_{r}^2\right)$. 
In this case we find
\begin{equation}\label{eq:kres_adia}
k_{z,\text{res}} = \pm \frac{k_{x}^2 \Delta u_{0}}{\sqrt{1 +N_{r}^2-\Delta u_0^2 k_{x}^2}},
\end{equation}
which is only marginally different from the isothermal case (\ref{eq:kres_iso}) on account of the dominance of disc rotation over radial buoyancy.
For general $\beta$ Eq.~(\ref{eq:krdi}) needs to be solved numerically, but
the deviations from the isothermal case are small.
Nevertheless, numerical solutions are plotted as dashed curves in all panels of Figure \ref{fig:si_cool}. The plots in the lower panels describe the pseudo-energy decomposition of the modes along the resonant curves as a function of $k_{z}$.
These curves confirm that buoyancy is the agent which suppresses the resonant SI modes at intermediate cooling times.

%

As noted in \S \ref{sec:res_drag_instab} the original RDI theory considered neutral gas waves. Cooling, however, not only affects the frequency of inertial waves, but also their growth (damping) rate, in the case of an unstable (stable) stratification, as given by Eq.~(\ref{eq:grate}). In the present situation with $N_{r}^2>0$ and $N_{z}^2=0$, pure gas inertial waves are damped. However, in order for the SI to occur the same waves are to be destabilized by relative dust-gas drift. Therefore, one may expect the SI to be ineffective, or even suppressed if the corresponding pure gas inertial waves decay at a sufficiently large rate.

\begin{figure}
\centering 
\includegraphics[width = 0.42\textwidth]{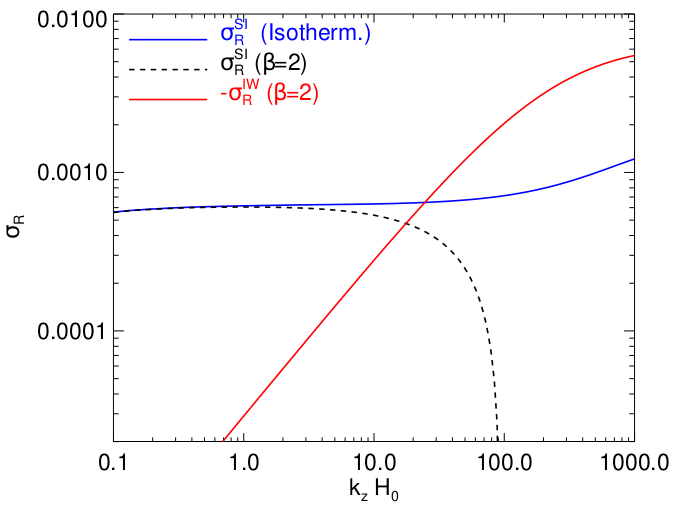}
\caption{Illustration of the damping mechanism of linear SI modes by radial buoyancy. Growth rates of resonant SI modes [i.e. those with wavenumbers fulfilling (\ref{eq:krdi})] of a nonisothermal dusty gas with $\beta=2$ and $N_{r}^2=0.03$ are displayed as dashed black curve (same as in Figure \ref{fig:si_cool}). These are compared to corresponding resonant \emph{isothermal} SI growth rates, drawn as blue solid curve. The difference between the two types of SI-modes is that in the non-isothermal case the underlying (pure gas) inertial waves are subject to buoyant damping (see the text and \S \ref{sec:gas_midplane}). The corresponding damping rate (for $\beta=2$) is drawn as red solid curve. Once this damping rate exceeds the isothermal SI growth rates we expect the nonisothermal SI growth rate to vanish, as explained in the text, in good agreement with the displayed curves.
}
\label{fig:si_cool2}
\end{figure}

In Fig. \ref{fig:si_cool2}, we illustrate this for the case $\beta=2$.
Here the dashed black curve is the numerically computed SI growth rate for $\beta=2$ (same as in Figure \ref{fig:si_cool}) for the resonant wavenumbers (dashed curve in Figure \ref{fig:si_cool}), numerically computed from (\ref{eq:krdi}), and plotted against the resonant wavenumber $k_{z,\text{res}} H_{0}$.
The solid red curve describes the decay rate (negative growth rate) of an inertial wave "$\sigma_{R}^{\text{IW}}$" computed using (\ref{eq:grate}) for  $\beta=2$. This decay rate is also evaluated for the same resonant wavenumbers.
The solid blue curve is the numerically computed SI growth rate for $\beta=10^{-3}$ (essentialy isothermal), adopting the resonant wavenumbers (\ref{eq:kres_iso}). As explained above, the latter are practically identical to the resonant wavenumbers for $\beta=2$. We expect the SI to be damped on wavenumbers for which the decay rate of the corresponding inertial wave surpasses the isothermal SI growth rate. Thus, for $\beta=2$, we expect the SI to be damped roughly for $k_{z} H_{0} \gtrsim 100$, which is in good agreement with the actual SI growth rate for $\beta=2$ which indeed rapidly drops for $k_{z} H_{0} \gtrsim 100$. 

\begin{figure*}
\centering 
\includegraphics[width = \textwidth]{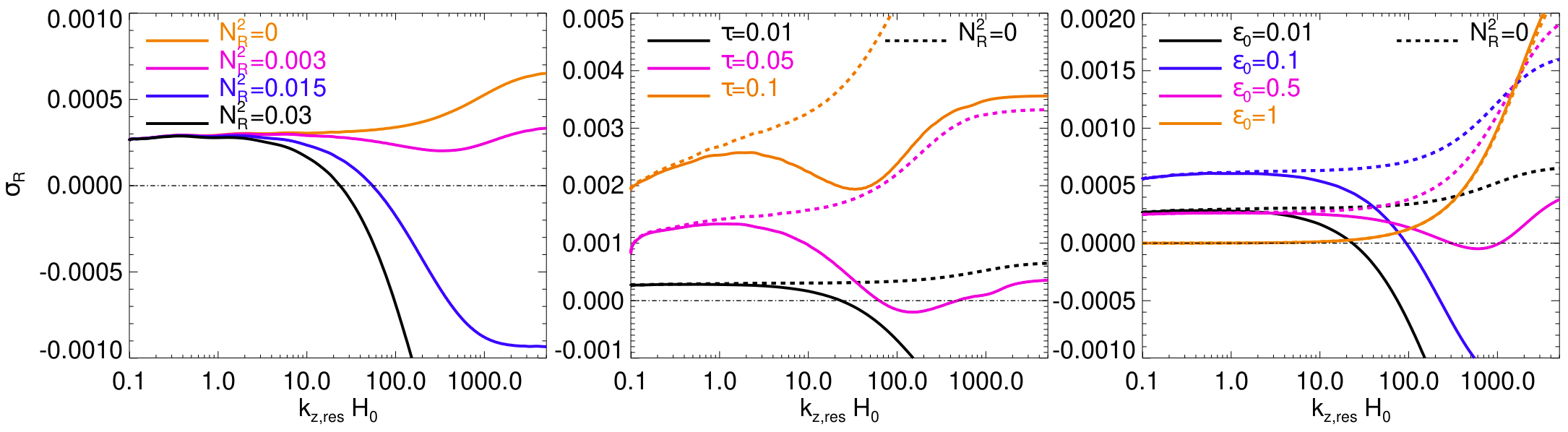}
\caption{Illustration of the effect of radial buoyancy on resonant linear SI growth rates (i.e. for those wavenumbers fulfilling (\ref{eq:krdi})) for different values of the dust prameters. \emph{left panel}: Different values of the radial buoyancy with fixed $\epsilon_{0}=0.01$ and $\tau=0.01$. \emph{middle panel}: Different values of the particle Stokes number with fixed $\epsilon_{0}=0.01$ and $N_{r}^2=0.03$. \emph{right panel}: Different values of the dust-to-gas ratio with fixed $\tau=0.01$ and $N_{r}^2=0.03$. The dashed curves in the right panels are computed with $N_{r}^2=0$. All plots assume $\eta=0.05$ and $H_{r}=0.16$}
\label{fig:si_nr2}
\end{figure*}

Furthermore, Figure \ref{fig:si_nr2} displays resonant SI growth rates for varying $N_{r}^2$ (left panel with fixed $\epsilon_{0}=0.01$ and $\tau=0.01$), varying $\tau$ (middle panel, with fixed $N_{r}^2=0.03$ and $\epsilon_{0}=0.01$) and varying $\epsilon_{0}$ (right panel, with fixed $\tau=0.01$ and $N_{r}^2=0.03$). The solid black curve in all panels describes the same case with $\epsilon_{0}=0.01$, $\tau=0.01$ and $N_{r}^2=0.03$. The dashed curves in the second and third panels are the corresponding growth rates for $N_{r}^2=0$. All curves assume $\beta=2$. These plots show that as long as $\epsilon_{0}<1$, radial buoyancy may in principle have a significant impact on SI growth rates.

%
%


We note that we also performed calculations of the SI in the presence of only vertical buoyancy. 
These reveal that a vertical buoyancy $N_{z}^2 \sim 0.1$ is in principle able to stabilize SI modes on sufficiently long length scales, already at small cooling times $\beta \gtrsim 0.05$. That is, even if the disc is only slightly non-isothermal. In our numerical examples, only modes with $k_z H_0\gtrsim 10^2$ survive against vertical buoyancy.
However, the SI is conventionally expected to operate in the midplane dust layer, where the gas-related vertical buoyancy is small ($N_{z}^2 \ll 1$). Instead, a dust-related vertical buoyancy, stemming from a vertical dust density gradient \citep{lin2017}, is expected to dominate the gaseous component close to the midplane. However, as our model does not account for this dusty component, the aforementioned results are unlikely of high relevance and are therefore not presented here. Instead, in \S \ref{sec:dsi_vbuo} we  present calculations of the effect of vertical buoyancy on the DSI, which can operate at distances from the midplane where the gaseous vertical buoyancy is expected to be significant.



\subsection{COS versus SI}\label{sec:cos_vs_si}

\begin{figure*}
\centering
\includegraphics[width = \textwidth]{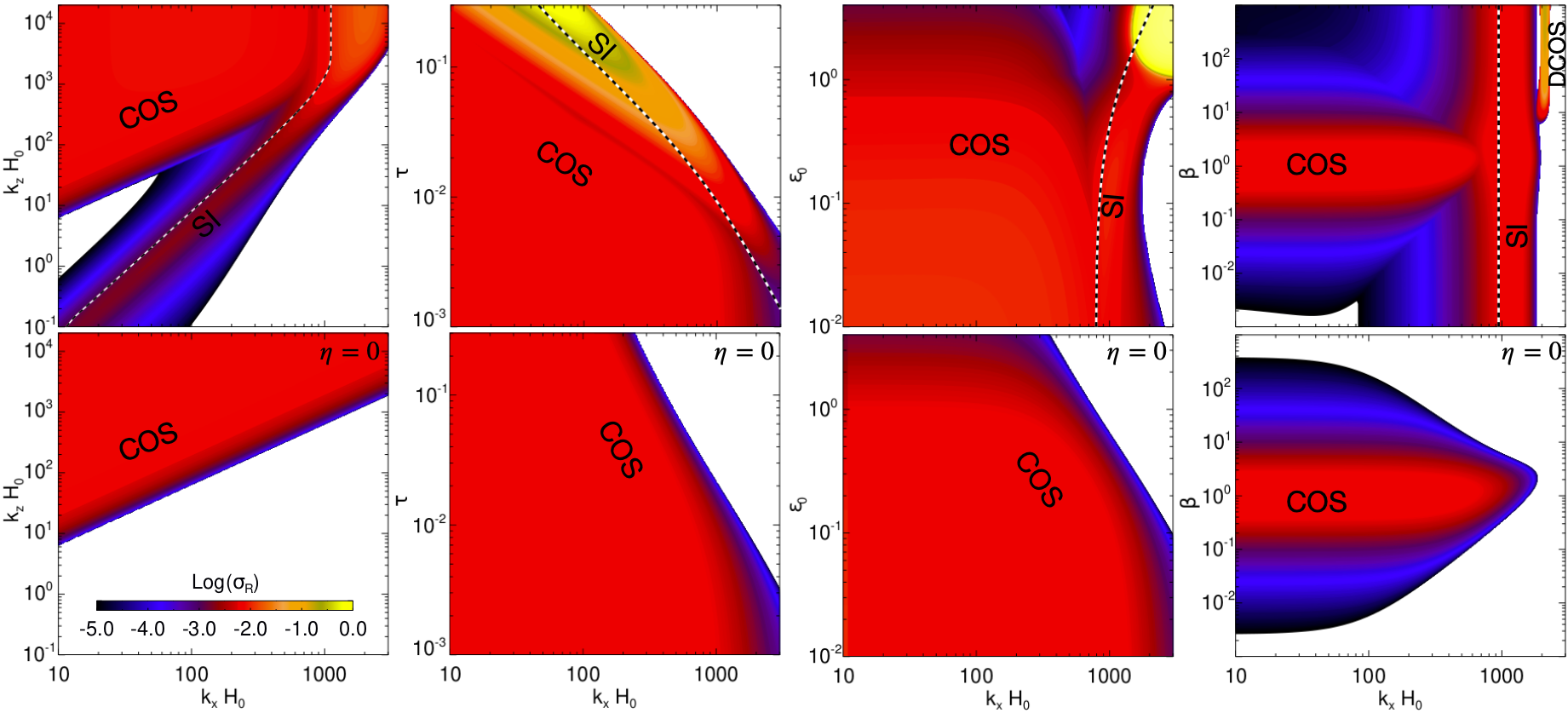}
\caption{Illustration of the co-existence of the COS and the SI in a dusty gas subject to a radial buoyancy frequency $N_{r}^2=-0.01$ ($H_{r}=-0.5$) and a radial pressure gradient $\eta=0.0025$. Presented are contours of growth rates for varying radial wavenumber $k_{x}$. From left to right panels we consider varying vertical wavenumbers $k_{z}$ (with $\epsilon_{0}=0.3$, $\tau=10^{-2}$, $\beta=1$), varying Stokes number $\tau$ (with $\epsilon_{0}=0.3$, $\beta=1$), varying dust-to-gas density ratio $\epsilon_{0}$ (with $\tau=10^{-2}$, $\beta=1$), and varying cooling time $\beta$ (with $\epsilon_{0}=0.3$, $\tau=10^{-2}$), respectively. In all panels apart from the first row we consider a fixed vertical wavenumber $k_{z} H_{0} =1000$. The bottom panels show the growth rates in absence of a radial pressure gradient $\eta=0$, such that the SI is suppressed and the COS behaves as described in \S \ref{sec:cos_nodp}. The dashed curve in each panel describes the resonant SI wavenumbers given by (\ref{eq:krdi}), solved for the quantity varied on the vertical axis in each panel, respectively. Note that in the upper right panel the DCOS appears for cooling times $\beta \gtrsim 10$, with growth rates exceeding those of the SI and the COS (see \S \ref{sec:cos_dp} for details).}
\label{fig:si_vs_cos}
\end{figure*}

We briefly consider the co-existence of the linear COS and SI within the disc midplane in the absence of vertical buoyancy. 
Figure \ref{fig:si_vs_cos} (upper panels) shows the contours of maximum growth rates of a dusty gas subject to both the COS and the SI. The lower panels display the corresponding contours in absence of radial dust-gas drift ($\eta=0$) and show the behaviour of COS modes as discussed in \S \ref{sec:cos_nodp}.  The first column compares different radial and vertical wavenumbers with fixed $\beta=1$, $\epsilon_{0}=0.3$, and $\tau=10^{-2}$.
The remaining panels consider a fixed vertical wavenumber $k_{z} H_{0}\sim 1000$.
 Generally, for $k_{x}\to 0$ only the COS exists, while for radial wavenumbers approaching the resonant wavenumber (\ref{eq:krdi}) the SI tends to dominate, depending on the dust parameters. The dashed curve in each panel represents Eq.~(\ref{eq:kres_iso}), in each case solved for the quantity that is varied along the vertical axis. As outlined in \S \ref{sec:si_cool}, this is an excellent approximation for the resonant wavenumber in the presence of only radial buoyancy.

The second column compares different Stokes numbers $\tau$ with fixed $\beta=1$ and $\epsilon_{0}=0.3$.
For large values $\tau\gtrsim 0.5$ such that dust and gas are not tightly coupled the COS is largely suppressed in favor of the SI. We also note that with increasing Stokes number the radial cutoff wavenumber for COS modes is increasingly reduced by the presence of the SI. This can be seen when comparing the upper and lower panels.

In the third column we compare different values of the dust-to-gas ratio $\epsilon_{0}$ with fixed $\beta=1$ and $\tau=10^{-2}$. As explained in \S \ref{sec:cos_nodp}, in the tight coupling limit of dust and gas ($\tau \ll 1$) order unity values of $\epsilon_{0}$ are required to significantly damp the COS. 
For such values of $\epsilon_{0}$ the largest SI growth rates exceed the COS growth rates by several orders of magnitude. 
For the parameter values adopted here the SI growth rates start to exceed the COS growth rates for $\epsilon_{0}\sim 0.5$. 

Finally, the fourth column compares different cooling times with fixed $\epsilon_{0}=0.3$ and $\tau=10^{-2}$. The COS growth rates (represented by the horizontal `branch' extending to $k_{x}\to 0$) show the expected `symmetrical' behaviour about the optimal cooling time $\beta=1$, 
In addition, the SI (represented by the vertical `branch' at $k_{x} H_{0} \sim 1000$) experiences a slight `boost', also in a symmetrical fashion about $\beta=1$ 
The latter effect can be understood in the same way as the damping effect on SI modes discussed in the previous section. In the case $N_{r}^2<0$ pure gas inertial waves are subject to the COS and are thus amplified. In the presence of dust, the same waves are resonantly amplified through drag forces. Hence, the resulting growth rates are larger than for a gas with $N_{r}^2=0$, which are in turn larger than those for a gas with $N_{r}^2>0$ (cf. Figure \ref{fig:si_cool2}). The fact that the (resonant) SI modes show almost no vertical dependence reflects the indepence of the resonant wavenumber on $\beta$ in this case. Moreover, the dust-induced version of the COS (the DCOS: see \S \ref{sec:cos_dp}) appears for cooling times $\beta \gtrsim 10$ around $k_{x} H_{0}\sim 2000$, with growth rates larger than those of the SI and the classical COS.


\section{Analysis: The DSI and the VSI}\label{sec:instab_z}

\begin{table*}\label{tab:pars2}
\centering
 \caption{Overview of parameter-values used in the various analyses presented in this work. The upper row indicates the instabilities and the corresponding section in the paper. A value of $H_{r/z}=\infty$ indicates that no entropy stratification exists in $r/z$-direction.}
 \begin{tabular*}{\textwidth}{@{}l@{\hspace*{25pt}}l@{\hspace*{35pt}}l@{\hspace*{35pt}}l@{\hspace*{35pt}}l@{\hspace*{35pt}}l@{\hspace*{35pt}}l@{\hspace*{35pt}}l@{\hspace*{35pt}}l@{\hspace*{35pt}}l@{}}
  \hline
       &   VSI (gas) & VSI (gas) &  DSI & DSI & VSI (gas+dust) & DSSI & COS+DSI\\
     & \S\ref{sec:vsi_iso} & \S\ref{sec:vsi_vbuo} & \S\ref{sec:dsi_vbuo} & \S\ref{sec:dsi_vshear} & \S\ref{sec:vsi_dust_mp} & \S\ref{sec:vsi_dust_z}+\S\ref{sec:vsi_kz0} & \hspace{0.cm} \S\ref{sec:cos_general} \\
      \hline
  $\beta$ & $0$ & $>0$ & $>0$ & $>0$ & $0$ & $0$ & \hspace{0.cm} $>0$      \\[2pt]  
  $N_{r}^2$    & $0$ & $0$ & $0$ & $0$ & $0$ & $0$ & \hspace{0.cm}$<0$\\[2pt]
  $H_{r}$  & $\infty$ & $\infty$ & $\infty$ & $\infty$ & $\infty$ & $\infty$ & \hspace{0.cm}$<0$\\[2pt]
  $N_{z}^2$    & $0$ & $>0$ & $>0$ & $>0$ & $0$ & $0$ & \hspace{0.cm}$>0$    \\[2pt] 
  $H_{z}$     & $\infty$ & $<0$ & $<0$ & $<0$ & $\infty$ & $\infty$ & \hspace{0.cm}$<0$    \\[2pt]
  $q_{z}$  & $>0$ & $>0$ & $0$ & $>0$ & $>0$ & $>0$ & \hspace{0.cm}$>0$    \\[2pt] \hline
   $\epsilon_{0}$   & $0$ & $0$ & $>0$ & $>0$ & $>0$ & $>0$ & \hspace{0.cm}$>0$   \\[2pt]
  $\tau$    & $0$ & $0$ & $>0$ & $>0$ & $>0$ & $>0$ & \hspace{0.cm}$>0$   \\[2pt]
   $\eta \,[\times 10^{-3}]$   & $0$ & $0$ & $1.0$ & $1.0$ & $1.0$ & $0$ & \hspace{0.cm}$1.0$ \\[2pt]
  \hline
 \end{tabular*}
\end{table*}

Compared to our analysis in \S \ref{sec:instab_midp}, we now include vertical shear ($q_z\neq0$), dust settling ($w_{d0}\neq0$), or both; which are generally applicable to locations off the disc midplane. We can thus expect the occurrence of the DSI and the VSI in addition to SI and the COS discussed above. In the following, we will investigate the conditions under which one or more of these instabilities are expected to develop. In analogy to \S \ref{sec:instab_midp}, our focus lies on the effect of buoyancy on the DSI, and the effect of dust on the VSI. Furthermore, we will consider the interaction of the VSI and the DSI. By the latter, we mean the effect of vertical shear on the DSI, as well as the effect of dust settling on the VSI. Table \ref{tab:pars2} summarises the analyses considered in this section and lists the relevant parameters involved in each analysis.

Unless otherwise stated, we adopt the fiducial values $\epsilon_{0}=10^{-3}$ and $\tau=10^{-2}$ for comparison with \citet{krapp2020}. Since dust is expected to be concentrated near the midplane in a PPD, larger heights should be more depleted of dust.

\subsection{Pure gas, the VSI}\label{sec:gas_z}

As in \S \ref{sec:gas_midplane}, we first restrict to a pure gas and consider Eqs. (\ref{eq:linrhog})-(\ref{eq:dpress}) with $\epsilon_{0}=\delta \rho_{d}=0$ yielding the dispersion relation
\begin{equation}\label{eq:detgas_z}
\beta \sigma^3 + \sigma^2 + \beta \widetilde{\xi} \sigma + \widetilde{\mu}_{z}^2 =0,
\end{equation}
where
\begin{align}
  \widetilde{\xi} & =   \mu_{z}^2 \left(N_{r}^2 - 2 \widetilde{q}_{z} \right) +\mu_{x}^2 N_{z}^2 -\mu_{x}\mu_{z} \left(\frac{H_{r}}{H_{z}}N_{r}^2 + \frac{H_{z}}{H_{r}} N_{z}^2 \right),\label{eq:xi_all} \\
\widetilde{\mu}_{z}^2  & = -2 \widetilde{q}_{z} \mu_{z}^2,\\
\widetilde{q}_{z} & = q_{z} \frac{\mu_{x}}{\mu_{z}}-\frac{1}{2}\label{eq:qzt},
\end{align}
now including radial and vertical buoyancy, as well as vertical shear, the latter parameterized through $q_{z}$. Eq.~(\ref{eq:detgas_z}) generalises Eq.~(\ref{eq:deteqz}) with $\xi\to \xi - 2q_zk_xk_z/k^2 \equiv  \widetilde{\xi} $ and $\mu_z^2\to \mu_z^2 - 2q_zk_x k_z/k^2 \equiv \widetilde{\mu}_{z}^2$. Hence, the gas can be subject to both the COS and the VSI.
%

\subsubsection{Solberg-H\o iland criteria}

Before we focus on the VSI, for completion we briefly derive the conditions for stability in the adiabatic limit $\beta \to \infty $.
In this limit the dispersion relation (\ref{eq:detgas_z}) yields
\begin{equation}
    \sigma^2 + \widetilde{\xi} =0,
\end{equation}
such that stability affords $\widetilde{\xi} \geq 0$.
Using Eq. (\ref{eq:xi_all}), this condition can be written as the quadratic form 
\begin{equation}\label{eq:quadratic}
 \begin{pmatrix}
\mu_{x}  \\
\mu_{z}
\end{pmatrix} ^T \cdot
\widehat{A}\, \cdot 
 \begin{pmatrix}
\mu_{x}  \\
\mu_{z}
\end{pmatrix} 
\geq 0
\end{equation}
with
\begin{equation}
  \widehat{A}=  \begin{pmatrix}
N_{z}^2 & -\frac{1}{2}\left(\frac{H_{r}}{H_{z}}N_{r}^2+ \frac{H_{z}}{H_{r}}N_{z}^2\right)-q_{z} \\
-\frac{1}{2}\left(\frac{H_{r}}{H_{z}}N_{r}^2+ \frac{H_{z}}{H_{r}}N_{z}^2\right) -q_{z} & 1+N_{r}^2 
\end{pmatrix} .
\end{equation}
Stability for all wavenumbers $\mu_{x},\mu_{z}$ is assured if the two eigenvalues  of $\widehat{A}$ are real and positive.
Using (\ref{eq:nr2})-(\ref{eq:nz2}) and (\ref{eq:dSdr})-(\ref{eq:dSdz}), these two conditions result in the two criteria
\begin{align}
    & 1-\frac{1}{\gamma \rho_{g}}\left(\frac{\partial P}{\partial r} \frac{\partial S}{\partial r} +\frac{\partial P}{\partial z} \frac{\partial S}{\partial z} \right) >0\label{eq:sh1},\\
   & - \frac{\partial P}{\partial z}\left(\frac{\partial S}{\partial z} - 2 q_{z} \frac{\partial S}{\partial r} \right)>0\label{eq:sh2},
\end{align}
i.e. the Solberg-H\o iland criteria \citep{tassoul1978}.

\subsubsection{Isothermal VSI}\label{sec:vsi_iso}

In the isothermal limit $\beta\to 0$, we directly find from (\ref{eq:detgas_z}):
\begin{equation}\label{eq:sig_vsi_iso}
\sigma_{\text{VSI,iso}}= \pm \mu_{z}\sqrt{2\widetilde{q}_{z}},
\end{equation}
which corresponds to\footnote{The symbol $q$ defined in \citet{latter2018} is the negative of $q_{z}$ defined here.} Eq. (13) of \citet{latter2018}.
Thus, modes with $\widetilde{q}_z>0$, or $k_{x} > k_{z}/(2 q_z)$, are growing and, since the stabilizing effect of rotation is eliminated by vertical shear, lacking a restoring force and are hence non-oscillatory. Modes with smaller $k_{x}$ are purely oscillatory since the stabilizing effect of rotation still dominates for these modes (cf. Figure \ref{fig:iwav_bifurc}, lower left panel). Since $|q_z|\ll 1$, growing modes have $k_x\gg k_z$.

\subsubsection{Effect of vertical buoyancy and cooling}\label{sec:vsi_vbuo}

\begin{figure}
\centering 
\includegraphics[width = 0.5\textwidth]{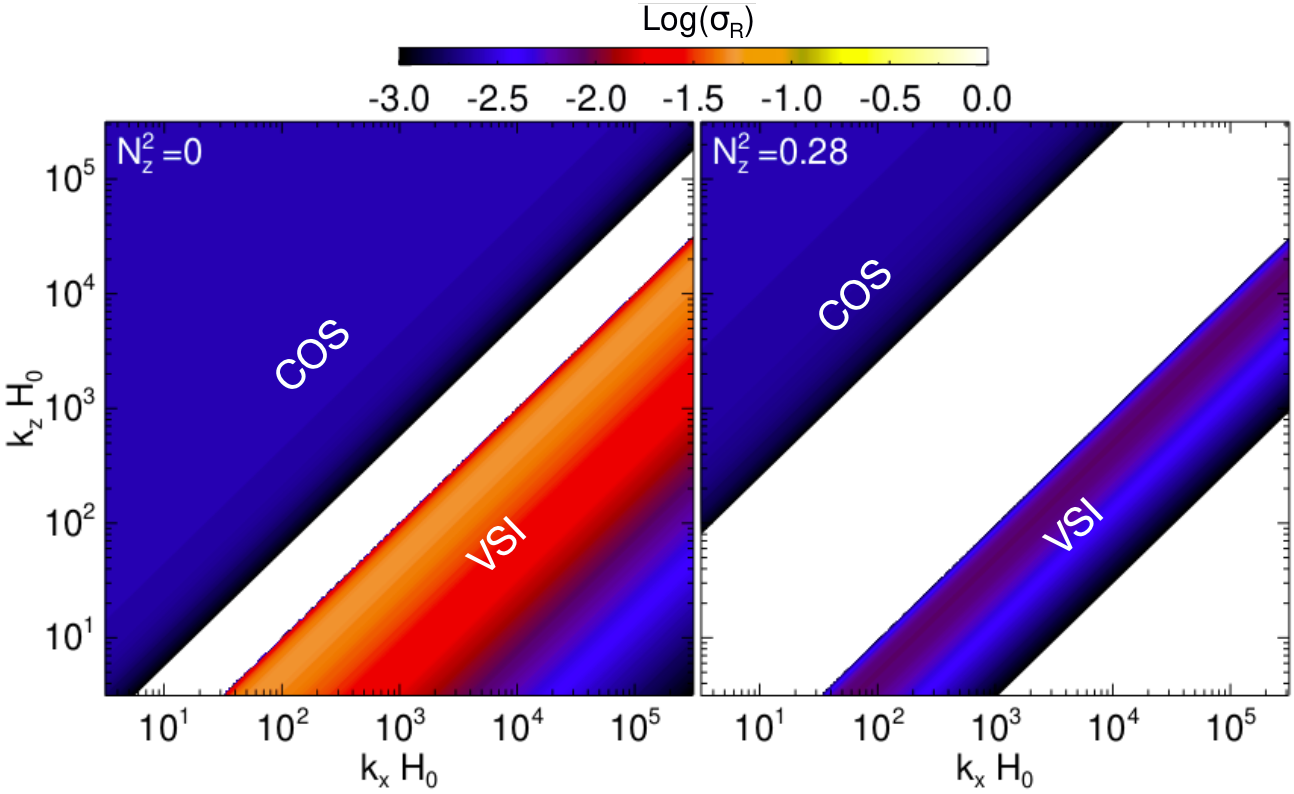}
\caption{Illustration of the co-existence of the COS and the VSI, away from the disc midplane at $z_{0}=h_{0}$. The parameters used in bioth panels are $H_{r}=-0.2$, $N_{r}^2=-0.01$,  and $\beta=1$. The left and right panel considers a vertical buoyancy frequency $N_{z}^2=0$ with $1/H_{z}=0$, and $N_{z}^2=0.28$ with $H_{z}=-0.11$, respectively. The plot in the left panel is somewhat inconsistent in that it assumes $q_{z}>0$ but $N_{z}^2=0$. Nevertheless, it shows that the VSI is more strongly affected by vertical buoyancy than the COS, which is expected as the VSI possesses smaller vertical wavenumbers $k_{z}$ than the COS for the same radial wavemumber $k_{x}$.  }
\label{fig:cos_vsi}
\end{figure}

We now consider buoyancy effects on the VSI. We find from the numerical solution of (\ref{eq:detgas_z}) that radial buoyancy has negligible effects on the VSI. This can be expected since in \S \ref{sec:gas_midplane} we have shown that radial buoyancy mainly affects modes with $k_{x}<k_{z}$ and since $N_z^2\gg |N_r^2|$ off of the disc midplane. Furthermore, even if $N_r^2<0$, the COS and the VSI would not overlap in wavenumber space for typical disc parameters\footnote{By overlapping we mean that both instabilities exist with significant (and comparable) growth rates}. Hence, we will focus here on the VSI including the effect of vertical buoyancy. For completeness, Figure \ref{fig:cos_vsi} shows example growth rates of the COS and the VSI for gas without (left panel) and with (right panel) vertical buoyancy. 

In what follows, we set $N_{r}^2\to 0$ and $H_{r}\to \infty$ such that $H_{r} N_{r}^2\to0$. Then
\begin{equation*}
\widetilde{\xi} \to \widetilde{\mu}_{z}^2 + \mu_{x}^2 N_{z}^2. 
\end{equation*}
%
The full effect of vertical buoyancy is realised in the adiabatic limit $\beta \to \infty$, where we find
\begin{equation}
    \sigma_{\text{VSI,adiab}} = \pm \mu_{z}\sqrt{2 \widetilde{q}_{z}  - \frac{\mu_{x}^2}{\mu_{z}^2} N_{z}^2}.
\end{equation}
One can easily show that this expression remains imaginary if $N_{z}^2>q_{z}^2$, which is expected to be the case under normal conditions. Hence, the VSI is fully stabilized by vertical buoyancy in the adiabatic limit. This can be explained through the increased restoring force on vertical gas motions induced by vertical buoyancy. This result holds also in more realistic vertically stratified disc models \citep{lin2015}.

If we now consider a small but finite cooling time $\beta\ll 1$,
we can assume a series solution
\begin{equation*}
    \sigma = \sigma_{0} + \sigma_{1} \beta + \sigma_{2} \beta^2 + \ldots.
\end{equation*} 
Inserting the series into (\ref{eq:detgas_z}) and solving the latter equation order by order in $\beta$ yields 
%
\begin{equation}
    \begin{split}\label{eq:vsi_bet}
        \sigma_{\text{VSI}} & =  \pm \mu_{z} \sqrt{2\widetilde{q}_{z}} -\frac{1}{2} \mu_{x}^2 N_{z}^2 \beta + \mathcal{O}\left(\beta^2\right),
    \end{split}
\end{equation}
i.e. the isothermal VSI modes with an additional damping term due to vertical buoyancy, since $N_{z}^2>0$. Note that this damping term is identical to the damping term by vertical buoyancy on COS modes [Eq. (\ref{eq:gr_cos_z})]

\begin{figure}
\centering 
\includegraphics[width = 0.45\textwidth]{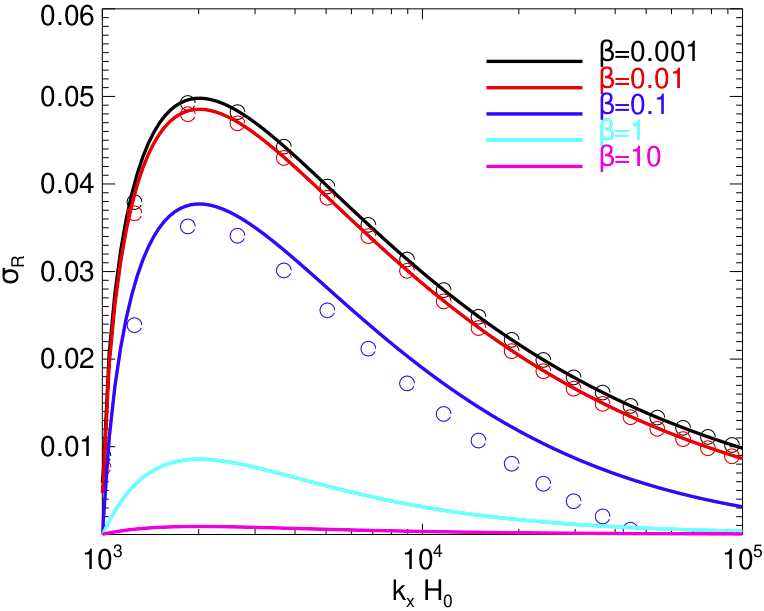}
\caption{VSI growth rates for different values of the cooling time $\beta$ with vertical buoyancy frequency $N_{z}^2=0.28$ with $H_{z}=-0.11$ and fixed $k_{z} H_{0}=100$, numerically obtained from (\ref{eq:detgas_z}), drawn as solid curves. The open circles represent the perturbation formula (\ref{eq:vsi_bet}). For the curves with $\beta\geq 1$, the latter incorrectly predicts a damping for all $k_{x}$. }
\label{fig:vsi_vbuo}
\end{figure}

It is worth noticing that the onset condition for the VSI
\begin{equation}\label{eq:vsi_crit}
    \widetilde{q}_{z}>0
\end{equation}
holds independently of the values of $N_{z}^2$ and $\beta$ as long as $\beta$ is finite. 
Since the dispersion relation (\ref{eq:detgas_z}) is a cubic with real coefficients and $\beta>0$, there exists a real, positive solution for $\sigma$ if the constant term is negative, i.e. $\widetilde{\mu}_z^2<0$, which translates to the above condition. This differs from the semi-global analysis carried out by \cite{lin2015}, where there exists a critical cooling time beyond which small-$k_x$ VSI modes are formally stabilized. Nevertheless, the VSI growth rates are strongly affected by the cooling rate $1/\beta$, which is illustrated in Figure \ref{fig:vsi_vbuo}.



\subsection{DSI in the presence of a finite cooling time and vertical shear}\label{sec:dsi}

We now include dust. In this case, the DSI is expected to become important at heights $\gtrsim h_{0} H_{0}$ away from the disc midplane, where dust settling becomes significant \citep{squire2018} (see also \S \ref{sec:prelim_DSI}). The isothermal DSI without vertical shear was considered by \citet{krapp2020}. However, in a disc subject to a finite cooling time, vertical buoyancy should be operating off of the disc midplane. In addition, vertical shear is generally present, depending on the radial temperature stratification in the disc. Therefore we first investigate the effect of vertical buoyancy on the DSI in \S \ref{sec:dsi_vbuo}. In \S \ref{sec:dsi_vshear} we consider the effect of vertical shear.

\subsubsection{Effect of vertical buoyancy}\label{sec:dsi_vbuo}

In this section, we consider the effect of vertical buoyancy on the DSI. Here we omit radial buoyancy and vertical shear, i.e. $H_{r}^{-1}=N_{r}^2= 0$ (such that $H_{r}N_{r}^2= 0$) and $q_{z}= 0$.

\begin{figure*}
\centering 
\includegraphics[width = \textwidth]{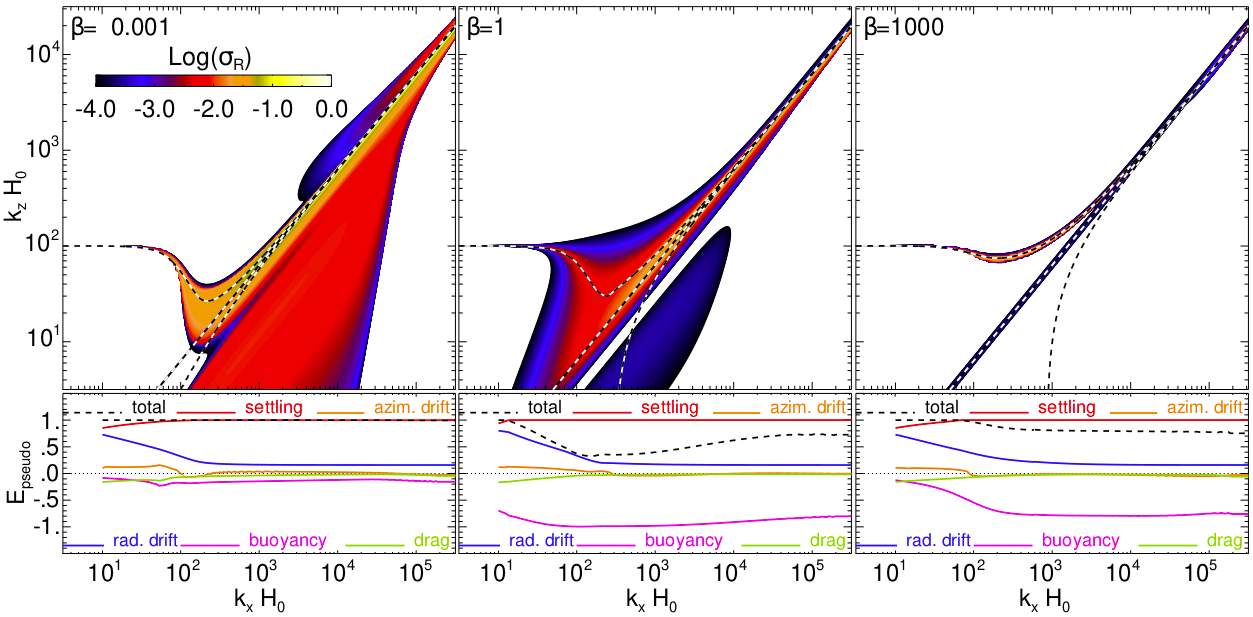}
\caption{Effect of vertical buoyancy $N_{z}^2=0.28$ with $H_{z}=-0.11$ on the DSI (its linear growth rates $\sigma_{R}$) with increasing cooling time $\beta$ from left to right panels and with $\eta=10^{-3}$,  $\epsilon_{0}=10^{-3}$ and $\tau=10^{-2}$. Apart from some modes that closely follow the upper and middle resonant `branches' (i.e. the upper and middle dashed curves, see \S \ref{sec:res_drag_instab}), the instability is increasingly damped with increasing $\beta$. The lower panels show the pseudo-energy decomposition along the upper resonant branch. For display purpose (see the text) the energy decomposition is shown only for $k_{x} H_{0} >10$. The energy decomposition also confirms that the DSI is mainly powered by vertical dust settling, and to a small extent by radial dust-gas drift.}
\label{fig:dsi_damp}
\end{figure*}

Figure \ref{fig:dsi_damp} shows contours of DSI growth rates in the presence of vertical buoyancy for increasing gas cooling time $\beta$ from left to right panels. The leftmost panel, corresponding to the isothermal limit, agrees well with Fig. 1 in \citet{krapp2020}. The value $N_{z}^2=0.28$ corresponds to (\ref{eq:nz_eval}) evaluated at $z_0=H_{0}$. We find (as expected) that vertical buoyancy has a mitigating effect on the DSI, as with increasing cooling time all modes are increasingly damped, apart from some modes that closely fulfill the RDI condition for inertial waves with positive frequency $\sigma_{I}>0$, described by the upper resonant `branch' [i.e. the upper dashed curve: cf. Eq. (\ref{eq:krdi_dsi})]. On the other hand, modes along the lower dashed curve, fulfilling the RDI condition for inertial waves with negative frequency $\sigma_{I}<0$ are effectively damped by vertical buoyancy.


In addition, we also find instability associated with cooling modes, which have $\sigma_I=0$. The corresponding RDI condition yields $k_z \propto k_x$ for the most unstable modes (see Eqs. (\ref{eq:krdi}) and (\ref{eq:rdi_zero}), noting that $\Delta u_{0}\leq 0$ ). These are represented by the dashed straight line in all panels, but this instability is most easily identified in the adiabatic limit ($\beta=10^3$) as the middle `branch', albeit with substantially smaller growth rates than the upper `branch'. In the absence of dust, the cooling modes typically have a small decay rate $\sigma_R\sim -10^{-3}$ (cf. Appendix \ref{app:iw_bifurc}); but are now rendered weakly unstable via resonance with the background dust-gas drift. 


%
The lower panels in Figure \ref{fig:dsi_damp} show the pseudo-energy decomposition (\S \ref{app:pseudo}) along the upper resonant curve. These plots confirm that the DSI is (as its name suggests) primarily driven by vertical dust settling (the red curve), and to a small amount by radial dust-gas drift (the blue curve). The damping due to vertical buoyancy is described by the magenta curve. Interestingly, and in stark contrast to the SI (Figure \ref{fig:si_cool}), the resonant DSI modes are not subject to any notable damping due to dust-gas drag. However, for small radial wavelengths $k_{x} H_{0} \lesssim 100$ the modes become increasingly powered by radial drift and decreasingly powered by dust settling. Also, the unstable region increasingly shrinks around  the resonant curve for decreasing $k_{x}$. Eventually, for $k_{x}\to 0$ the contributions due to radial drift and settling drop and the instability disappears\footnote{For practical reasons this is not shown here, since with decreasing $k_{x}$ the required resolution increases substantially in order to properly resolve the ever-shrinking region of unstable modes.}.

In the following, we derive a simplified model for the DSI which allows us to present an analytical expression for its linear growth rate and to better understand the effect of vertical buoyancy on this instability in the limit of long cooling times $\beta \gg 1$.

We start with the complete set of linearised two-fluid Eqs. (\ref{eq:linrhog})-(\ref{eq:linwd}) as well as the pressure perturbation (\ref{eq:dpress}). Although the DSI modes are powered by both radial and vertical drift (cf. Figure \ref{fig:dsi_damp}), the latter is dominant. To ease the analysis we, therefore, neglect radial drift by setting the radial pressure gradient $\eta=0$.  Using our fiducial parameters we find by numerical experiment that the drag force terms (radial, azimuthal and vertical) from the dust onto the gas do not alter the DSI growth rates and can hence be neglected for this analysis.
This neglect is most likely invalid for order unity dust-to-gas ratios $\epsilon_{0}$. However, we assume that $\epsilon_{0}\ll 1$ in regions where the settling velocity $w_{d0}$ of small dust grains is dynamically relevant.
Thus, we consider the following set of equations:
\begin{align}
    \sigma \delta u_{g} & = 2 \mu_{z}^2 \delta v_{g} + \mu_{x} \mu_{z} \left( H_{z}N_{z}^2 \delta \rho_{g} - \frac{w_{d0}}{\tau} \delta \rho_{d}\right),\label{eq:dsi_ug_p}\\
    \sigma \delta v_{g} & = -\frac{1}{2} \delta u_{g},\label{eq:dsi_vg_p}\\
    \sigma\delta \rho_{g} & = -\frac{1}{\beta}\delta \rho_{g} -\frac{k_{x}}{k_{z}}\frac{1}{H_{z}}\delta u_{g}\label{eq:dsi_rhog_p},\\
    \sigma \delta u_{d} & = 2 \delta v_{d} -\frac{1}{\tau} \left(\delta u_{d} -\delta u_{g}\right) - i k_{z} w_{d0} \delta u_{d}, \label{eq:dsi_ud_p}\\
\sigma \delta v_{d} & = -\frac{1}{2}\delta u_{d} -\frac{1}{\tau} \left(\delta v_{d} -\delta v_{g}\right) - i k_{z} w_{d0} \delta v_{d}\label{eq:dsi_vd_p},\\
\sigma \delta w_{d} & = -\frac{1}{\tau}\left(\delta w_{d} +\frac{k_{x}}{k_{z}} \delta u_{g} \right) - i k_{z} w_{d0} \delta w_{d}\label{eq:dsi_wd_p},\\
 \sigma  \delta \rho_{d} & = -i \epsilon_{0}\left( k_{x}\delta u_{d} + k_{z} \delta w_{d} \right) -  i k_{z} w_{d0} \delta \rho_{d} \label{eq:dsi_rhod_p}.
\end{align}
Notice that buoyancy only enters the problem via the product $H_z\delta\rho_g$ and $N_z^2$. 

In Appendix \ref{app:dsi} we show that the linear growth rate of the DSI resulting from above Eqs. in the adiabatic limit $\beta\to \infty$ and assuming $|\sigma \tau| \ll1$ and $|a_{d,z}\tau|  \ll 1$, 
where
\begin{equation}\label{eq:adv_dz}
   a_{d,z} = k_{z} w_{d0},
\end{equation}
which describes vertical dust advection,
reads
\begin{equation}
\begin{split}\label{eq:gr_dsi_p}
    \sigma_{R,\text{DSI}} & = \frac{2^{\frac{2}{3}} \sqrt{3}}{12} \chi^{\frac{1}{3}}\left[1- \frac{C }{\left(\frac{\chi}{2}\right)^{\frac{2}{3}}}   \right],
    \end{split}
\end{equation}
with 
\begin{align}
 \chi & = A + \sqrt{|B|},\\
    A & = a_{d,z} \left[ 27 \epsilon_{0} \mu_{x}^2 + 18 \left(\mu_{z}^2 + \mu_{x}^2 N_{z}^2 \right) -2 a_{d,z}^2\right],\\
    B & = A^2-4 C^3  \leq 0,\\
    C & = 3 \left(\mu_{z}^2 + \mu_{x}^2 N_{z}^2\right) + a_{d,z}^2.
\end{align}
It is also shown that if $B\geq 0$ then generally $\sigma_{R,\text{DSI}}=0$.

\begin{figure}
\centering 
\includegraphics[width = 0.5\textwidth]{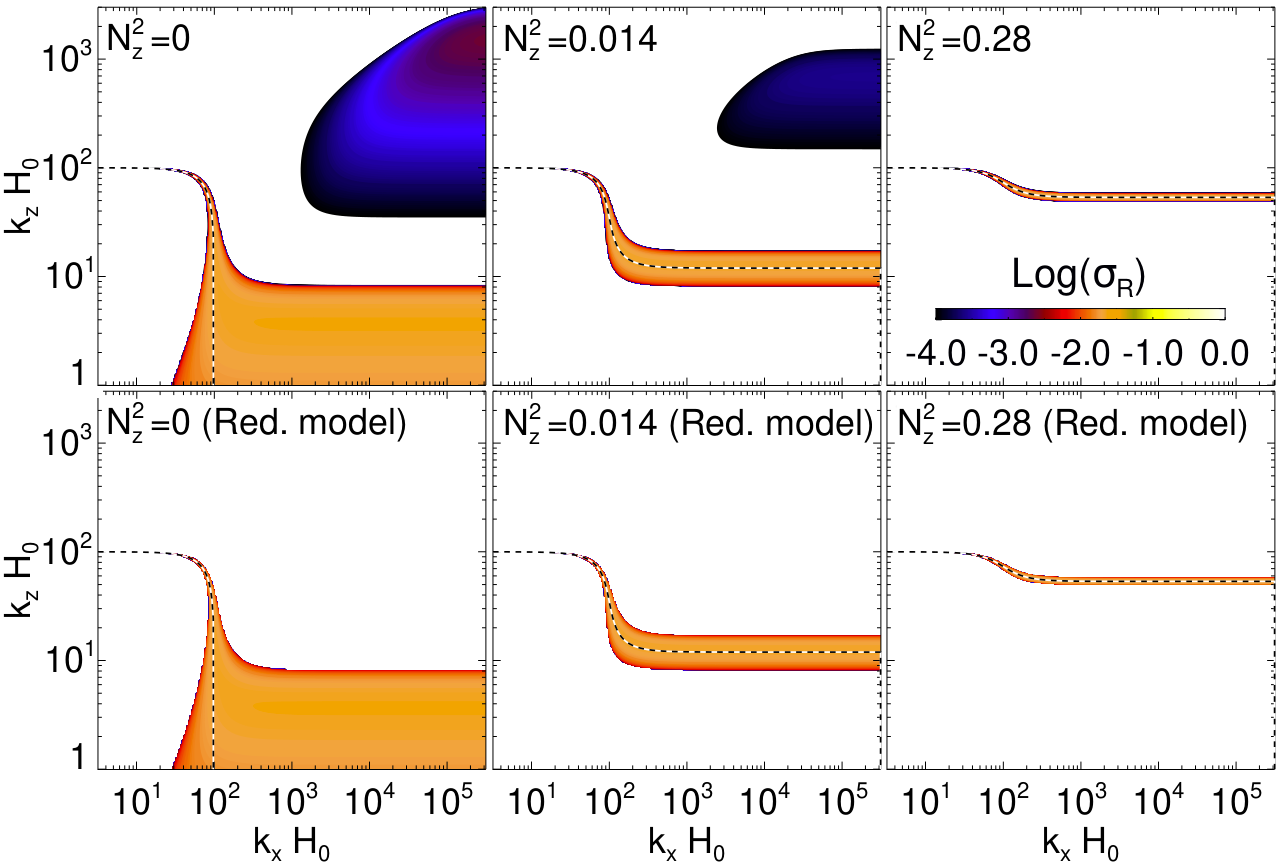}
\caption{The effect of vertical buoyancy on linear DSI growth rates in the adiabatic limit $\beta\to \infty$, resulting from the analytical expression (\ref{eq:gr_dsi_p}) (lower panels) and the numerical solution of the full eigenvalue problem (\ref{eq:eigenproblem}) with $\eta=0$ (upper panels). From left to right panels the vertical buoyancy frequency $N_{z}^2$ is increased as indicated, with $H_{z}=-0.11$ in all cases. The dashed curves describe the resonant wavenumbers (\ref{eq:kres_dsi}). As seen from the plots, and as also derived in Appendix \ref{app:dsi}, with increasing vertical buoyancy the unstable region in wavenumber space shrinks and concentrates around the resonant wavenumbers. Interestingly, the maximum growth rate of the modes is independent of $N_{z}^2$ and is given by (\ref{eq:gr_dsi_lim}). The agreement between the full numerical solution and our reduced model is excellent, apart from some modes with negligible growth rates appearing in the full numerical solution.}
\label{fig:dsi_vbuo}
\end{figure}

Figure \ref{fig:dsi_vbuo} compares growth rates of the DSI in the adiabatic limit resulting from numerical solution of the full eigenvalue problem with $\eta=0$ (upper panels) with those resulting from (\ref{eq:gr_dsi_p}), drawn in the lower panels.
From left to right different values of the vertical buoyancy frequency $N_{z}$ are compared, the smallest being vanishing buoyancy, and the largest being the fiducial value. The agreement is excellent, apart from some modes with subdominant growth rates which are present in the full solution, showing that the terms neglected in (\ref{eq:dsi_ug_p})-(\ref{eq:dsi_rhod_p}) indeed have no significant effect on the DSI growth rates.
Similar as in Figure \ref{fig:dsi_damp} (where $\eta>0$) with increasing cooling time $\beta$, we find here that with increasing $N_{z}^2$ the unstable region shrinks and becomes increasingly concentrated around the resonant wavenumber, indicated by the dashed curve. Interestingly, the growth rate of the resonant modes is not affected by buoyancy, see below.

The drawn resonant wavenumber follows from (\ref{eq:krdi}) using
\begin{align}
    \Delta \vec{\bm{v}}& =\tau z_{0}\vec{\mathbf{e}}_{z},\\
    \sigma_{I}& =\sqrt{\mu_{z}^2 + N_{z}^2 \mu_{x}^2},
\end{align}
and reads\footnote{Since we only consider positive values of $k_{x}$ and $k_{z}$ we discard the negative root.}
\begin{equation}\label{eq:kres_dsi}
    k_{z,\text{res}}  = \sqrt{ \frac{1-k_{x}^2 \tau^2 z_{0}^2}{2 \tau^2 z_{0}^2}  + \frac{1}{2\tau^2 z_{0}^2} \sqrt{\left(k_{x}^2\tau^2 z_{0}^2-1\right)^2 + 4 \tau^2z_{0}^2 N_{z}^2k_{x}^2} }.
\end{equation}
As seen in Figure \ref{fig:dsi_vbuo}, the resonant wavenumber settles on two distinct values for $k_{x}\ll 1/(\tau z_{0})$ and $k_{x} \gg 1/(\tau z_{0})$, respectively, both of which can be obtained from (\ref{eq:kres_dsi}). The former limit reads
\begin{equation}
    k_{z,\text{res}} \approx \frac{1}{\tau z_{0}}.
\end{equation}
On the other hand, for large radial wavenumbers $k_{x} \gg 1/(\tau z_{0})$ we have
\begin{equation}\label{eq:kzrdi}
\begin{split}
        k_{z,\text{res}} \approx \frac{N_{z}}{\tau z_{0}}.
    \end{split}
\end{equation}

Note that the resonant `branch' in Figure \ref{fig:dsi_vbuo} corresponds to the upper `branch' in Figure \ref{fig:dsi_damp}. The remaining branches do not appear in Figure  \ref{fig:dsi_vbuo}, since there we consider the adiabatic limit, such that the cooling mode is non-existent (eliminating the middle `branch'), and neglect radial drift  (which eliminates the lower `branch'). When comparing the adiabatic case in Figure \ref{fig:dsi_damp} (right panel) with the equivalent case of Figure \ref{fig:dsi_vbuo} (right panel), one may conclude that radial drift (which is absent in the latter) has a major impact on the unstable (resonant) wavenumbers, which substantially deviate between the two figures for large $k_{x}$. This appears to contradict the dropping energy contribution related to radial drift, as seen in the lower panel in Figure \ref{fig:dsi_damp}. This can be explained as follows. The energy decomposition as constructed in \S \ref{app:pseudo} corresponds to the real part of the pseudo-energy. Therefore, it measures the energies that determine the growth rates of the linear modes. When taking the imaginary part instead, one retains the energies that determine the oscillation frequencies of the modes. Indeed, when inspecting the latter we find (not shown) that the contribution due to radial drift becomes substantial for large $k_{x}$, explaining its large impact on the unstable wavenumbers in Figure \ref{fig:dsi_damp}.

Furthermore, it is worth noting that in the limit of vanishing vertical buoyancy $N_{z}^2 \to 0$ we have $k_{z,\text{res}}\to 0$ for large $k_{x}$. According to the RDI theory, we expect the largest growth rates of the DSI to occur at the resonant wavenumbers. However, the growth rate of the DSI drops to zero in the limit $k_z\to 0$. This can be seen from (\ref{eq:gr_dsi_p}) using $A=0$ and hence $\chi=4 C^3>0$, which readily yields $\sigma_{\text{DSI}}=0$. 
Thus, for $N_{z}^2=0$, small vertical wavenumber modes are no longer classified as RDIs.

In Appendix \ref{app:dsi}, we show in the limit of large radial wavenumbers $k_{x}\gg k_{z}$ that with increasing $N_{z}^2$ the region in $k_{z}$-space within which the DSI exists becomes increasingly small and indeed converges to (\ref{eq:kzrdi}). The growth rate in this limit is found to be
\begin{equation}\label{eq:gr_dsi_lim}
    \sigma_{R,\text{DSI}}  = \sqrt{\frac{\epsilon_{0}}{2}}, \quad\text{at } k_{z}=\frac{N_z}{\tau z_0} 
\end{equation}
which is indeed independent of $N_{z}$ (and also $\tau$) and yields $ \sigma_{R,\text{DSI}}=0.022$ for $\epsilon_{0}=10^{-3}$ in good agreement with the numerical result.

\subsubsection{Effect of vertical shear}\label{sec:dsi_vshear}

\begin{figure*}
\centering 
\includegraphics[width = \textwidth]{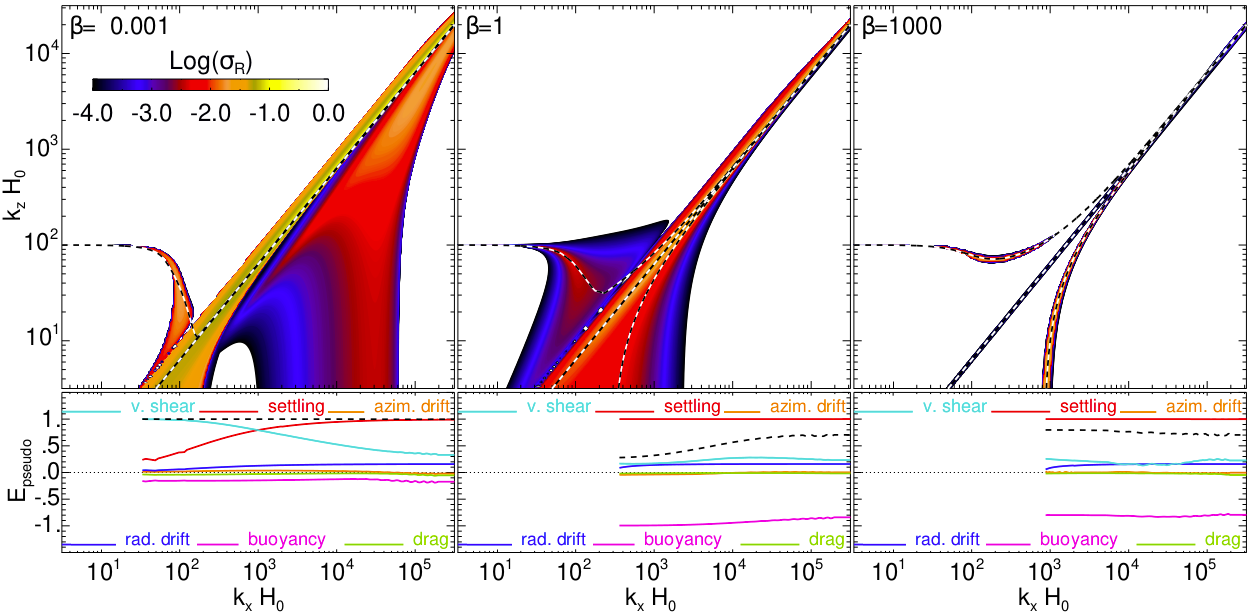}
\caption{Same as Figure \ref{fig:dsi_damp}, but now including vertical shear $q_{z}=0.05$. In contrast to Figure \ref{fig:dsi_damp}, here we plot the pseudo-energies along the \emph{lower} resonant `branch' (i.e. the lower dashed curves in the upper panels), the latter experiencing amplification by vertical shear, whereas the upper `branch' is slightly damped, which can be seen by comparison with Figure \ref{fig:dsi_damp}. The adiabatic limit with vertical shear clearly reveals the three distinct resonant branches of the DSI (cf. \S \ref{sec:prelim_DSI}), as all non-resonant modes are effectively damped by vertical buoyancy.}
\label{fig:dsi_vsi_damp}
\end{figure*}

We find under general conditions the effect of vertical shear on the DSI to be complex. This is not necessarily surprising, as the DSI and the VSI, both of which should be operating, overlap in wavenumber space (cf. Figures \ref{fig:cos_vsi} and \ref{fig:dsi_damp}). Figure \ref{fig:dsi_vsi_damp}, which we discuss in the following, is equivalent to Figure \ref{fig:dsi_damp}, with the difference that in the former we include vertical shear with  $q_{z}=0.05$.
While some modes with smaller $k_{z}$ (including those on the upper resonant `branch') are now damped, modes along the lower resonant `branch' appear to be slightly amplified by vertical shear at larger cooling times $\beta \gtrsim 1$. This amplification is confirmed by the energy-decomposition, now drawn along the lower resonant `branch', as displayed in the lower panels. Particularly, for the case with a small cooling time $\beta=0.001$ the energy contribution by vertical shear even exceeds the one by dust settling for  radial wavenumbers $k_{x} H_{0} \lesssim 1000$, suggesting that the corresponding modes constitute a dusty VSI, rather than the DSI modified by vertical shear.
Indeed, in \S \ref{sec:vsi_dust_z} we will show (in absence of a radial pressure gradient $\eta=0$) that dust settling and vertical shear in combination  result in an augmented version of the VSI.

We note that in the isothermal limit $\beta \ll 1$, pure gas inertial waves, as considered in this section, are subject to the VSI, such that their eigenvalues are given by (\ref{eq:sig_vsi_iso}) and become purely real-valued for $\widetilde{q}_{z}\geq 0$. These non-oscillatory modes are either growing or decaying. Therefore, the lower branch (which requires the existence of pure gas waves with negative frequency) vanishes in this limit, and the DSI formally (according to the RDI concept) cannot exist for radial wavenumbers $k_{x}>\text{min}\left[-(w_{d0}/\Delta u_{0}) k_{z},k_{z}/(2 q_{z})\right]$, as given by (\ref{eq:rdi_zero}) and (\ref{eq:vsi_crit}), respectively. For the parameter values used here, the latter value is slightly smaller. Thus, we conclude that the modes with $k_{x}> k_{z}/(2 q_{z})$ (including those on the dashed straight line) in the left panel are not the DSI, but rather a modified (damped) VSI. However, for larger cooling times $\beta\gtrsim 1$ waves with positive and negative frequencies exist also for $k_{x}> k_{z}/(2 q_{z})$, exceeding the VSI threshold (see Appendix \ref{app:iw_bifurc} for an explanation). This explains the occurrence of the lower resonant branch for cooling times $\beta \gtrsim 1$.

\subsection{VSI in the presence of dust}\label{sec:vsi_dust_p}

In this section, we investigate the effect of dust on the VSI modes described in \S \ref{sec:gas_z}. Here we will consider the isothermal limit, such that the energy equation can be omitted. The effect of buoyancy on VSI modes has been described in \S \ref{sec:gas_z}.
It should be noted at this point that vertical buoyancy resulting from a (vertically) stratified dust layer near the disc midplane, which has a mitigating effect on the VSI \citep{lin2017}, is not accounted for in our local model. This mitigation is most effective for perfectly coupled dust ($\tau\to 0$). Instead, here we study the effect of dust-gas drift on the VSI, which requires $\tau>0$. 
It turns out that while radial dust-gas drift (stemming from a radial pressure gradient) has a damping effect, vertical drift (due to dust settling) has an amplifying one. Drag forces in the perturbed state are also accounted for, which are in fact critical in the following analysis.

\subsubsection{Effect of radial dust-gas drift close to the disc midplane: dusty damping of the VSI}\label{sec:vsi_dust_mp}

We consider a sheet of the disc close to the midplane such that $q_{z}>0$, but dust-settling $w_{d0}$ is negligible as compared to radial dust-gas drift. 
That is, 
\begin{equation*}
    \left|\Delta w_{0}\right| \lesssim \left|\Delta u_{0}\right|.
\end{equation*}
Using Eqs. (\ref{eq:ug0}), (\ref{eq:ud0}), and (\ref{eq:wd0}) with $\tau \ll 1$ and $\eta\sim h_{0}^2$, and neglecting the small contribution from vertical shear $q_{z}$, we find the condition
\begin{equation*}
        z_{0}\lesssim h_{0}^2,
\end{equation*}
or $z_0\lesssim h_0 H_0$ in dimensional terms.

Based on numerical experimentation, we find that dust-gas drift terms involving dust-density variations (which are responsible for the SI) are negligible as long as $\epsilon_{0}\lesssim 1$, and are discarded. However, radial advection terms [the terms $\propto u_{g0}$ in Eqs. (\ref{eq:linrhog})---(\ref{eq:linwg}) and $\propto u_{d0}$ in Eqs. (\ref{eq:linrhod})---(\ref{eq:linwd})], which also describe radial dust-gas drift, are kept.
This setup allows us to study the damping effect of radial dust-gas drift, which mainly stems from the radial pressure gradient $\eta$ on the VSI. We, therefore, consider the following reduced set of equations:
\begin{equation}
    \begin{split}
\left(\sigma + i k_{x} u_{g0} \right) \delta u_{g} & =  2 \mu_{z}^2 \delta v_{g} + \frac{\epsilon_{0}}{\tau} \left( \delta u_{d} -\delta u_{g} \right), \\
\quad & -\frac{\epsilon_{0}}{\tau} \left( \mu_{x}^2 \delta u_{d} +   \mu_{x}\mu_{z} \delta w_{d} \right),\label{eq:ug4p}
\end{split}
\end{equation}
\begin{align}
\left(\sigma + i k_{x} u_{g0} \right) \delta v_{g} & = \frac{\epsilon_{0}}{\tau} \left(\delta v_{d}-\delta v_{g} \right) + \left(q_{z}\frac{k_{x}}{k_{z}} -\frac{1}{2}\right) \delta u_{g},\label{eq:vg4p}\\
\left(\sigma + i k_{x} u_{d0} \right)\delta u_{d} & = 2 \delta v_{d} -\frac{1}{\tau} \left(\delta u_{d} -\delta u_{g}\right),\label{eq:ud4p}\\
\left(\sigma + i k_{x} u_{d0} \right)\delta v_{d} & = -\frac{1}{2}\delta u_{d} - q_{z}\delta w_{d} -\frac{1}{\tau} \left(\delta v_{d} -\delta v_{g}\right),\label{eq:vd4p}
\end{align}
where we used (\ref{eq:dpress}) to replace the pressure perturbation in (\ref{eq:ug4p}). 

In Appendix \ref{app:vsi_rad}, we use the above Eqs. and the approximation $\epsilon_{0}\ll1$ to derive the approximate eigenvalue of the (growing) dusty VSI mode:
\begin{equation}\label{eq:sig_vsi_1_p}
\begin{split}
    \sigma & =  \mu_{z} \sqrt{2 \widetilde{q}_{z}}  - \frac{ \epsilon_{0}\tau \, a_{d,x}^2  }{ \left(1+ a_{d,x}^2 \tau^2 \right)}  - i \frac{\epsilon_{0} \,a_{d,x}}{\left(1+ a_{d,x}^2\tau^2\right)}\\
    \quad & \approx  \mu_{z} \sqrt{2 \widetilde{q}_{z}}  -  \epsilon_{0}\tau \, a_{d,x}^2    - i \epsilon_{0} \,a_{d,x},
    \end{split}
\end{equation}
where the last equality is a good approximation (with relative error $\lesssim 10^{-2}$) for  $k_{x} H_{0} \lesssim 10^4$, and
where
\begin{equation}\label{eq:adv_dx}
    a_{d,x} = k_{x} u_{d0} 
\end{equation}
describes the radial advection of dust with the equilibrium velocity $u_{d0}$, [given by (\ref{eq:ud0})] and therefore, as explained above (see also Appendix \ref{app:vsi_rad}), radial dust-gas drift. Thus, compared to the pure gas case, radial dust-gas drift results (via dust-gas drag) in a damping of VSI modes. Furthermore, it induces a finite frequency of the modes (also via drag).

\begin{figure*}
\centering 
\includegraphics[width = 0.95\textwidth]{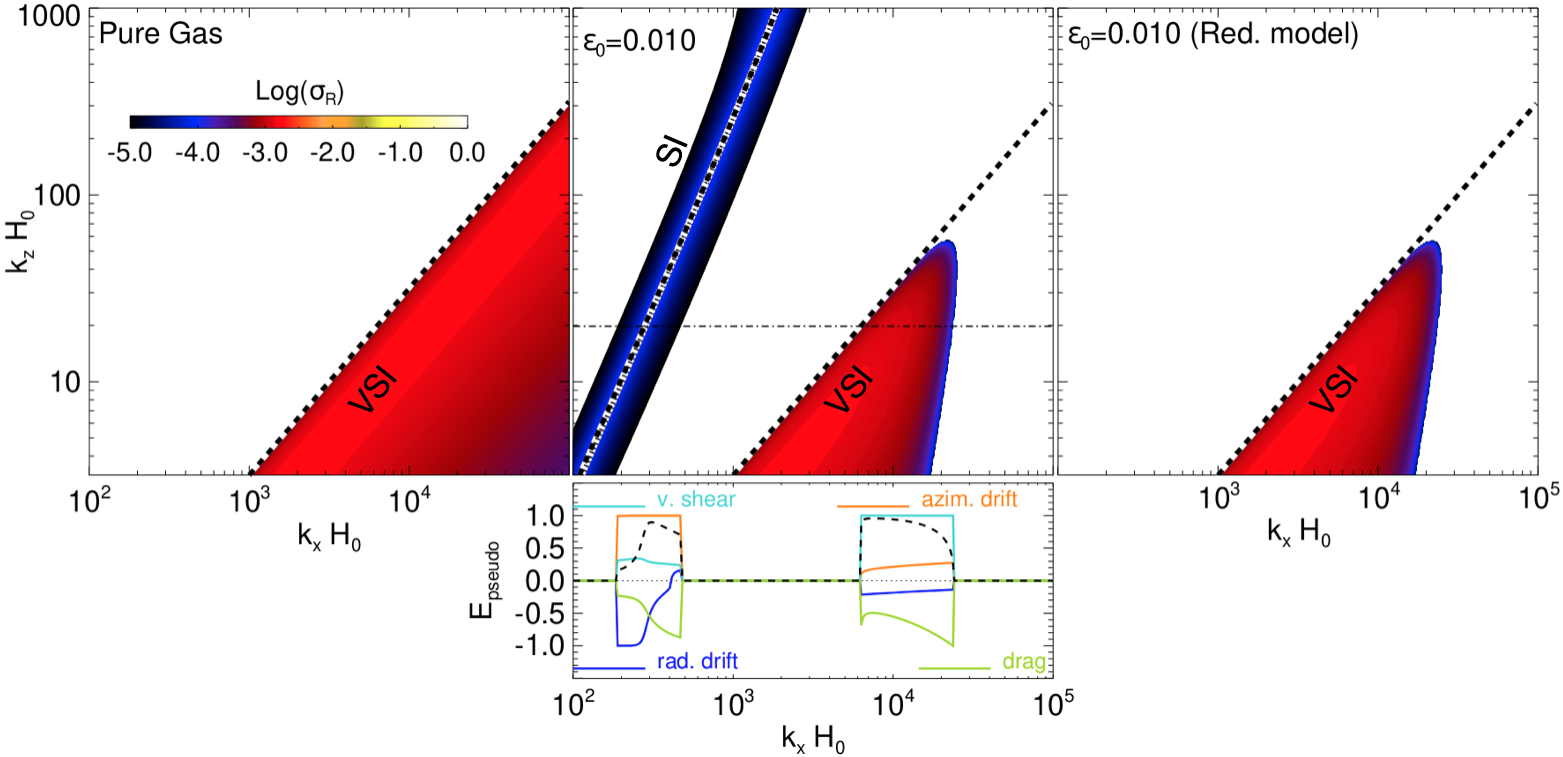}
\caption{Illustration of the damping of linear VSI modes due to radial dust-gas drift, stemming effectively from the equilibrium advection of dust with radial velocity $u_{d0}$ in Eqs. (\ref{eq:ud4p})-(\ref{eq:vd4p}). Compared are VSI growth rates of a pure gas (left panel) with those resulting from numerical solution of (\ref{eq:eigenproblem}), where dust settling has been neglected (middle panel) and the analytical expression (\ref{eq:sig_vsi_1_p}), resulting from the reduced model derived in Appendix \ref{app:vsi_rad}. The parameters used for these plots are $\eta=0.001$, $\epsilon_{0}=0.01$, $\tau=0.004$ and $q_{z}=0.0016$. The latter value corresponds to the vertical shear at a height $z_{0}=\sim h_{0} H_{0}$ (cf. \S \ref{sec:pars}).
The dashed curves in all panels trace the VSI threshold (\ref{eq:vsi_crit}). The additional dashed curve in the middle panel represents the resonant SI wavenumbers (\ref{eq:krdi_si_qz}). The lower panel displays the pseudo-energy decomposition (\S \ref{app:pseudo}) for fixed $k_{z} H_{0}=20$ (indicated by the dashed horizontal line), showing that damping of the VSI is dominated by drag forces, as explained in the text.}
\label{fig:vsi_rad}
\end{figure*}

Figure \ref{fig:vsi_rad} shows the damping of the VSI due to radial dust-gas drift at $z_{0}\approx 0.03 H_{0}$, by comparing linear VSI growth rates of a pure gas (left panel) with those following from the numerical solution of the full eigenvalue problem, where only dust-settling has been neglected (middle panel) and the results using Eq.~(\ref{eq:sig_vsi_1_p}) in the right panel. The plots correspond to $\beta=0.001$ (quasi-isothermal), $\eta=0.001$, $\epsilon_{0}=0.01$, $\tau=0.004$ and $q_{z}=0.0016$. The latter value corresponds to a height $z_{0}\sim h_{0}H_{0}$ (cf. \S \ref{sec:pars}). The dashed lines trace the isothermal VSI criterion (\ref{eq:vsi_crit}) in all panels. The second dashed curve in the middle panel represents the resonant SI wavenumbers, given by
\begin{equation}\label{eq:krdi_si_qz}
    k_{z,\text{res}} = \frac{q_{z} k_{x}}{1-k_{x}^2 \Delta u_{0}^2}\left[1+\sqrt{1+\frac{k_{x}^2\Delta u_{0}^2\left(1-k_{x}^2\Delta u_{0}^2\right)}{q_{z}^2}}\right],
\end{equation}
which follows from (\ref{eq:krdi}) using (\ref{eq:sig_vsi_iso}) with $\widetilde{q}_{z}<0$ 
 In the limit $q_{z}\to 0$, Eq. (\ref{eq:krdi_si_qz}) is identical to (\ref{eq:kres_iso}). For the value of $q_{z}=0.0016$ the difference to the latter is negligible.
 The energy decomposition (bottom panel), which is drawn for fixed $k_{z} H_{0}=10$ (indicated by the dashed horizontal line) confirms that for a small value $\epsilon_{0}=0.03$ the dust-gas drift contribution that is related to dust density perturbations is sub-dominant, justifying its neglect in the reduced model. 
 
 Instead, most of the damping occurs through the drag-force terms $\propto (\delta \vec{\mathbf{v}}_{d}-\delta\vec{\mathbf{v}}_{g})$ (\S \ref{app:pseudo}).
 Note that the difference in the dust and gas advection velocities 
 (here $\propto a_{d,x}$, since the gas drift $a_{g,x} \propto \epsilon_{0} \,a_{d,x}$ is neglected because $\epsilon_0\ll1$), which also describes radial dust-gas drift as explained above, 
eventually contributes via the drag terms. Indeed, by switching off the radial advection terms in the full numerical solution one essentially recovers the pure gas VSI, with a negligible amount of drag-related damping. 
If $\epsilon_{0}\gtrsim 1$ dust density perturbations are no longer negligible, and the approximation breaks down.

\begin{figure}
 \centering 
 \includegraphics[width = 0.5\textwidth]{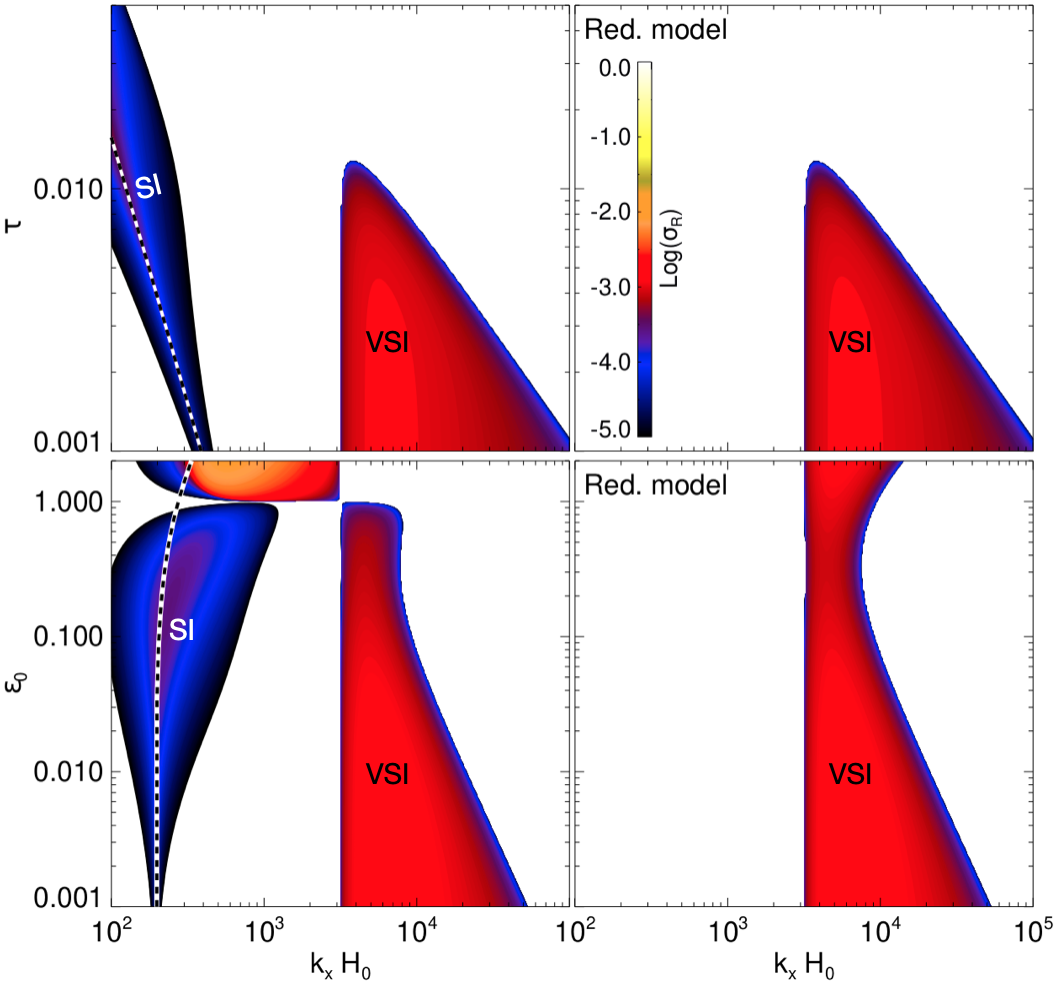}
 \caption{Growth rates of the VSI, similar to those described in Figure \ref{fig:vsi_rad}, but now for varying Stokes number $\tau$ with fixed $\epsilon_{0}=0.01$ (upper panels) and varying dust-to-gas ratio $\epsilon_{0}$ and fixed $\tau=0.004$ (lower panels). All plots consider a fixed $k_{z} H_{0}=10$ and adopt $\eta=0.001$ and $q_{z}=0.0016$. The latter value corresponds to the vertical shear at a height $z_{0}\sim h_{0} H_{0}$ (cf. \S \ref{sec:pars}). The left and right panels describe numerically obtained growth rates and those resulting from (\ref{eq:sig_vsi_1_p}), respectively. The reduced model (Appendix \ref{app:vsi_rad}) turns out to be valid for $\epsilon_{0}<1$. For larger values, the neglected terms in reduced model equations involving dust density perturbations become significant and the SI starts to repress the VSI. The dashed curves in the left panels are the resonant SI wavenumbers, resulting from (\ref{eq:krdi_si_qz}).}
 \label{fig:vsi_rad_2}
 \end{figure}

This is further illustrated in Figure \ref{fig:vsi_rad_2}, where we show numerically obtained growth rates of the dusty VSI for varying $\tau$ (top panel) and $\epsilon_{0}$ (bottom panel) and fixed $k_{z} H_{0}=100$. These are again compared to those resulting from (\ref{eq:sig_vsi_1_p}). Interestingly, the damping of the VSI appears to be nearly independent of $\epsilon_{0}$ for $\epsilon_{0}<1$. On the other hand, for $\epsilon_{0}\gtrsim 1$ the VSI is overwhelmed by the SI, which the reduced model fails to describe. The dependence on $\tau$ in the upper panel shows a clear decrease of the unstable region in $k_{x}$-space. This decrease is accurately captured by the reduced model equation (\ref{eq:sig_vsi_1_p}).

\subsubsection{Effect of vertical dust settling away from the disc midplane: the DSSI}\label{sec:vsi_dust_z}

We now consider typical heights $z_{0} \sim H_{0}$ away from the disc midplane such that the settling of dust becomes dominant over radial dust-gas drift. At the same time, for simplicity, we omit radial and azimuthal dust-gas drift ($\eta=0$). 
The radial and azimuthal drift induced by dust settling in combination with vertical shear [cf. Eqs. (\ref{eq:ud0})---(\ref{eq:wg0})] is very small and can be neglected for our purposes. Furthermore, since the problem considered here is very similar to the one of the DSI in \S \ref{sec:dsi_vbuo}, we again neglect radial and azimuthal drag force terms from the dust onto the gas.
The difference to the problem in \S \ref{sec:dsi_vbuo} is that here we consider the isothermal limit instead of the adiabatic limit and include vertical shear rather than vertical buoyancy. 
\\ 
Thus, we have
\begin{align}
    \sigma \delta u_{g} & = 2 \mu_{z}^2 \delta v_{g} - \mu_{x} \mu_{z} \frac{w_{d0}}{\tau} \delta \rho_{d}\label{eq:ug5p},\\
    \sigma \delta v_{g} & = \frac{\epsilon_{0}}{\tau} \left(\delta v_{d}-\delta v_{g} \right) + \widetilde{q}_{z}  \delta u_{g}\label{eq:vg5p},\\
    \left(\sigma + i k_{z} w_{d0} \right)\delta u_{d} & = 2 \delta v_{d} -\frac{1}{\tau} \left(\delta u_{d} -\delta u_{g}\right)\label{eq:ud5p},\\
\left(\sigma + i k_{z} w_{d0} \right)\delta v_{d} & = -\frac{1}{2}\delta u_{d} - q_{z}\delta w_{d} -\frac{1}{\tau} \left(\delta v_{d} -\delta v_{g}\right)\label{eq:vd5p},\\
 \left(\sigma + i k_{z} w_{d0} \right) \delta \rho_{d} & = -i \epsilon_{0}\left( k_{x}\delta u_{d} + k_{z} \delta w_{d} \right)\label{eq:rhod5p}.
\end{align}

In Appendix \ref{app:vsi_set} we derive (by assuming $|\sigma \tau| \ll 1$ and $\left|a_{d,z}\tau \right| \ll 1$) the approximate cubic dispersion relation
\begin{equation}\label{eq:det_dssi}
    \sigma^3 + i k_{z} w_{d0} \sigma^2 - 2 \mu_{z}^2 \widetilde{q}_{z} \sigma -2 i \left(\epsilon_{0} \mu_{x}^2 + \mu_{z}^2 \right) k_{z} w_{d0} \widetilde{q}_{z}=0,
\end{equation}
as well as an analytic expression for the growth rate of the VSI mode resulting from (\ref{eq:det_dssi}), which is given by Eq. (\ref{eq:sig_dssi}). 

\begin{figure*}
\centering 
\includegraphics[width = 0.9\textwidth]{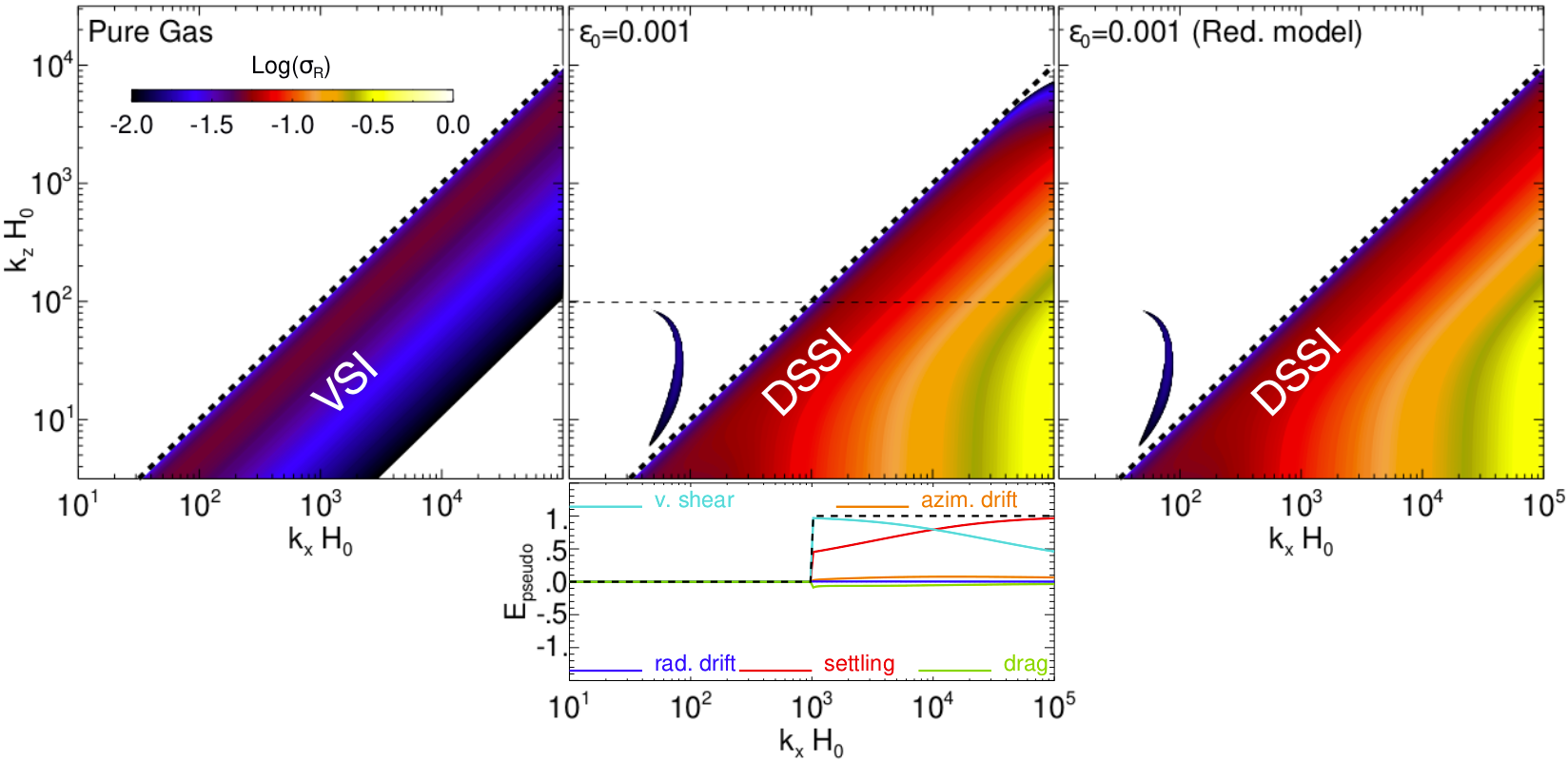}
\caption{Illustration of the DSSI, a combination of the VSI and the DSI. The left panel shows the isothermal pure gas VSI growth rates, obtained from numerical solution of (\ref{eq:deteqz}) with $q_{z}=0.05$ and $\beta=10^{-3}$ (quasi-isothermal) and at (scaled) distance $z_{0}=h_{0}$ away from the disc midplane. The second panel shows growth rates of a dusty gas resulting from numerical solution of (\ref{eq:eigenproblem}) with aforementioned parameter values, $\tau=10^{-2}$, $\epsilon_{0}=10^{-3}$ and $\eta=0$ (no radial pressure gradient). The third panel describes growth rates resulting from the analytical expression (\ref{eq:sig_dssi}), in very good agreement with the numerically obtained values. The plots clearly show how dust settling amplifies the pure gas VSI growth rates. This is confirmed by the pseudo-energy decomposition of the modes, shown in the bottom panel for fixed $k_{z} H_{0}=100$. The DSSI modes here are subject to very little damping by dust. This is in contrast to the situation with $\eta\neq 0$, where the DSSI is substantially suppressed (see Figure \ref{fig:dsi_vsi_damp}).}
\label{fig:dssi}
\end{figure*}

Figure \ref{fig:dssi} compares contours of the analytic growth rate (\ref{eq:sig_dssi}) (right panel) with those obtained using the full eigenvalue problem with $\eta=0$ and $\beta=10^{-3}$ (essentially isothermal, middle panel) and those of a pure gas (left panel). 
Small deviations occur between the growth rates in the middle and right panel
for large $k_{z}\gtrsim 10^4$ since we assumed $\left|a_{d,z}\tau \right| \ll 1$ in our derivation. The energy decomposition (lower panel) for $k_{z} H_{0}=100$ (indicated by the dashed horizontal line) reveals that the growing modes are powered by vertical shear as well as dust settling, in contrast to the pure gas case. We thus refer to this instability as the Dust Settling Shearing Instability (DSSI). Indeed, the resulting growth rates of the DSSI are substantially larger than those of the pure gas VSI. 
The VSI threshold $\widetilde{q}_{z}=0$ is delineated by a dashed line in all panels.

Note that the DSSI is not an RDI. Since here $\eta=0$ we have $\Delta u_{0}\approx 2\tau^3 q_{z} z_{0}>0$, following from Eqs. (\ref{eq:ug0}), (\ref{eq:ud0}). Thus, the RDI condition, which in the present case reads $k_{x} \Delta u_{0} + k_{z} w_{d0}=0$ (since the pure gas frequency following from (\ref{eq:sig_vsi_iso}) vanishes for $\widetilde{q}_{z}>0$), cannot be satisfied for positive values of $k_{x,z}$ and $w_{d0}$ as considered here. 
 Nevertheless, a very similar energy composition to that of the DSSI presented here is also found for \emph{near}-resonant DSI modes in the presence of vertical shear with $\eta\neq 0$ (see Figure \ref{fig:dsi_vsi_damp}, the left panel). In \S \ref{sec:dsi_vshear} we argued that the latter modes are actually not the DSI. Instead, these can be considered as DSSI modes in the presence of a radial pressure gradient $\eta\neq 0$. Figure \ref{fig:dsi_vsi_damp} suggests that a radial dust-gas drift (induced by $\eta \neq 0$) has a mitigating effect on DSSI modes with $k_{x}\gtrsim k_{z}/(2 q_{z})$. In the following section, we will confirm in a different way that the DSSI is indeed a unique instability.

\subsubsection{Limit of small vertical wavenumbers}\label{sec:vsi_kz0}

Rather than exploiting the exact solution of the cubic dispersion relation (\ref{eq:det_dssi}) we here derive analytical expressions for the growth rate and frequency of the DSSI in the limit of small vertical wavenumbers $k_{z}$. In order to better understand this instability, we formally set $k_{z}=0$. This consideration eliminates the pure gas VSI as well as the DSI, since both instabilities vanish as $k_{z}\to 0$ (cf. \S \ref{sec:res_drag_instab}). 

In Appendix \ref{app:vsi_kz0} we use the full set of linearised equations in the isothermal limit to derive growth rate and frequency for the DSSI in the limit $k_{z}\to 0$:
 
 \begin{equation}\label{eq:gr_dssi_p}
    \sigma_{R}  =   \sqrt{3}\left(\frac{1}{4}  k_{x} q_{z}  w_{d0}\frac{\epsilon_{0}}{1+\epsilon_{0}}\right)^{\frac{1}{3}}
\end{equation}
and
\begin{equation}\label{eq:freq_dssi_p}
    \sigma_{I}  =  -2 k_{x} q_{z} w_{d0} \frac{\epsilon_{0}}{1+\epsilon_{0}} -\left(\frac{k_{x} q_{z} w_{d0}}{4 } \frac{\epsilon_{0}}{1+\epsilon_{0}}\right)^{\frac{1}{3}}.
\end{equation}
These results show that the here-discussed instability is indeed novel.
It independently vanishes with vanishing dust settling (described via the factor  $\tau z_{0}$) and vertical shear $q_{z}$, distinguishing it from both the VSI and the DSI.

\begin{figure}
\centering 
\includegraphics[width = 0.5\textwidth]{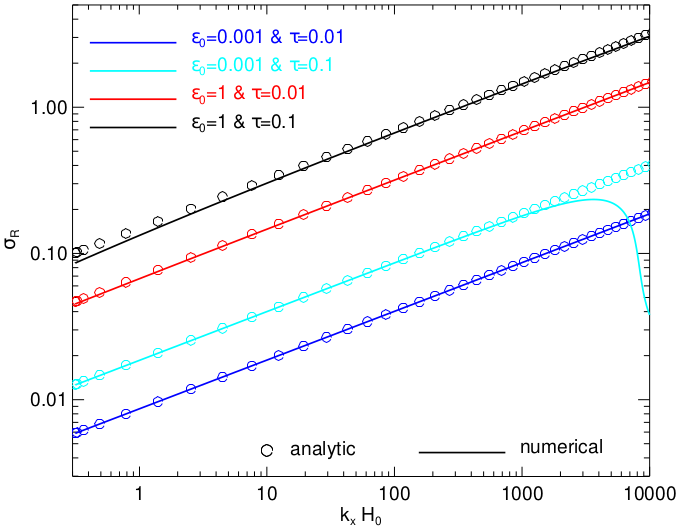}
\caption{Linear growth rates of the DSSI in the limit $k_{z}\to 0$, obtained from numerical solution of (\ref{eq:eigenproblem}) with $q_{z}=0.05$, $\eta=0$ and $\beta=10^{-3}$ (quasi-isothermal) are displayed as solid curves for different values of $\tau$ and $\epsilon_{0}$. The circles describe corresponding growth rates resulting from the analytical expression (\ref{eq:gr_dssi_p}), which is derived from a reduced model in Appendix \ref{app:vsi_kz0}. The agreement is generally excellent for Stokes numbers $\tau \lesssim 10^{-1}$ and $k_{x} H_{0} \lesssim 1000$.}
\label{fig:dssi_kz0}
\end{figure}

We verified that (\ref{eq:gr_dssi_p}) agrees accurately with the full numerical solution for values of $10^{-3} \leq \epsilon_{0} \lesssim 1$
and $\tau \lesssim 10^{-1}$, for a large range of wavenumbers, which is illustrated in Figure \ref{fig:dssi_kz0}. However, large values of $\epsilon_{0}\sim 1$ are not typically expected away from the disc midplane anyway. The reason for the good agreement exhibited in Figure \ref{fig:dssi_kz0} is that (\ref{eq:gr_dssi_p}) [and also (\ref{eq:freq_dssi_p})] is derived from the exact solution of the cubic dispersion relation. In the derivation, we only assumed $|\sigma \tau| \ll 1$ and neglected the contribution due to azimuthal drift, which is subdominant (see Appendix \ref{app:vsi_kz0} for details). Note that for $\epsilon_{0}\tau\to 0$, also the frequency (\ref{eq:freq_dssi_p}) vanishes, in agreement with the pure gas VSI.

\subsection{Dusty damping of the COS away from the disc midplane, generalised growth rate}\label{sec:cos_general}

We do not find any notable effect of radial buoyancy on the DSI or the VSI. This is not surprising since radial buoyancy mainly affects modes with $k_{x}\lesssim k_{z}$ [cf. Eq.~(\ref{eq:grate_cos})], while the aforementioned instabilities typically exist for $k_{x} \gg k_{z}$ . In principle, however, the COS can co-exist with the DSI, the VSI (cf. Figure \ref{fig:cos_vsi}), or their combination (the DSSI). Here, we briefly explore the COS with vertical shear and dust settling.

\begin{figure*}
\centering 
\includegraphics[width = \textwidth]{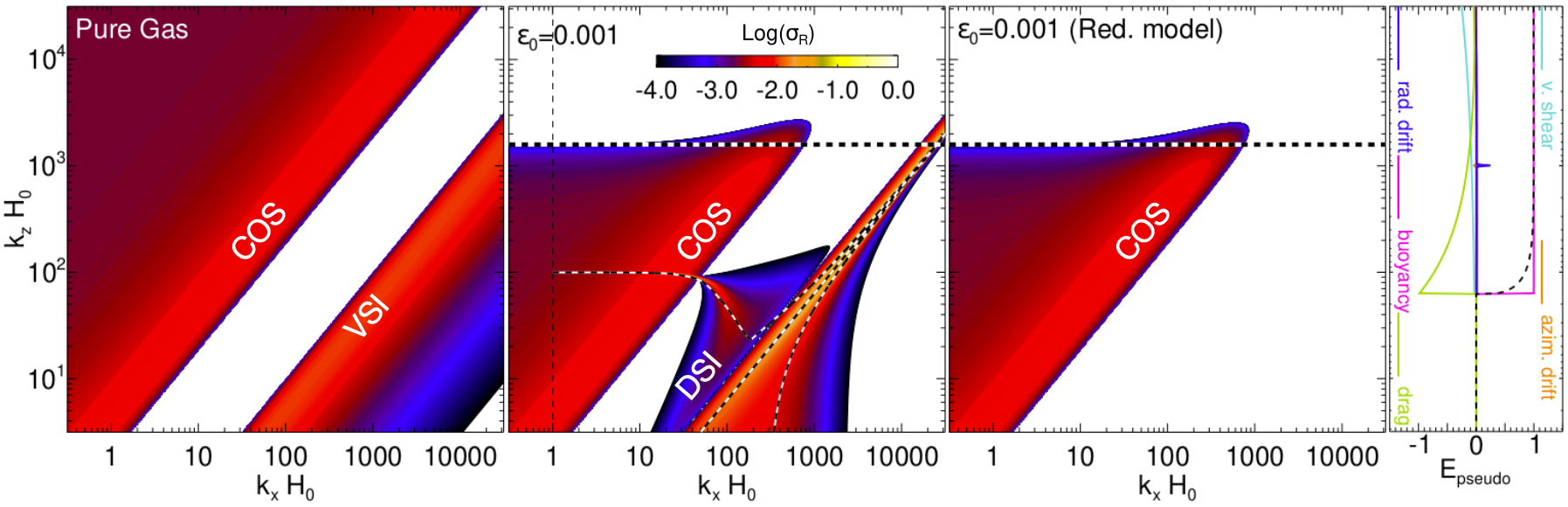}
\caption{Illustration of the damping of the COS by dust away from the disc midplane. The left panel shows numerically obtained maximum growth rates of a pure gas with $N_{r}^2=-0.01$, $H_{r}=-0.2$, $q_{z}=0.05$, $\beta=1$, $N_{z}^2=0.28$ and $H_{z}=-0.11$, at distance $z_{0}=h_{0}$ away from the disc midplane, i.e. the COS subject to vertical buoyancy (cf. Figure \ref{fig:cos_gas}) and vertical shear (which has a negligible effect). The middle panel shows corresponding maximum growth rates of a dusty gas with $\epsilon_{0}=10^{-3}$ and $\tau=10^{-2}$. Here the COS experiences, compared to the situation within the disc midplane (Fig. \ref{fig:cos_dust}), an additional cutoff for a given \emph{vertical} wavenumber, indicated by the dashed horizontal line, which is given by (\ref{eq:kz_cut}). Moreover, in this case the DSI appears (\S \ref{sec:dsi}). The third panel shows the growth rates of the COS resulting from Eq.  (\ref{eq:gr_cos_general}), in good agreement with those in the second panel. Finally, the fourth panel shows the pseudo-energy deomposition of the modes traced by the dashed vertical line in the second panel. These (COS) modes are powered by radial buoyancy and are mainly damped by dust-gas drag, as explained in the text.}
\label{fig:dsi_cos_damp}
\end{figure*}

Figure \ref{fig:dsi_cos_damp} (first and second panel) shows the co-existence of the COS with aforementioned instabilities at a distance $z_{0}=h_{0}$ from the midplane (recall that $z_{0}$ is scaled with $r_{0}$). We use $N_r^2=-0.01$ to enable the COS and include vertical buoyancy, dust settling, vertical shear, and a radial pressure gradient, the latter of which enables a background dust-gas drift. Compared to the situation within the disc midplane (Figure \ref{fig:cos_dustdamp}), the presence of dust now also results in a \emph{vertical} wavenumber cutoff for the COS growth rates (seen in the second panel). In Appendix \ref{app:dusty_iw} we derive an expression for this dust-induced damping effect, which is related to the vertical advection of dust with the equilibrium velocity $w_{d0}$.  The remaining panels will be discussed below

Combined with our description of the COS within the disc midplane in \S \ref{sec:cos_nodp} we can now formulate an analytical expression for the growth rate of the COS, which includes the effects of dust and (vertical) buoyancy for a general distance $z_{0}\lesssim h_{0}$ away from the disc midplane.
This generalised growth rate reads 
\begin{equation}\label{eq:gr_cos_general}
    \sigma_{R,\text{COS}}  =  -\frac{\left(\xi-\mu_{z}^2\right)\beta}{2\left(1+\epsilon_{0}\right)\left(1+\beta^2 \xi\right)}
   -\frac{\epsilon_{0} \mu_{x}^2\tau}{2 \left(1+\epsilon_{0}\right)} -\frac{\epsilon_{0} k_{z}^2 z_{0}^2 \tau^3}{\left(1+\epsilon_{0}\right)^3} ,
\end{equation} 
where $\xi$ is given by (\ref{eq:xi}) 
and where the factor $1/(1+\epsilon_{0})$ in the first term has been added as compared with (\ref{eq:grate_cos}) to account for the effect of dust loading, as described in \S \ref{sec:cos_nodp}. The above expression for the growth rate does not include the effect of vertical shear. However, based on numerical solutions we find its effect on the COS to be negligible.
In the limit $z_{0}\to 0$, $\epsilon_{0}\to 0$, $N_{z}^2\to 0$ (midplane, no dust, no vertical buoyancy), we recover the growth rate (\ref{eq:grate_cos}), originally derived by \citet{lyra2014} and \citet{latter2016}. 

The expression (\ref{eq:gr_cos_general}) is constructed by hand and is not derived from the underlying equations in a rigorous manner. Nevertheless, the first term $\propto \beta $ with $\epsilon_{0}\to 0$, which corresponds to the dust-free limit, is exact (at least in practical terms, see Figure \ref{fig:cos_gas}). Furthermore, in Section \ref{sec:cos_nodp} we have shown that the second term, which is independent of buoyancy, accurately describes the sharp dust-induced drop of the growth rates beyond some $k_{x}$. Moreover, as shown in Figure \ref{fig:cos_dustdamp}, the inclusion of the factor $1/(1+\epsilon_{0})$ in the first term accurately describes the dust-induced reduction of the growth rates for small $k_{x}$ for  $0<\epsilon_{0}\lesssim 1$. 

Figure \ref{fig:dsi_cos_damp} (the third panel) shows growth rates of the COS resulting from (\ref{eq:gr_cos_general}), in good agreement with the full numerical solution. The DSI does not appear in the reduced model, as the latter ignores the vertical dust-gas drift term in the vertical gas momentum equation, which triggers the DSI.
The energy decomposition (for fixed $k_{x}H_{0}=1$) shows that drag forces lead to the cutoff. As outlined in Appendix \ref{app:dusty_iw}, it is the vertical advection of dust with the equilibrium velocity $w_{d0}$, which is the origin of the vertical cutoff wavenumber $k_{z,\text{cutoff}}$. 
In order to obtain an estimate for $k_{z,\text{cutoff}}$ we solve (\ref{eq:gr_cos_general}) for vanishing growth rate in the limit $k_{x}\to 0$.
This results in a quadratic equation for $k_{z}$, readily yielding (the positive root)
\begin{equation}\label{eq:kz_cut}
    k_{z,\text{cutoff}} = \sqrt{\frac{-\beta N_{r}^2}{2  \left[1+\beta^2\left(1+N_{r}^2\right)\right]} \frac{\left(1+\epsilon_{0}\right)^2}{\epsilon_{0} \tau w_{d0}^2} },
\end{equation}
where the first factor in the square root equals the growth rate (\ref{eq:grate_cos}) evaluated at $k_{x}=0$, and which only exists for $N_{r}^2<0$. This vertical cutoff wavenumber is indicated by the dashed horizontal line in Figure \ref{fig:dsi_cos_damp} (middle and right panels).
Since (as outlined in Appendix \ref{app:dusty_iw}) we find that the dust-gas drift terms involving dust density perturbations are negligible for this damping effect, the only remaining energy contribution describing a coupling between dust and gas is that due to drag force terms. Indeed, the vertical drift of dust relative to the gas, expressed by the dust's vertical advection terms [the terms $\propto w_{d0}$ in Eqs. (\ref{eq:linrhod})---(\ref{eq:linwd})], acts onto the gas via drag forces.


It should be noted that the applicability of (\ref{eq:gr_cos_general})
has some limitations within our model. First of all, in its construction, we neglected the radial gas pressure gradient. However, this neglect only affects those modes in a notable manner which become susceptible to the SI for larger values of $\epsilon_{0}$. Moreover, the derivation of the damping terms away from the disc's midplane assumes small values of $\epsilon_{0}$, such that $\epsilon_{0}\lesssim 0.01$ is required to keep relative deviations from the exact solution within a few percent. As stated earlier, this assumption is reasonable away from the disc's midplane.

\section{Discussion}\label{sec:summary}



\subsection{SI in non-isothermal discs}\label{sec:disc_si}


Our analysis of the SI in the disc midplane in \S \ref{sec:si_cool} revealed that the instability behaves very similar in the adiabatic and the standard, isothermal limits. This can be understood within the RDI framework by noting that the corresponding inertial waves, which are fundamental for the SI, have similar frequencies and are neutrally stable in these limits. This suggests that properties of SI-driven turbulence and dust clumping should be similar in the inner, optically thick regions of PPDs as it would in the outer parts \citep{lin2015,barranco2018,pfeil2019,fukuhara2021}.

However, at intermediate disc radii, the gas may cool on a dynamical timescale ($\beta\sim 1$). If the radial stratification is stable ($N_r^2>0$), then radial buoyancy can effectively stabilize the SI on small scales ($k_{x,z}H_0\gtrsim 10^2$ in our numerical examples). This is because the corresponding pure gas inertial waves decay sufficiently rapid. The stabilized SI modes include the fastest growing ones that would occur in an isothermal or adiabatic gas. Thus, overall the SI is weakened and restricted to larger scales than in aforementioned limits.

What could be the potential impact of radial buoyancy 
on planetesimal formation in PPDs? In order to address this question within the limitations of our local, unstratified model, in the following we consider a range of possible gas density slopes $-2<p<-0.5$ and gas temperature slopes $-1.5<q<-0.1$ of the underlying global model for a PPD, as these determine the value of $N_{r}^2$ (see \S \ref{sec:pars}).
We note that observations suggest that in \emph{smooth} disc regions typical values of the temperature slope are  $-0.75<q<-0.5$ \citep{andrews2009}. The gas density profile, on the other hand, is generally hard to constrain from observations. For the above values of $p$ and $q$ and using $h_{0}=0.05$ we find values for the radial buoyancy frequency $-0.006 \lesssim N_{r}^2 \lesssim 0.003$ in the disc midplane $z_{0}=0$ at $r=r_{0}$.
As for the dust parameters $\tau$ and $\epsilon$, it is useful to distinguish between such  parameters that have been found to cause strong dust clumping due to SI in nonlinear simulations, and those which lead to only weak clumping, and which are thus expected to be less relevant for planetesimal formation in smooth disc regions. For instance, for  $\tau=10^{-1}$ it has been found that values $\epsilon_{0}\gtrsim 1$ are required to cause a strong clumping in nonlinear (isothermal) unstratified shearing box simulations \citep{johansen2007}. That being said, for such parameter values we do not find any influence of radial buoyancy on the linear SI growth rates for the (positive) values of $N_{r}^2$ given above. It should be noted at this point that recent vertically stratified shearing box simulations of the SI \citep{li2021} revealed that strong clumping due to the SI can already occur for substantially smaller values of $\epsilon_{0}$. 
Nevertheless, we conclude that if radial buoyancy has any significant effect on dust concentration in PPDs, it must be at special disc locations, such as edges of deadzones, snowlines, pressure bumps, etc. . But considering that such special locations are also preferable sites for dust to concentrate, it may thus be worthwile to provide estimates for $N_{r}^2$ at such locations.

\begin{figure}
\centering 
\includegraphics[width =0.45  \textwidth]{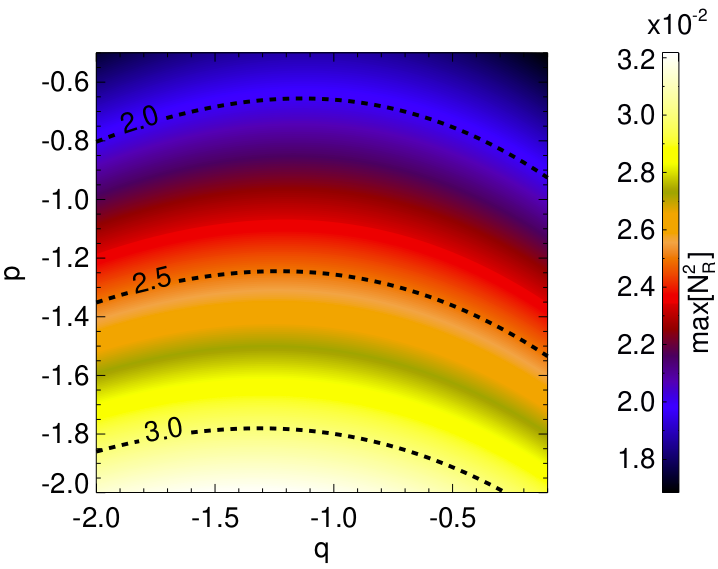}
\caption{Maximal values of the radial buoyancy frequency in the vicinity of a Gaussian pressure bump with dimensionless amplitude $A=0.4$, embedded  at $r=r_{0}$ in the global disc (\S \ref{sec:hydromodel}) with $h_{0}=0.05$, radial temperature profile (\ref{eq:plaw_cs}) and density profile (\ref{eq:plaw_dens}), for varying power law parameters $p$ and $q$ (see Figure \ref{fig:nr2_pbump} for an example).}
\label{fig:nr2_pq}
\end{figure}
\begin{figure}
\centering 
\includegraphics[width =0.4  \textwidth]{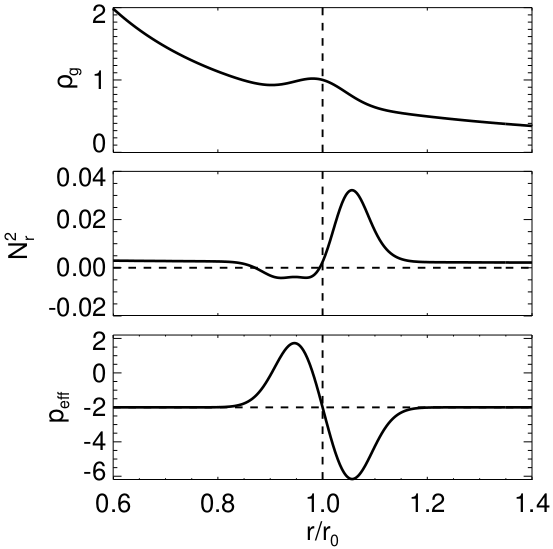}
\caption{Illustration of the effect of a pressure bump on the radial buoyancy frequency embedded in a PPD. The top panel shows the radial profiles of the gas density (\ref{eq:plaw_dens}) with a Gaussian pressure bump with dimensionless amplitude $0.4$ superposed at $r=r_{0}$. The middle and lower panels display the resulting radial buoyancy frequency (\ref{eq:nr2_hat}) with $p$ replaced by $p_{\text{eff}}$, which takes into account the pressure bump, and which is displayed in the bottom panel. }
\label{fig:nr2_pbump}
\end{figure}

Indeed, one scenario that has been found to bear direct relevance for planetesimal formation via the SI is the concentration of dust at a pressure bump. Specifically, using isothermal shearing box simulations \citet{carrera2021a,carrera2021b,carrera2022} studied the process of planetesimal formation of grains with $10^{-2} \lesssim \tau \lesssim 10^{-1}$ at initially solar abundance. As such, they found that for grains with $\tau\sim 10^{-1}$ planetesimal formation is a robust process at a Gaussian pressure bump with dimensionless amplitude $\gtrsim 0.2$, whereas \citet{carrera2022} found that this is not the case for grains with $\tau\lesssim 10^{-2}$.
Nevertheless, the presence of a pressure bump can affect the slopes $p$ and $q$ in such a way that buoyancy effects may be important, not captured in an isothermal model. 

Figure \ref{fig:nr2_pq} shows the \emph{maximum} value of $N_{r}^2$ in the vicinity of a Gaussian pressure bump with dimensionless amplitude\footnote{For a disc with $h_{0}=0.05$ \citet{lehmann2022} found such a bump to be stable with respect to the Rossby wave instability.} $0.4$ and radial width $H_{0}$ (see for instance Eq. (5) in \citealt{carrera2021a}). Thus, larger values of the density slope $p$ generally lead to a larger maximum value of $N_{r}^2$. Specifically, we find a maximum value $N_{r}^2\approx 0.03$ for $p=-2$ and $q=-1.35$, but significant values of $N_{r}^2\gtrsim 0.01$ are found within the entire range of $p$ and $q$ values.
The case with $N_{r}^2=0.03$ is illustrated in Figure \ref{fig:nr2_pbump}, showing the midplane radial profiles of the gas density $\rho_{g}$, radial buoyancy frequency $N_{r}^2$ (with maximum value $0.03$), and the effective density slope 
\begin{equation}
    p_{\text{eff}} \equiv \frac{\partial \ln  \rho_{g}}{\partial \ln r},
\end{equation}
which has been used in (\ref{eq:nr2_hat}) instead of $p$ to compute $N_{r}^2$. We note that \citet{carrera2021a} adopted $p=2.25$ and $q=0.5$, yielding even larger values of $N_{r}^2$ than the largest values considered here.

How radial buoyancy in practice affects the growth of SI modes and the related clumping of dust at a pressure bump needs to be clarified in non-isothermal simulations.
One may speculate that in disc regions where gas cooling occurs roughly on the dynamical time scale, radial buoyancy could increase the minimum particle size that can lead to planetesimal formation at a pressure bump via the SI.

\subsection{DSI in non-isothermal discs}\label{sec:disc_dsi}
The DSI operates away from the disc midplane \citep{squire2018} and has been suggested to seed the (standard) SI. However, simulations carried out by \cite{krapp2020} revealed that the DSI only drives weak dust clumping, putting its potential relevance in planetesimal formation into question. These studies adopted an isothermal gas and therefore did not include buoyancy effects.

Our generalisation of the DSI to non-isothermal discs shows that it is progressively stabilised by vertical buoyancy (Figures \ref{fig:dsi_damp}, \ref{fig:dsi_vsi_damp}) with increasing cooling time. In the adiabatic limit, all modes, except those with precisely the wavenumbers that satisfy the RDI condition, are suppressed. It is unclear if this narrow band of modes can drive significant turbulence or clumping. Our results, therefore, suggest that the DSI may be even less relevant for planetesimal formation than previously shown.

\subsection{Effect of dust on the COS}\label{sec:disc_cos}

The COS requires an unstable radial stratification ($N_r^2<0$), which might be realized at special radial locations in PPDs, such as pressure bumps, snow lines, dead zone edges, etc. , or away from the disc midplane, where the radial gas density profile becomes flatter (see \S \ref{sec:pars}). In the former cases, dust can also be expected to accumulate.


We find that dust generally has a damping effect on the linear COS, where two aspects can be distinguished. For one, tightly coupled dust added to a non-isothermal gas effectively decreases the buoyancy frequency of the dusty gas when viewed as a single mixture. This damping effect, however, is only noticeable for dust-to-gas ratios $\epsilon_{0} \gtrsim 1$. 


Secondly, additional damping occurs if there is a finite coupling time between dust and gas, which originates from the dust-modification of inertial waves (see Appendix \ref{app:dusty_iw}). These damping effects result in a disappearance of the COS at sufficiently large radial wavenumbers, and, if away from the disc midplane, in addition at sufficiently large vertical wavenumbers. 
The radial cutoff is described by (\ref{eq:kxcut}). As such, within the midplane, dust raises the minimum radial length scale of growing modes (with growth rates $\gtrsim 10^{-3} \Omega_{0}$) by a factor $\sim 10$ for $\epsilon_{0}\sim 1$ and Stokes numbers $\tau \sim 10^{-2}$ and assuming the optimal cooling time $\beta=1$ for the COS. For smaller/larger cooling times and larger values of the Stokes number this effect is more severe.
In our numerical example away from the midplane, the COS is suppressed for $k_zH_0\gtrsim 10^3$ (and consequently for $k_xH_0\gtrsim 100$, see Fig. \ref{fig:dsi_cos_damp}). 

Consequently, in dusty discs away from the midplane there is a lower vertical lengthscale limit on the COS, given by (\ref{eq:kz_cut}). For optimal cooling times 
 $\beta\sim 1$ and assuming $\epsilon_{0}\lesssim 1$  and $N_{r}^2 \sim -h_{0}^2$ (\S \ref{sec:pars}) it approximately reads 
 \begin{equation}
    k_{z,\text{cut}} H_{0} \sim \frac{h_{0}}{\sqrt{\epsilon_{0} \tau^3}} \frac{ H_{0}}{ z_{0}},
\end{equation}
where $z_{0}$ is now the dimensional height away from the disc midplane. 
The existence of minimum radial and vertical scales suggests that numerical simulations of the dusty COS should converge with respect to resolution, as long as these scales are resolved, even in the absence of dissipation. At the same time, the vertical cutoff length scale sets a lower limit for the vertical extent of the simulation region.


\subsection{Instabilities resulting from the combined effect of dust-gas drift and buoyancy: the DCOS}\label{sec:disc_dcos}


The DCOS (\S \ref{sec:cos_dp}, \S \ref{sec:cos_onef_dcos}) is a new, unique instability in the presence of radial dust-gas drift and an unstable radial stratification $N_{r}^2<0$. Unlike the COS, however, the DCOS is most relevant to gas subject to slow cooling ($\beta\gg1$, where the COS is negligible, see Fig. \ref{fig:si_vs_cos}). In this limit, there is no physical cooling. However, a dusty gas in the presence of a radial pressure gradient  exhibits a radial drift between the pure gas (i.e. the entropy carrying species) and the center of mass of the dust-gas mixture, which does not drift. This relative drift results in an effective transport of entropy perturbations of the dusty gas (Eq. \ref{eq:contrhog1f}), which in turn constitutes an effective cooling, as expressed by Eq. (\ref{eq:beta_eff_full}). This effective cooling drives instability if $N_r^2<0$. Like the COS, the DCOS grows most rapidly if the effective cooling occurs on a dynamical timescale (see Figs. \ref{fig:bet_phas}, \ref{fig:bet_phas_2}). 


In our examples, the DCOS dominates over the SI for small grains ($\tau\lesssim 10^{-1}$) at low abundances ($\epsilon_0\lesssim 1$). The DCOS may thus provide an alternative pathway to planetesimal formation where the SI and COS are either inoperative or inefficient, for example in the inner regions of a dust-poor disc. On the other hand, DCOS modes are characterized by small lengthscales ($k_{x,z}H_0\gtrsim 10^3$), which may be subject to dissipation. They are also constrained to a rather narrow "strip" in wavenumber space. Nonlinear simulations are required to assess the relevance of DCOS for dust dynamics.

\subsection{Effect of dust in vertically-shearing discs: the VSI and DSSI}\label{sec:disc_dssi}




We find that, near the disc midplane, radial dust-gas drift stabilises VSI modes below a critical radial lengthscale (corresponding to $k_xH_0\gtrsim 10^3-10^4$ in our examples, see Figs. \ref{fig:vsi_rad}, \ref{fig:vsi_rad_2}). If the removal of these small-scale modes weakens VSI-turbulence such that grain growth is promoted (e.g. by reducing the collisional velocities), then radial drift would accelerate, and the VSI would be further weakened. This would then produce a self-sustaining process to facilitate planetesimal formation.

On the other hand, we find that vertical dust settling in combination with vertical shear produces a new, unique instability, the DSSI. At least in the absence of radial dust-gas drift, DSSI growth rates considerably exceed those of the VSI and the DSI. In pinciple, the DSSI can be considered either an enhanced version of the VSI, or an enhanced version of the DSI. Since it has the same stability boundary as the VSI (see Figure \ref{fig:dssi} and Appendix \ref{app:vsi_set}), and since it is --- unlike the DSI --- not an RDI (\S \ref{sec:vsi_dust_z}), it may rather be seen as an enhanced VSI. Our analysis reveals that (for the parameter values used here) modes with $k_{x}/k_{z} \lesssim 100$ are dominated by vertical shear, whereas modes with larger ratios are dominated by settling.  On the other hand, the DSSI exists even in the limit $k_{z}=0$, unlike either the VSI or the DSI, suggesting that the instability is unique. Nevertheless, which of the two descriptions is more adequate needs to be clarified in nonlinear simulations. If the DSSI is an enhanced DSI, which has been shown to lead to dust clumping (albeit weak), the former might actually be more capable of forming clumps than the latter.  




However, if the DSSI behaves like an enhanced VSI in its nonlinear state, we can expect it to generate increased turbulence, and hence increased vertical stirring of dust. In this case, dust settling may be prevented, which would make planetesimal formation more difficult.

\subsection{Caveats and outlook}

\subsubsection{Thermodynamic equilibrium of dust and gas}


A dusty gas, strictly speaking, cannot generally be in a steady, thermodynamic equilibrium, as the radial gas flow in response to dust drag would transport the background entropy and drive evolution. In our model (\S \ref{sec:local}) we only account for the advection of the background due to the perturbed flow. That is, we neglect the term $\vec{\mathbf{v}}_{{g}0}\cdot\vec{\nabla}S_{0}$, which should appear on the right hand side of Eq. (\ref{eq:conteg}) . We can estimate the evolution of the background due to gas drift as 
\begin{equation}
\begin{split}
  \frac{\partial S}{\partial t} \sim\frac{u_{g0}}{H_{r}}  \sim \frac{\epsilon_{0}\tau \eta r \OmK}{H_{r}}.
   \end{split}
\end{equation}
In the limit $\epsilon_{0}\to 0$, exact equilibrium is possible since then $u_{g0}=0$.
The evolution of the equilibrium state should not affect our linear analyses, as long as computed growth rates are $\gg |\epsilon_{0}\tau \eta r \OmK/H_{r}|$. For instance, with typical growth rates of linear COS modes given by $\sigma \sim \left|N_{r}^2\right| \OmK$ (cf. \S \ref{sec:gas_midplane}), this requirement translates to
\begin{equation}
    |H_{r}| \gg \epsilon_{0}\tau r,
\end{equation}
where we used $N_{r}^2 \sim h^2$ and $\eta \sim h^2$. For typical dust parameters $\tau\sim 10^{-2}$ and $\epsilon_{0}\sim 10^{-2}$, this condition is fulfilled by a large margin. Nevertheless, future work should account for this background evolution in global analyses and simulations. 



\subsubsection{Dust-gas frictional heating}
We only couple dust and gas dynamically via the momentum equations. Strictly speaking, dust-gas friction also causes heating, which should be accounted for in the gas energy equation.  This frictional heating of the gas via dissipation of kinetic energy is described by the term \citep{laibe2014}
\begin{equation}\label{eq:drag_heat}
    \Lambda_{h} = \frac{\rho_{d} \rho_{g}}{\gamma(\gamma-1)\tau\left(\rho_{g}+\rho_{d} \right)H_{0}r_{0}\Omega_{0}}\left|\vec{\mathbf{v}}_{d}-\vec{\mathbf{v}}_{g}\right|^2,
\end{equation}
which should be formally added to Eq.~(\ref{eq:conteg}). 

Also dust-gas heating generally prevents the dusty gas from attaining a steady thermodynamic equilibrium. However, this equilibrium frictional heating can, in principle, be balanced by the advection of the background entropy. 
If both aspects are included, the energy equilibrium reads 
\begin{equation}
    \partial_{r}S_0 u_{g0} + 
 \epsilon_{0}  \frac{(\vec{\mathbf{v}}_{g0}-\vec{\mathbf{v}}_{d0})^2}{(1 + \epsilon_{0}) (-1 + 
    \gamma) h_0 \Omega_{0} r_{0}^2 \tau} = 0.
\end{equation}
Using the solutions for the equilibrium drift (Eqs. \ref{eq:ug0}, \ref{eq:ud0}), the above equation can be written as
\begin{equation}
N_{r}^2 = -\frac{
    \eta^2 (4 + 8 \epsilon_{0} + 4 \epsilon_{0}^2 + \tau^2)}{(1 + \epsilon_{0})\gamma (\gamma-1)  h_0 (1 + 2 \epsilon_{0} + \epsilon_{0}^2 + \tau^2)}\sim - h_{0}^3.
\end{equation}
Interestingly, equilibrium can, strictly, only be maintained in the presence of an unstable entropy gradient $N_{r}^2<0$. 

In the perturbed state, dust-gas heating should also be active if there is an equilibrium drift. However, we performed example calculations including the perturbed heating term and found no effect on the local instabilities discussed in this work. Nevertheless, the role of this heating term should be investigated in more detail, particularly in global settings, and whether or not it can drive instabilities by enhancing pressure gradients, such that dust drift is accelerated, which would in turn lead to further heating.


\subsubsection{Gas cooling}

In this study we adopted a Newtonian cooling prescription for the gas, so that temperature perturbations on all scales relax at the same rate. This prescription is useful in controlling the disc's thermodynamic response. Physically, it is applicable to the optically thin cooling regime, where perturbation length scales are smaller than the photon mean free path. For a Minimum Mass Solar Nebula \citep{chiang2010}, \cite{lin2015} estimate that the optically thin regime applies to lengthscales 
\begin{align}\label{optically_thin_criterion}
l \lesssim 10^{-3}\mathrm{e}^{z^2/2H}r_\mathrm{AU}^{33/14}H,
\end{align} 
where $r_\mathrm{AU}$ is the distance from the central star in au. However, in the optically thin regime, the dimensionless cooling time is usually $\beta\ll 1$, unless the disc is depleted of small grains \citep[see Fig. 18 in][]{lin2015}.

To examine intermediate or slow cooling ($\beta \gtrsim 1$), typically applicable to scales larger than that given by Eq. (\ref{optically_thin_criterion}) in the optically thick regime, one should use radiative diffusion. In this case, we expect $\beta\to \beta/k^2$, so that cooling times are scale-dependent. Future work should incorporate a physically-motivated cooling prescription, accounting for both optically thin and thick cooling regimes, to self-consistently examine how the discussed instabilities depend on wavenumbers.

\section{Summary}\label{sec:final_summary}

In this paper, we study the local stability of a PPD comprised of a non-isothermal gas and a single species of dust. Our axisymmetric model incorporates gas cooling, buoyancy, vertical shear, and drift of dust and gas. As such, it captures a number of known instabilities, namely the COS, VSI, SI, and DSI.  We examine how the gaseous instabilities, COS and VSI, are affected by dust; and how the drag instabilities, SI and DSI, are modified in a non-isothermal gas. Results obtained for the SI and the DSI in the isothermal limit, and for the COS and the VSI in the dust-free limit, agree well with those of previous studies. 

We show that dust introduces a minimum length scale for the COS, below which the instability is fully suppressed. At high dust loading, COS growth rates are also reduced because of the reduction of the effective radial buoyancy of the dust-gas mixture. Similarly, we find that radial dust-gas drift sets a minimum radial length scale for the VSI.

For the SI, we find that radial buoyancy suppresses sufficiently small-scale modes when the gas cools on a timescale comparable to the dynamical one. 
We show that vertical buoyancy stabilises the DSI, except for modes along a narrow band of `resonant' wavenumbers.

Finally, we identify two new instabilities unique to the above system. First, the DCOS operates in a dusty, adiabatic gas with an unstable radial stratification. This instability arises from the fact that gas drift, induced by dust, is able to transport temperature perturbations, which constitutes effective cooling, even if the gas in isolation is not subject to cooling. Second, the DSSI arises when both vertical shear (responsible for the VSI) and dust settling (responsible for the DSI) are present. In the absence of radial dust-gas drift (e.g. near a pressure bump), the DSSI is found to have significantly larger growth rates than either the VSI or DSI alone .

Whether or not the above instabilities primarily drive turbulence or can promote dust clumping, and hence their role in the planetesimal formation, will need to be addressed with nonlinear hydrodynamical simulations.

\section*{Acknowledgements}
This work is supported by the National Science and Technology Council (grants 107-2112-M-001-043-MY3, 110-2112-M-001-034-, 111-2112-M-001-062-, 111-2124-M-002-013-) and an Academia Sinica Career Development Award (AS-CDA-110-M06).

\section*{Data Availability}

The data underlying this article will be shared on reasonable
request to the corresponding author.




\clearpage

\appendix

 \section{One-Fluid formulation for dusty non-isothermal Gas}\label{app:onefl}

In addition to our two-fluid model (\S \ref{sec:hydromodel}) of dusty gas, we also consider a one-fluid description \citep{laibe2014,lin2017,lovascio2019}. 
Here we work with the total density
\begin{equation}\label{eq:rho}
    \rho = \rho_{g} + \rho_{d},
\end{equation}
the center of mass velocity
\begin{equation}\label{eq:cms}
    \vec{\mathbf{v}}=\frac{\rho_{g} \vec{\mathbf{v}}_{g} + \rho_{d} \vec{\mathbf{v}}_{d}}{\rho},
\end{equation}
and the relative velocity
\begin{equation}
    \Delta \vec{\mathbf{v}}=\vec{\mathbf{v}}_{d}-\vec{\mathbf{v}}_{g}.
\end{equation}
This implies
\begin{align}
    \vec{\mathbf{v}}_{g}=\vec{\mathbf{v}}-f_{d}\Delta \vec{\mathbf{v}},\\
    \vec{\mathbf{v}}_{d}=\vec{\mathbf{v}}+f_{g}\Delta \vec{\mathbf{v}},
\end{align}
where
\begin{align}
    f_{d}=\frac{\rho_{d}}{\rho} = \frac{\epsilon}{1+\epsilon}\label{eq:fd},\\
    f_{g}=\frac{\rho_{g}}{\rho}=\frac{1}{1+\epsilon}\label{eq:fg},   
\end{align}
are the dust and gas mass fractions, respectively. 
Furthermore, we work with the dust-to-gas ratio (\ref{eq:eps}).
 The incompressibility condition (\ref{eq:contrhog}) in the one-fluid formulation reads
 \begin{equation}\label{eq:incomp1f}
     \vec{\nabla} \cdot \vec{\mathbf{v}} = \vec{\nabla} \cdot (f_{d} \Delta \vec{\mathbf{v}}),
 \end{equation}
 so the dusty gas is no longer incompressible, unlike the gas in isolation.
%
The evolution Eqs. for $\epsilon$ and $\vec{\mathbf{v}}$ read
\begin{equation} \label{eq:contrho1f}
\left(\frac{\partial}{\partial t} + \vec{\mathbf{v}}\cdot\vec{\nabla}\right) \, \epsilon  =  -(1+ \epsilon) \left( \vec{\nabla} \cdot \vec{\mathbf{v}} \right), 
\end{equation}
\begin{equation}
\begin{split}
 \left(\frac{\partial}{\partial t} + \vec{\mathbf{v}}\cdot\vec{\nabla}\right) \, \vec{\mathbf{v}}  & =  - 2 \Omega_{0} \vec{\mathbf{e}}_{z}\times \left[\vec{\mathbf{v}}-\left(-\frac{3}{2}\Omega_{0} x + q_{z} \Omega_{0} z\right)\vec{\mathbf{e}}_{y} \right] - \frac{1}{\rho} \vec{\nabla} \delta P \\
 & \quad +\left(2 \eta \, \Omega_{0}^2 \frac{\rho_{g0}}{\rho}r_{0}  -H_{r}N_{r}^{2} \frac{\delta \rho_{g}}{\rho} \right) \vec{\mathbf{e}}_{x}\\
 & \quad +\left( \frac{\delta \rho_{d}}{\rho} z_{0} \Omega_{0}^2 - H_{z}N_{z}^{2}\frac{\delta \rho_{g}}{\rho}  \right)\vec{\mathbf{e}}_{z} . \label{eq:contv1f}  
 \end{split}
\end{equation}
To arrive at (\ref{eq:contv1f}) we multiply Eq. (\ref{eq:contvg}) by $\rho_{g}$ and (\ref{eq:contvd}) by $\rho_{d}$, add the two equations and use (\ref{eq:rho})-(\ref{eq:cms}). We also drop terms of $O(\Delta \vec{\mathbf{v}}^2)$, assuming it is negligibly small for tightly coupled dust. 
Furthermore, from the equation for the velocity difference \citep[cf. Eq.~(14) in][]{laibe2014} $\Delta \vec{\mathbf{v}}$ in the terminal velocity approximation (TVA) we find
\begin{equation}\label{eq:tva}
\begin{split}
  \Delta \vec{\mathbf{v}} & = t_{s}\Bigg(\frac{1}{\rho}\nabla \delta P -2 \eta \, \Omega_{0}^2 r_{0} f_{g} \vec{\mathbf{e}}_{x}  + \frac{\delta \rho_{g}}{\rho} \left(H_{r} N_{r}^2 \vec{\mathbf{e}}_{x} + H_{z} N_{z}^2 \vec{\mathbf{e}}_{z}\right)  \\
  & \quad   + \left(1+\epsilon_{0} \right) z_{0}\Omega_{0}^2 f_{g}   \vec{\mathbf{e}}_{z} \Bigg) .
  \end{split}
\end{equation}
In order to arrive at (\ref{eq:tva}) we subtract Eqs. (\ref{eq:contvg}) and (\ref{eq:contvd}). In the resulting equation for $\Delta \vec{\mathbf{v}}$, we expand (since $\tau\ll 1$) $\Delta\vec{\bm{v}} \to \Delta\vec{\bm{v}} \tau + \mathcal{O}(\tau ^2)$ and balance terms to zeroth order in $\tau$ \citep{lovascio2019}. 

Thus, in the TVA, the dust is assumed to relax to its terminal velocity instantly and dust-gas drift is attributed to the differential forces that act on the dust and gas.  
Note that for axisymmetric systems there is no azimuthal velocity drift in the TVA. The above expression contains additional terms compared to the original expression provided in \citet{youdin2005}. These terms are related to buoyancy and vertical dust settling. 

Using this expression and the above definitions, we can rewrite the gas energy equation in the one-fluid formalism as
\begin{equation}\label{eq:contrhog1f}
\begin{split}
 \left(\frac{\partial}{\partial t} + \vec{\mathbf{v}}\cdot\vec{\nabla}\right) \, \delta \rho_{g}    & = 
  -\frac{1}{ t_{c}}\delta \rho_{g} + \rho_{g0} \left(\frac{1}{H_{r}} \delta u + \frac{1}{H_{z}} \delta w \right)\\
 & \quad + \frac{t_{s} \Omega_{0}^2 \epsilon_{0}}{(1+\epsilon_{0})^2} \left( -2 \eta  r_{0} \partial_{x} \delta\rho_{g}   + (1+\epsilon_{0})z_{0} \partial_{z} \delta\rho_{g}  \right)\\
 & \quad -\rho_{g0} \left( \frac{1}{H_{r}} \delta(f_{d} \Delta u ) + \frac{1}{H_{z}} \delta(f_{d} \Delta w ) \right),
 \end{split}
\end{equation}
where
\begin{align}
\begin{split}
   \delta(f_{d} \Delta u )  = & t_{s} \Bigg(  \frac{f_{d0}}{\rho_{0}} \partial_{x} \delta P -2 \eta \Omega_{0}^2 r_{0} \frac{(1-\epsilon_{0})}{(1+\epsilon_{0})^3} \delta \epsilon \\
   \quad & +\frac{f_{d0}}{\rho_{0}} \delta \rho_{g} H_{r} N_{r}^2 \Bigg),\label{eq:dfddu}
   \end{split}\\
   \begin{split}
      \delta(f_{d} \Delta w )  = & t_{s} \Bigg(  \frac{f_{d0}}{\rho_{0}} \partial_{z} \delta P +  \Omega_{0}^2 z_{0} \frac{(1-\epsilon_{0})}{(1+\epsilon_{0})^2}\delta \epsilon\\
      \quad & +\frac{f_{d0}}{\rho_{0}} \delta \rho_{g} H_{z} N_{z}^2 \Bigg).\label{eq:dfddw} 
      \end{split}
\end{align}

\subsection{One-fluid equilibrium}
The above set of Eqs. yield the equilibrium 
\begin{align}
\epsilon_{0} & = \text{const.}, \\
u_{0} & = -\frac{2 \epsilon_{0} q_{z} \tau z_{0} \Omega_{0}}{1 + \epsilon_{0}}, \\
v_{0} & = -\frac{\eta r_{0} \Omega_{0}}{1 + \epsilon_{0}},\\
w_{0} & = \frac{\epsilon_{0} \tau z_{0}\Omega_{0}}{1 + \epsilon_{0}},
\end{align}
and 
\begin{align}
    \Delta u_{0} & =-\frac{2\eta\tau \Omega_{0} r_{0}}{1+\epsilon_{0}}\label{eq:u01f},\\
    \Delta v_{0} & = 0\label{eq:v01f},\\
    \Delta w_{0} & = \tau z_{0}\Omega_{0}\label{eq:w01f},
\end{align}
which are the velocities relative to the background shear flow
 $\left(-\frac{3}{2}x + q_z z\right)\Omega_0$.
Compared to the two-fluid equilibrium (\S \ref{sec:equilibrium}), terms which are $\propto \tau^2$ are neglected. Consequently, also $\Delta v_{0}=0$ in the one-fluid formulation.

\subsection{One-fluid linearised equations}

The (scaled) linear perturbation Eqs. are obtained in the same fashion as those of the two-fluid Eqs. in \S \ref{sec:linpert}.
From the linearized incompressibility condition (\ref{eq:incomp1f}), we obtain the pressure perturbation 
\begin{equation}\label{eq:dpres1f}
\begin{split}
\delta P & = \frac{i}{k^2}\Big[ (H_{r} N_{r}^2 k_{x} + H_{z} N_{z}^2 k_{z}) \delta \rho_{g} - \frac{
    (1 + \epsilon_{0})^2  }{\epsilon_{0}  \tau}(k_{x} \delta u + k_{z} \delta w)\\
    & \quad  - \frac{
    (-1 + \epsilon_{0})  (-2 \eta k_{x} + (1 + \epsilon_{0}) k_{z} z_{0})}{ \epsilon_{0} (1 + \epsilon_{0}) }\delta \epsilon \Big]
\end{split}
\end{equation}

We note that Eq.~(\ref{eq:dpres1f}) is not valid in the limit $\tau\epsilon_0 \to 0$. 
That is, in this limit, Eq.~(\ref{eq:incomp1f}) cannot be used to compute the pressure perturbation $\delta P$. In that case, the dusty gas then becomes incompressible with
\begin{equation}\label{eq:dincompmod}
    \delta w = -\frac{k_{x}}{k_{z}} \delta u.
\end{equation}
Using this relation we can linearise the $x-$ and $z$-components of the momentum Eq.~(\ref{eq:contv1f}) and combine these to obtain$\delta P$
The resulting expression can then be used in Eqs. (\ref{eq:dcontu1f})-(\ref{eq:dcontrhog1f}).

The remaining linearised equations are
\begin{equation}\label{eq:dconteps}
 \sigma \delta \epsilon  = -i  (1+\epsilon_{0}) (k_{x} \delta u + k_{z} \delta w) -i (k_{x} u_{0} + k_{z} w_{0}) \delta \epsilon  ,   
\end{equation}
\vskip -0.5cm
\begin{equation}\label{eq:dcontu1f}
\begin{split}
 \sigma \delta u & = -\frac{H_{r}N_{r}^2}{1+\epsilon_{0}} \delta \rho_{g} + 2 \delta v -i (k_{x} u_{0} + k_{z} w_{0})\delta u -\frac{i k_{x}}{1+\epsilon_{0}}\delta P \\
 & \quad - \frac{2 \eta}{(1+\epsilon_{0})^2} \delta \epsilon,
 \end{split}
\end{equation}
\vskip -0.25cm
\begin{equation}\label{eq:dcontv1f}
 \sigma \delta v  = -\frac{1}{2}\delta u -i( k_{x} u_{0} + k_{z} w_{0}) - q_{z} \delta w,
\end{equation}
\vskip -0.5cm
\begin{equation}\label{eq:dcontw1f}
\begin{split}
\sigma \delta w & = -\frac{H_{z}N_{z}^2}{1+\epsilon_{0}} \delta \rho_{g} -i (k_{x} u_{0} + k_{z} w_{0})\delta w -\frac{i k_{z}}{1+\epsilon_{0}}\delta P + \frac{z_{0}}{1+\epsilon_{0}}\delta \epsilon,
 \end{split}
\end{equation}
\vskip -0.5cm
\begin{equation}\label{eq:dcontrhog1f}
\begin{split}
\sigma \delta \rho_{g} & =
\frac{1}{H_{r}}\delta u + \frac{1}{H_{z}}\delta w +\delta \rho_{g}\bigg(-\frac{1}{\beta} -\frac{\epsilon_{0}\tau}{(1+\epsilon_{0})^2}  (N_{r}^2 + N_{z}^2 + 2 i \eta k_{x}) \\
\quad & -i(k_{x} u_{0} + k_{z} w_{0}) - \frac{i\epsilon_{0} \tau k_{z} z_{0}}{1+\epsilon_{0}}\bigg) - \frac{i\epsilon_{0}\tau}{(1+\epsilon_{0})^2}\left(\frac{k_{x}}{H_{r}} + \frac{k_{z}}{H_{z}} \right)\delta P \\
\quad & -\frac{(-1+\epsilon_{0})\tau}{(1+\epsilon_{0})^2}\left(\frac{2\eta}{(1+\epsilon_{0})H_{r}}-\frac{z_{0}}{H_{z}} \right)\delta \epsilon.
 \end{split}
\end{equation}

From the energy Eq.~(\ref{eq:dcontrhog1f}) one can define an \emph{effective} cooling time of the dusty gas (see also \S \ref{sec:cos_onef_cool})
 \begin{equation}\label{eq:beta_eff_full}
 \begin{split}
     & \beta_{\text{eff}} = \Bigg[ \frac{1}{\beta} +  i \left( \frac{2 \eta  \epsilon_{0} \tau k_{x}}{\left(1+\epsilon_{0}\right)^2} + k_{x} u_{0} + k_{z} w_{0} +\frac{\epsilon_{0}\tau k_{z}z_{0}}{1+\epsilon_{0}} \right) \\
     \quad & +\frac{\epsilon_{0} \tau}{(1+\epsilon_{0})^2}\left( 
 \left[ \mu_{z}^2-\mu_{x}\mu_{z} \frac{H_{r}}{H_{z}}\right]N_{r}^2 
 +\left[ \mu_{x}^2-\mu_{x}\mu_{z}\frac{H_{z}}{H_{r}} \right]N_{z}^2\right)
      \Bigg]^{-1}, 
 \end{split}    
 \end{equation}
 where we used (\ref{eq:dpres1f}) to replace the pressure perturbation $\delta P$.
This effective cooling time is generally different from the pure gas cooling time $\beta$, and in particular, is generally complex-valued. In \S \ref{sec:cos_nodp} we show that the presence of dust can either give rise to a quasi thermal instability, or enhance the cooling rate compared to a pure gas, depending on the sign of $N_{r}^2$. Furthermore, a complex cooling time implies a phase shift between the gas density perturbation and the cooling term, and can give rise to the DCOS, as  discussed in \S \ref{sec:cos_dp} and Appendix \ref{app:dcos}.


\section{Inertial waves in the presence of vertical buoyancy, vertical shear, and cooling}\label{app:iw_bifurc}

In this section, we investigate linear modes in a pure gas disc with vertical buoyancy, vertical shear, and cooling. These  are governed by the cubic dispersion relation, Eq. (\ref{eq:detgas_z}), and we shall find that they have a non-trivial dependence on the cooling time $\beta$. The solution method outlined in \S\ref{sec:gas_midplane} assumes complex roots, but there also exist purely growing (or decaying) modes with $\sigma_I=0$. For one, this is the cooling mode. Generally, if we assume sufficiently small values of $\beta$, such that $\left|\sigma \beta\right| \ll 1$, Eq. (\ref{eq:detgas_z}) can be approximated by the quadratic dispersion relation
\begin{equation*}
    \sigma^2 + \beta \widetilde{\xi} \sigma + \widetilde{\mu}_{z}^2=0,
\end{equation*}
such that the cooling mode has been discarded, but the effect of cooling on the remaining solutions is retained.
The latter solutions read
\begin{equation}\label{eq:quadsol}
    \sigma = -\frac{1}{2}\mu_{x}^2 N_{z}^2 \beta  \pm \frac{1}{2}\sqrt{\mu_{x}^4 N_{z}^4 \beta^2 - 4 \mu_{z}^2\left(1-2 q_{z} \frac{k_{x}}{k_{z}}\right)}.
\end{equation}
These are either real- or complex-valued, depending on the sign of the discriminant. We conclude that if the discriminant is negative, the solutions describe inertial waves modified by buoyancy and vertical shear. On the other hand, the solutions are non-oscillatory if it is positive. Note that (\ref{eq:quadsol}) agrees with (\ref{eq:vsi_bet}) to first order $\beta$.
Thus, a bifurcation should occur for a specific ratio of radial and vertical wavenumbers, $k_{x,\text{bif}}/k_{z}$, for which the square root vanishes. At this bifurcation, the complex conjugate pair describing inertial waves with identical decay rates will become a pair of non-oscillatory modes with different decay rates.  In the case $q_{z}=0$ we readily find 
\begin{equation}\label{eq:kxbif}
    \frac{k_{x,\text{bif}}}{k_{z}} = \sqrt{\frac{4+2\beta N_{z}^4}{\beta^2 N_{z}^4}} \approx \frac{2}{\beta N_{z}^2}.
\end{equation}
In the limit $k_{x} \gg k_{z}$, Eq. \ref{eq:quadsol} yields $\sigma=0$ and $\sigma=-\beta N_{z}^2$.
The case $q_{z}\neq 0$ yields a quartic equation for $k_{x,\text{bif}}$, such that solutions must be obtained numerically.

\begin{figure*}
\centering 
\includegraphics[width = \textwidth]{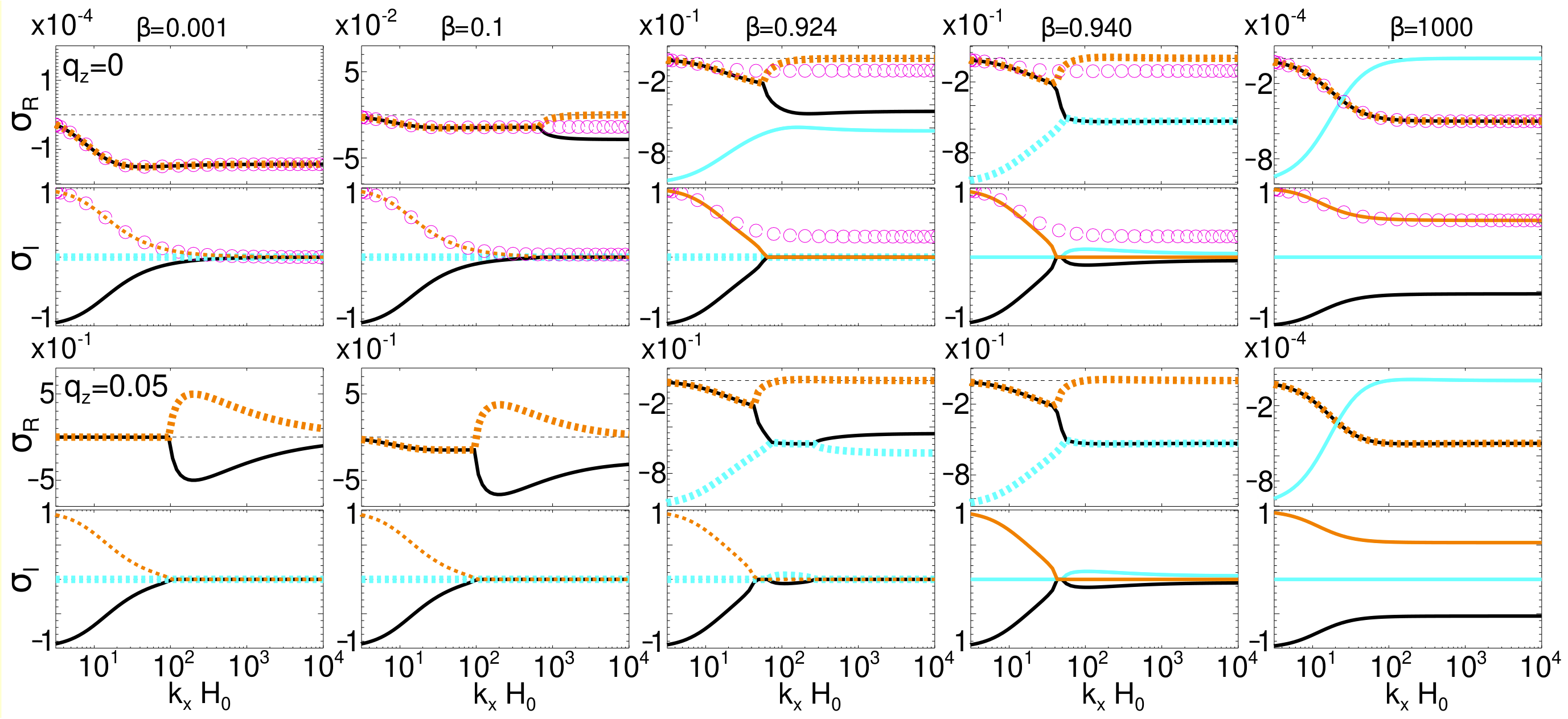}
\caption{Numerical solutions for the three pure gas eigenvalues resulting from the cubic dispersion relation Eq. (\ref{eq:detgas_z}) with $N_{z}^2=0.28$, $H_{z}=-0.11$ and $N_{r}^2=0$ (growth rate $\sigma_{R}$ and frequency $\sigma_{I}$). Generally, these are the cooling mode (cyan curve) and a pair oscillatory modes (black and orange curves), which for small $k_{x}$ are inertial waves. The upper (lower) two rows consider solutions without (with) vertical shear $q_{z}=0.05$. The open circles represent the analytical solutions (\ref{eq:fr_nz})-(\ref{eq:gr_nz}) for $\sigma_{I}$ and $\sigma_{R}$, respectively. For cooling times $\beta\lesssim 1$ the inertial wave eigenvalues undergo a bifurcation at a given $k_{x}\equiv k_{x,\text{bif}}$, where the growth rates `fork' into two different values and the frequencies drop to zero. We find that for $\beta \lesssim 0.6$ (including the first two columns) the value of $k_{x,\text{bif}}$ is well described by (\ref{eq:kxbif}).
For $\beta\sim 1$ the cooling mode starts to drop to smaller values and eventually vanishes with increasing $k_{x}$, via a prior merging with one of the inertial waves modes (third and fourth columns). In the adiabatic limit (the last column), the three modes are again clearly distinguishable. 
Note that for $q_{z}=0.05$ and $\beta \gtrsim 1$ there exist oscillatory modes (with $\sigma_{I}\neq 0$). This is in contrast to the isothermal limit and explains the occurrence of a lower resonant DSI `branch' (Figure \ref{fig:dsi_vsi_damp}).}
\label{fig:iwav_bifurc}
\end{figure*}

To anchor our discussion, we compute numerical solutions with vertical buoyancy $N_{z}^2=0.28$ and possibly with vertical shear $q_{z}=0.05$ (see \S \ref{sec:pars}). Figure \ref{fig:iwav_bifurc} shows all of the three growth rates and corresponding frequencies resulting from the cubic dispersion relation (\ref{eq:detgas_z}), for increasing cooling time $\beta$ from left to right as indicated. Furthermore, the plots in the upper and lower rows are computed without and with vertical shear, respectively. 

Let us first consider the panels with $\beta=0.001$, corresponding to the isothermal limit. The calculation with $q_{z}=0$ produces a pair of decaying inertial waves with oscillation frequencies of $\pm1$ in the limit $k_x\to0$, which are represented by the black and orange curves. In addition, the cooling mode appears (the dashed cyan curve), with vanishing frequency $\sigma_{I}=0$  and with a large decay rate $\sigma_{R}=-10^{3}$, far outside of the displayed region. In this and all other panels with $q_{z}=0$ the circles represent the analytical expressions 
%
%
 \begin{align}\label{eq:fr_nz}
     \sigma_{I} & = \pm \sqrt{\mu_{z}^2 + \mu_{x}^2 N_{z}^2 + \frac{2}{\beta}\sigma_{R} + 3 \sigma_{R}^2},\\
     \sigma_{R} & = -\frac{\beta N_{z}^2 \mu_{x}^2}{2 \left[ \beta^2 \left(\mu_{z}^2 + N_{z}^2 \mu_{x}^2\right) +1\right]}\label{eq:gr_nz}, 
 \end{align} 
 which are the frequency and growth rate of pure gas inertial waves following from (\ref{eq:freq}) and (\ref{eq:grate}) using 
 \begin{equation}\label{eq:xiz}
     \xi = \frac{k_{z}^2 + N_{z}^2 k_{x}^2}{k^2}.
 \end{equation}
In this case, these expressions agree well with the numerical result. The same holds for the solutions described by (\ref{eq:quadsol}) (not shown). 
If $q_{z}=0.05$ (lower panels) the two VSI modes (one growing and one decaying) emerge for sufficiently large $k_{x}$, the latter determined by the condition $\widetilde{q}_{z}>0$ [Eq. (\ref{eq:vsi_crit}]. We note that for small $k_{x}$ the frequencies with and without vertical shear are identical. Only if $\widetilde{q}_{z}\gtrsim0$ the frequencies with $q_{z}=0.05$ sharply drop to zero (local VSI modes are non-oscillatory \citep{latter2018}).


The panels corresponding to $\beta=0.1$ show an interesting result. We first consider the case $q_{z}=0$. The eigenvalues of the pair of inertial waves appear to undergo a bifurcation at a given $k_{x}$. This is most clearly seen in the growth rates $\sigma_{R}$, which start to `fork' into two distinct curves around $k_{x} H_{0} \sim 800$. For these wavenumbers, the frequencies drop to zero. With $q_{z}=0.05$ this bifurcation results in the two VSI modes, similar to the case with $\beta=0.001$. In particular, this bifurcation is not described by the analytical expressions (\ref{eq:fr_nz}) and (\ref{eq:gr_nz}). The reason is that the method used to derive these expressions requires modes to be oscillatory. However, it is accurately captured by the solutions (\ref{eq:quadsol}) (not shown).

Around cooling times $\beta\sim 1$, the eigenvalues show an even more complicated behaviour. The third column shows the eigenvalues with and without vertical shear for $\beta=0.924$. Here we see for $q_{z}=0$ (the upper panels) that the cooling mode (the cyan curve) decay rate decreases, such that it approaches the decay rate of one of the two inertial waves (the black curve). In the case with $q_{z}=0.05$ (lower panels) the two curves already start to merge in a complex manner for this value of $\beta$, resulting in another bifurcation. The same happens for the curves with $q_{z}=0$ for a slightly larger $\beta$. For an even (slightly) larger value $\beta\sim 0.940$ the cooling mode decay rate becomes identical to the decaying inertial wave's decay rate for a given $k_{x}$, as seen in the fourth row. 

Our interpretation of the situation with $q_{z}=0$ is that for sufficiently large $k_{x}$, a pair of decaying inertial waves exists, whereas the cooling mode disappears. That is, the cooling mode attains $\sigma_{R}=\sigma_{I}=0$ at large $k_{x}$. This is more clearly seen in the case with $\beta=10^3$ (the last column). We speculate that the cooling mode disappears once its decay rate coincides with the decay rate of one of the inertial waves. In the presence of vertical shear $q_{z}=0.05$, we speculate that the cooling mode becomes the (non-oscillatory) growing VSI mode for $\beta \gtrsim 0.940$. It is also noteworthy that for $\beta=10^3$ the frequencies and decay rates of the inertial wave pair are again correctly described by the analytical expressions
(\ref{eq:fr_nz}) and (\ref{eq:gr_nz}). Thus, deviations from these expressions occur at intermediate cooling times $\beta\sim 1$. The complex behavior around $\beta=1$ is not captured by the quadratic solutions (\ref{eq:quadsol}) either, which we find to be accurate for $\beta \lesssim 0.6$.

Nevertheless, while a thorough analysis of these modes, which might or might not confirm our speculation is beyond the scope of this paper, we notice that for $\beta \gtrsim 1$ modes with positive and negative frequencies exist for large $k_{x}$ in the presence of vertical shear. This is not the case for smaller $\beta$ (see the left three columns). The presence of these modes explains the occurrence of a \emph{lower} resonant `branch' of DSI modes in Figure \ref{fig:dsi_vsi_damp}, which indeed requires the presence of pure gas waves with negative frequency based on the RDI concept (\S \ref{sec:res_drag_instab}).

\section{Damping of inertial waves by dust-gas drag}\label{app:dusty_iw}

In \S \ref{sec:cos_dp} we described the effect of dust on linear COS modes in the disc midplane and argued that two effects can be distinguished. One effect is that dust leads to an effectively reduced buoyancy frequency, resulting in an overall reduction of the growth rates. In addition, COS growth rates experience a cutoff beyond a critical radial wavenumber. Furthermore, in \S \ref{sec:cos_general} we found that away from the disc midplane an additional cutoff occurs beyond a critical vertical wavenumber. 

Here, we derive analytical expressions for the damping rates of inertial waves in a dusty gas, explaining the radial and vertical wavenumber cutoffs. Since the damping mechanism leading to these cutoffs does not rely on buoyancy or cooling, we can neglect these aspects here, i.e. we set $N_{r/z}^2\to 0$ and $H_{r/z}\to \infty$, such that $H_{r/z} N_{r/z}^2\to 0$, and discard the energy equation, which simplifies the analysis. 

To proceed, we  consider a region at distance $z_{0}\geq 0$ away from the disc midplane.
Numerical experimentation with the full eigenvalue problem (\ref{eq:eigenproblem}) reveals that it is not the vertical dust-drag term [the term $\propto w_{d0}$ in Eq. (\ref{eq:dpress})], 
but the dust's vertical advection terms [the terms $\propto w_{d0}$ in Eqs. (\ref{eq:linrhod})---(\ref{eq:linwd})], which result in the vertical wavenumber cutoff. 
Therefore, the vertical drag-force term in (\ref{eq:dpress}) is ignored. It should be noted that the advection of dust in the vertical direction (with velocity $w_{d0}$) describes a drift between dust and gas, which is mediated via drag forces, i.e. terms $\propto \delta \vec{\mathbf{v}}_{d}-\delta\vec{\mathbf{v}}_{g}$.

Furthermore, we omit the radial pressure gradient ($\eta=0$), which would otherwise result in a radial drift between dust and gas, and which would also trigger the SI. This approximation is valid in the vicinity of a pressure bump or generally for $\epsilon_{0}\ll 1$. This also implies that at sufficiently large heights $z_{0}$, where we indeed expect $\epsilon_{0}\ll 1$, the effect of the pressure gradient becomes negligible compared to the effect of vertical dust advection \citep{squire2018} 

We thus consider the equations
\begin{align}
    \sigma \delta u_{g} & = 2 \delta v_{g} +\frac{\epsilon_{0}}{\tau} \left(\delta u_{d} -\delta u_{g} \right) -i k_{x} \delta P,\label{eq:ug2}\\
    \sigma \delta v_{g} & = -\frac{1}{2}\delta u_{g} +\frac{\epsilon_{0}}{\tau}\left(\delta v_{d} -\delta v_{g} \right),\label{eq:vg2}\\
    \sigma \delta u_{d} & = 2 \delta v_{d} +\frac{1}{\tau}\left(\delta u_{g} -\delta u_{d} \right)- i k_{z} w_{d0} \delta u_{d},\label{eq:ud2}\\
    \sigma \delta v_{d} & = -\frac{1}{2}\delta u_{d} +\frac{1}{\tau} \left(\delta v_{g} -\delta v_{d} \right)- i k_{z} w_{d0} \delta v_{d}\label{eq:vd2}.
\end{align}
The pressure perturbation is now given by 
\begin{equation}\label{eq:dp2}
    \delta P = \frac{i}{k^2}\left[-2 k_{x} \delta v_{g} -\frac{\epsilon_{0}}{\tau}\left(k_{x}\delta u_{d} + k_{z} \delta w_{d}\right)\right].
\end{equation}
From the vertical dust momentum equation (\ref{eq:linwd}) we obtain
\begin{equation}\label{eq:wd2}
    \delta w_{d} = -\frac{k_{x}}{k_{z}}\frac{\delta u_{g}}{\widehat{\sigma}_{d,z}},
\end{equation}
where we define
    \begin{equation}\label{eq:sighat_z}
    \widehat{\sigma}_{d,z}  = \sigma \tau + i a_{d,z} \tau +1
\end{equation}
with $a_{d,z}$ given by (\ref{eq:adv_dz}), describing vertical dust advection.
From Eqs. (\ref{eq:ug2})---(\ref{eq:wd2})
we then derive the fifth order dispersion relation
\begin{equation}
\begin{split}\label{eq:disprel_cos_z}
   0 & = \left(\sigma_{\epsilon}\widehat{\sigma}_{d,z} -\epsilon_{0}\mu_{x}^2\right)\left[\left(\widehat{\sigma}_{d,z}^2+\tau^2\right)\sigma_{\epsilon}-\epsilon_{0}\widehat{\sigma}_{d,z}\right]\\
   \quad & -\mu_{z}^2\left[-\tau^2\left(\widehat{\sigma}_{d,z}^3 + \left(2 \epsilon_{0} +\tau^2\right)\widehat{\sigma}_{d,z}\right) + \epsilon_{0} \widehat{\sigma}_{d,z}^2 \sigma_{\epsilon} -\epsilon_{0}^2 \widehat{\sigma}_{d,z}\right],
    \end{split}
\end{equation}
 where we additionally defined
 \begin{equation}
     \sigma_{\epsilon}  = \sigma \tau +\epsilon_{0}
     \end{equation}
     to shorten the notation.

Similar to \S \ref{sec:cos_nodp}, we apply a series solution 
for the eigenvalue
\begin{equation*}
    \sigma = \sigma_{0} + \sigma_{1}\tau + \sigma_{2}\tau^2 + \sigma_{3} \tau^3+ \ldots 
\end{equation*}
which we insert into (\ref{eq:disprel_cos_z}), and the latter is then 
solved order by order in $\tau$, yielding 
the solutions
\begin{align}
    \sigma_{0} &= \pm i \mu_{z},\label{eq:s0}\\
    \sigma_{1} &=-\frac{\epsilon_{0} \left(\mu_{x}^2 -2 i k_{z} z_{0}\right)}{2\left(1+\epsilon_{0}\right)},\\
    \sigma_{2} &= \mp i \frac{\epsilon_{0} \left(4 + \epsilon_{0} \left(4-5 \mu_{x}^2 \right) -4 \mu_{x}^2 \right) \mu_{x}^2}{8 \left(1+\epsilon_{0}\right)^2 \mu_{z}},
    \end{align}
    \begin{equation}
 \begin{split}
    \sigma_{3} &= \frac{\epsilon_{0}}{2 \left(1+\epsilon_{0}\right)^3}\bigg[-\epsilon_{0} \left(3+2 \epsilon_{0}\right) \mu_{x}^4 -2 k_{z}^2 z_{0}^2 \\
    \quad & + \mu_{x}^2 \left(1+\epsilon_{0}^2 -3 i k_{z} z_{0} + \epsilon_{0}\left(2-i k_{z} z_{0}\right)\right) \bigg].\label{eq:s3}
\end{split}
\end{equation}

For perfectly coupled dust ($\tau\equiv0$), we recover inertial waves with oscillation frequencies $\pm\mu_z$. The terms $\sigma_1$, $\sigma_2$, $\sigma_3$, etc. then represent the effect of finite dust-gas drag on inertial waves. 
From these we obtain the damping rate
\begin{equation}\label{eq:gr_iw_dust}
   \sigma_{R} \approx -\frac{\epsilon_{0} \mu_{x}^2\tau}{2 \left(1+\epsilon_{0}\right)} -\frac{\epsilon_{0} k_{z}^2 z_{0}^2 \tau^3}{\left(1+\epsilon_{0}\right)^3},
\end{equation}
where we neglected relative corrections of $\mathcal{O} \left(\tau^2\right)$ to the first term which are independent of $z_{0}$.
In the limit $z_{0}\to 0$, we recover the dusty damping term appearing in the approximate COS growth rates (\ref{eq:cos_tau}) and (\ref{eq:cos_tau_2}),  resulting in the radial cutoff (\ref{eq:kxcut}). On the other hand, the damping term $\propto z_{0}$ results in the vertical cutoff (\ref{eq:kz_cut}). 



Furthermore, the wave frequency is given by 
\begin{equation}\label{eq:fr_cos_z}
    \sigma_{I} = \pm \mu_{z} \mp \frac{\epsilon_{0}\tau^2  \left[4 + \epsilon_{0} \left(4-5 \mu_{x}^2 \right) -4 \mu_{x}^2 \right] \mu_{x}^2}{8 \left(1+\epsilon_{0}\right)^2 \mu_{z}}.
\end{equation}
We note that (\ref{eq:s0})---(\ref{eq:fr_cos_z}) are not valid in the limit $k_{z}\to 0$. In this limit the entire last bracket in (\ref{eq:disprel_cos_z}) vanishes and we find that $\sigma_{0}$ and $\sigma_{2}$, and hence $\sigma_{I}$ vanish, since inertial waves cannot exist in this limit.
Nevertheless, for small $k_{z}$ the frequency $\sigma_{I}$ can in principle take large values. However, we expect the fluid approximation for dust adopted throughout this paper to break down once $|\sigma_{I} \tau| \gtrsim 1$ \citep{jacquet2011}.

\section{Simplified model for the COS and the DCOS}\label{app:dcos}

In \S \ref{sec:cos_dp}, we discovered an instability of a dusty gas possessing long cooling times $\beta \gg 1$, requiring both an unstable entropy gradient $N_{r}^2<0$ and a background radial pressure gradient $\eta \neq 0$. We hypothesized that this instability is a dust-induced version of the COS, i.e. the DCOS.

In this appendix we show using a simple `toy' model - which captures the physics essential to the classic COS in a pure gas - that the latter is actually amplified if one allows for a phase lag between the cooling term and the density (or entropy) perturbation $\delta \rho_{g}$. Indeed, Eq.~(\ref{eq:beteff}), which is the \emph{effective} cooling time of a \emph{dusty} gas within the parameter regime considered in \S \ref{sec:cos_dp}, implies
\begin{equation}\label{eq:beteff2}
    \beta_{\text{eff}} = A_{\beta} \exp\left[ i \varphi_{\beta} \right],
\end{equation}
such that the effective cooling term becomes
\begin{equation}\label{eq:lambda_eff}
    \Lambda_{\text{c,eff}} = -\frac{1}{A_{\beta}}\exp\left[-i \varphi_{\beta} \right] \delta \rho_{g},
\end{equation}
with a cooling time $A_{\beta}$ and a phase lag $\varphi_{\beta}$ of the cooling term with respect to the density perturbation $\delta \rho_{g}$.

\begin{figure}
\centering 
\includegraphics[width = 0.49\textwidth]{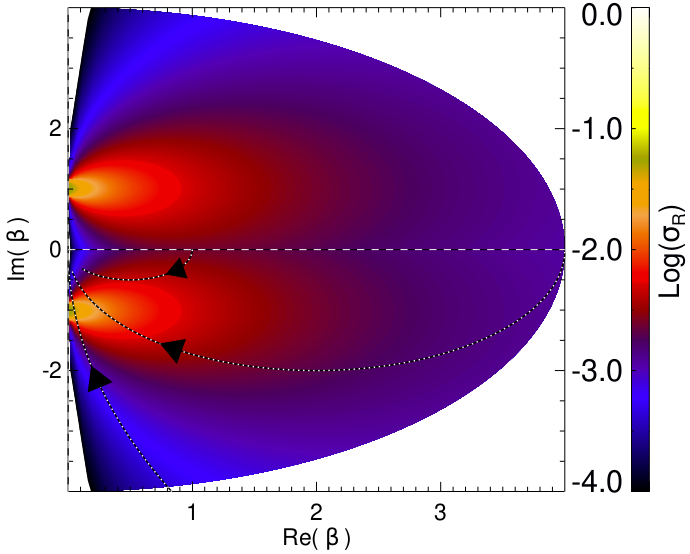}
\caption{Contours of maximum linear growth rates of a pure gas (across the same wavenumbers as in Figure \ref{fig:adiab_instab}) obtained from (\ref{eq:deteqz}) with $N_{r}^2=-0.01$ and an \emph{effective}, complex-valued cooling time (\ref{eq:beteff2}). As explained in the text, the COS is increasingly enhanced for a cooling time $A_{\beta}\equiv\sqrt{\text{Re}(\beta)^2 + \text{Im}(\beta)^2}=1$ 
 and an increasing phase delay $0<\varphi_{\beta}\equiv \arctan \text{Im}(\beta)/\text{Re}(\beta)< \pi/2$ between the cooling term (\ref{eq:lambda_eff}) and the gas entropy perturbation $\delta \rho_{g}$ (or decreasing to $-\pi/2$ for the corresponding other inertial wave), occurring in the one-fluid energy equation (\ref{eq:contrhog1f}). The over-plotted curves show the evolution of the effective cooling time (\ref{eq:beteff}) of a dusty gas for increasing $0<k_{x} H_{0} <10^5$ (indicated by the arrows) for three different values $\beta=1,4,20$ and $\tau=10^{-2}$, $\epsilon_{0}=0.03$. These curves illustrate how with increasing $\beta$ the maximum growth rates of the DCOS increase, and at the same time the domain of largest growing modes in $k_{x}$-space shrinks.}
\label{fig:bet_phas}
\end{figure}

In Figure \ref{fig:bet_phas} we present maximum growth rates of inertial waves in a pure gas obeying Eq.~(\ref{eq:deteqz}) in absence of vertical buoyancy, and with $\beta$ replaced by Eq.~(\ref{eq:beteff2}). We consider a range of cooling times $0<A_{\beta}<10$, and phase lags $-\pi/2<\varphi_{\beta}<\pi/2$. In this figure the real positive axis corresponds to the classical (pure gas) COS without a phase lag and where the growth rates are largest if $\beta=1$, as expected. However, we find that the growth rates are enhanced compared to the classical values with $A_{\beta}=\beta=1$ in case of a phase shift $0\leq\varphi_{\beta}<\pi/2$. This applies to one of the two COS modes, while the other mode is being damped. For a negative phase lag, the situation is reversed. The growth rate of the amplified mode maximizes for $\varphi_{\beta}\to \pi/2$.
This is in agreement with simplified model calculations presented below. We note that the COS is entirely extinguished for $3\pi/2 \leq \varphi_{\beta} \leq \pi/2$ where the real part $\text{Re}[\beta_{\text{eff}}]\leq 0$. In this parameter region the cooling mode would give rise to thermal instability (cf. \S \ref{sec:cos_nodp}).

The trajectories in Figure \ref{fig:bet_phas} show the evolution of $\beta_{\text{eff}}$ corresponding to (\ref{eq:beteff}) with increasing radial wavenumber $k_{x}$ for three different values $\beta=1,4,20$. Hence, the curves trace the corresponding \emph{maximum} (across all wavenumbers) COS growth rates with $\beta$ given by (\ref{eq:beteff2}). This figure explains qualitatively the behavior of the DCOS displayed in Figure \ref{fig:adiab_instab} for increasing $k_{x}$ through an increasing phase lag between gas cooling and entropy perturbation. Particularly, the larger $\beta$, the closer the achieved phase lag is to $\pi/2$, which gives rise to a larger growth rate. The phase lag also increases with increasing particle Stokes number, and maximizes for a dust-to-gas density ratio $\epsilon_{0}=1$.

In order to show that a phase lag between the gas density perturbation and the cooling term does indeed amplify the COS we consider here a simplified linearised entropy equation for a pure gas  
\begin{equation}\label{eq:toy}
 \frac{\partial}{\partial t}\delta \rho_{g}    = -\frac{1}{\beta_{\text{eff}}}\delta \rho_{g} -  \delta u_{g} ,
\end{equation}
 where we assumed a negative entropy gradient ($H_{r}<0$) and absorbed its magnitude into $\delta \rho_{g}$ for notational brevity.
Furthermore, we assume a plane wave solution
\begin{equation}\label{eq:toy_u}
    \delta u_{g} = \exp{\left[i \omega t\right]},
\end{equation}
with frequency $\omega$. For the density perturbation we use the ansatz
\begin{equation}
    \delta \rho_{g} = A \exp{\left[i \left(\omega t + \varphi \right)\right]}.
\end{equation}
Inserting the above expressions into (\ref{eq:toy})
and solving for real and imaginary parts separately results in
\begin{align}
    A & =  \left[ \left(\omega - \frac{\sin\varphi_{\beta}}{A_{\beta}} \right)^2 + \left(\frac{\cos\varphi_{\beta}}{A_{\beta}}\right)^2 \right]^{-1/2}\label{eq:cos_amp},\\
    \varphi & = \arccos\left[-\frac{A}{A_{\beta}} \cos \varphi_{\beta}\right]\label{eq:cos_phas}.
\end{align}

Now let us recall that the COS relies on the circumstance that a gas parcel that moves radially outward from its original radius will get in contact with colder material such that it will deposit some of its excess entropy during the first half of the epicycle. When it has returned to its original radius it is colder (and hence denser) than its surroundings and will therefore experience an inward buoyancy force. One can now argue that the instability should generally be promoted whenever a gas parcel is denser than its surroundings during its inward motion and when it is less dense than its surroundings during its outward motion. That is, $\re(\delta\rho_g)$ and $\re(\delta u_g)$ should be anti-correlated (cf. \S \ref{sec:prelim_cos}). 

We therefore propose that the optimal cooling time for the COS is the same cooling time for which the action integral
\begin{equation}\label{eq:action}
    \mathcal{W} \equiv -\int\limits_{0}^{2 \pi/\omega} \mathrm{d}t \, \text{Re}\left[\delta \rho_{g}\right] \text{Re}\left[\delta u_{g}\right] 
\end{equation}
will be maximal.

\begin{figure}
\centering 
\includegraphics[width = 0.49\textwidth]{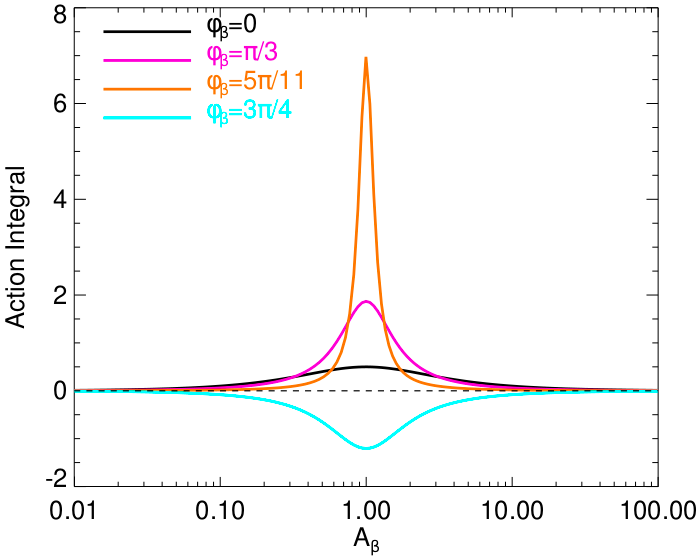}
\caption{Values of the action integral (\ref{eq:action}) for different values of the phase shift $\varphi_{\beta}$ and for varying amplitude $A_{\beta}$, appearing in (\ref{eq:cos_amp}) and (\ref{eq:cos_phas}). The plot shows how with $\varphi \to \pi/2$ the maximum value of the action integral (and hence the COS growth rate) increases, and at the same time the curve sharpens. Note that for $\varphi\gtrsim 5/11 \pi$ the integral rapidly increases. On the other hand, phase shifts $\varphi>\pi/2$ result in negative values, signifying a decay of COS modes, as illustrated by the curve with $\varphi=3\pi/4$. The classic COS with vanishing phase shift $\varphi=0$ is represented by the black curve. The plots assume a dimensionless wave frequency $\omega=1$.}
\label{fig:bet_phas_2}
\end{figure}

Figure \ref{fig:bet_phas_2} shows the value of $\mathcal{W}$ for a number of complex cooling times and dimensionless frequency $\omega=1$. The interpretation is the same as described above. Namely, the amplitude is the actual cooling time and the complex phase corresponds to the phase lag between the gas density perturbation and the cooling term in Eq.~(\ref{eq:toy}). Thus, for increasing $\varphi$ cooling will be increasingly delayed. Note that for simplicity here we describe only the behavior of one of the two inertial waves present in the gas, determined by our ansatz (\ref{eq:toy_u}) with $\omega>0$. Therefore, only positive phase lags lead to an amplified growth rate, in contrast to Figure \ref{fig:bet_phas}, which encompasses both inertial wave modes. Nevertheless, Figure  \ref{fig:bet_phas_2} illustrates, similarly to Figures \ref{fig:adiab_instab} and \ref{fig:bet_phas}, that a phase lag of $\varphi_{\beta}\sim\pi/2$ is optimal for the COS, together with a cooling time $A_{\beta}=1$.  Moreover, the "symmetry" of COS growth rates around the optimal value $A_{\beta}=1$ [which is clearly seen in Figure \ref{fig:si_vs_cos} (right most panels)]
is also explained using this simple model, as well as the occurrence of a peak for $\varphi_{\beta} \to \pi/2$. The latter corresponds to the narrowness of the unstable `branch' in Figure \ref{fig:adiab_instab}. Moreover, the model also predicts that for $\varphi\geq \pi/2$ the COS is extinguished 
, in agreement with (\ref{eq:grate_cos}). Note that the optimal cooling time for the COS is actually $A_{\beta}=1/\omega$, such that slower oscillations require slower cooling and vice versa. This is the reason why the optimal $k_{x}$ for the DCOS (Figure \ref{fig:adiab_instab}) becomes smaller for decreasing $k_{z}$, as the inertial wave frequency drops to values $<1$. Note also that all growth rates drawn in Figure \ref{fig:bet_phas} correspond to a wave frequency $\omega=1$ by selection of modes with appropriate wavenumbers $k_{z} \gg k_{x}$.


\section{Simplified Model for the DSI in presence of vertical buoyancy}\label{app:dsi}

In this appendix we use Eqs. (\ref{eq:dsi_ug_p})-(\ref{eq:dsi_rhod_p}) to derive an analytical expression for the growth rate of the DSI including the effect of vertical buoyancy. As explained in Section \ref{sec:dsi_vbuo} we set $\eta=0$, and ignore drag force terms $\propto \delta \vec{\mathbf{v}}_{d}-\delta\vec{\mathbf{v}}_{g}$. Furthermore, radial buoyancy is omitted. We then rewrite the aforementioned equations as:
\begin{align}
    \sigma \delta u_{g} & = 2 \mu_{z}^2 \delta v_{g} + \mu_{x}\mu_{z} \left(H_{z} N_{z}^2 \delta \rho_{g} -\frac{w_{d0}}{\tau} \delta \rho_{d}\right)\label{eq:dsi_ug},\\
    \sigma \delta v_{g} & = -\frac{1}{2}\delta u_{g}\label{eq:dsi_vg},\\
    \sigma_{\beta} \delta \rho_{g} & =  -\frac{k_{x}}{k_{z}}\frac{1}{H_{z}}\delta u_{g},\label{eq:dsi_rhog}\\
    \widehat{\sigma}_{d,z} \delta u_{d} & = \delta u_{g} + 2 \tau \delta v_{d}\label{eq:dsi_ud},\\
    \widehat{\sigma}_{d,z} \delta v_{d} & = \delta v_{g} -\frac{1}{2}\tau \delta u_{d}\label{eq:dsi_vd},\\
    \widehat{\sigma}_{d,z}\delta w_{d} &= -\frac{k_{x}}{k_{z}}\delta u_{g}\label{eq:dsi_wd},\\
    \left(\sigma + i a_{d,z} \right) \delta \rho_{d} & = -i \epsilon_{0} \left(k_{x} \delta u_{d} + k_{z} \delta w_{d}\right)\label{eq:dsi_rhod},
\end{align}
where we used (\ref{eq:sighat_z}) and (\ref{eq:adv_dz}) and defined
\begin{equation*}
    \sigma_{\beta}  = \sigma + \frac{1}{\beta}.
\end{equation*}

From these equations we directly derive the dispersion relation
\begin{equation}\label{eq:disp5}
\begin{split}
    & \left(\sigma + i a_{d,z}\right) \left[ \sigma^2 + \mu_{z}^2  +\mu_{x}^2 N_{z}^2 \frac{\sigma}{\sigma_{\beta}} \right]\left[\widehat{\sigma}_{d,z}^3 + \tau^2\widehat{\sigma}_{d,z}\right]\\
    \quad & +\mu_{x}^2 i \epsilon_{0} k_{z} w_{d0} \left[\widehat{\sigma}_{d,z} +\sigma\tau\right] =0.
    \end{split}
\end{equation}
This equation is too complex to proceed analytically.
We therefore assume that the gas is adiabatic ($\beta \to \infty$) such that
$\sigma_{\beta}\to \sigma$, and that $\left|a_{d,z}\tau\right| \ll 1$ and $\left|\sigma \tau\right| \ll 1$, on account of tight coupling between dust and gas. This implies $\widehat{\sigma}_{d,z}\to 1$ and allows us to write
\begin{equation}\label{eq:disp6}
  \sigma^3 + i a_{d,z} \sigma^2 + \left(\mu_{z}^2 + \mu_{x}^2 N_{z}^2 \right) \sigma + i a_{d,z} \left[\mu_{x}^2 \left(N_{z}^2+\epsilon_{0}\right)+\mu_{z}^2\right]=0,
\end{equation}
which is a cubic equation in $\sigma$, such that analytical solutions can be found (for instance by using computer algebra).

The pair of modified inertial waves representing the DSI following from (\ref{eq:disp6}) is described by the eigenvalues
\begin{equation}\label{eq:sig_dsi}
    \sigma_{\text{DSI}}  = \frac{1}{12}\Bigg[-4 i a_{d,z} + \frac{2^{\frac{4}{3}}\left(1\pm i\sqrt{3}\right) C }{X^{\frac{1}{3}}} +2^{\frac{2}{3}}\left(-1\pm i\sqrt{3}\right) X^{\frac{1}{3}}\Bigg],
\end{equation}
where
\begin{equation}
X = -i A + \sqrt{B}
\end{equation}
with
\begin{align}
    A & = a_{d,z} \left[ 27 \epsilon_{0} \mu_{x}^2 + 18 \left(\mu_{z}^2 + \mu_{x}^2 N_{z}^2 \right) -2 a_{d,z}^2\right]\label{eq:A},\\
    B & = 4 C^3 -A^2\label{eq:B},\\
     C & = 3 \left(\mu_{z}^2 + \mu_{x}^2 N_{z}^2\right) + a_{d,z}^2.
\end{align}
It can be easily verified that Eq.~(\ref{eq:sig_dsi}) reduces to $\sigma = \pm i\mu_{z}$ (as expected) in the limit $N_{z}^2\to0$ and $a_{d,z}\to 0$. 

We now proceed to compute the growth rate $\sigma_{R,\text{DSI}}\equiv \text{Re}\left[\sigma_{\text{DSI}}\right]$.
By applying basic algebra on (\ref{eq:sig_dsi}) we find 
\begin{equation}
\begin{split}\label{eq:gr_dsi}
    \sigma_{R,\text{DSI}} & = \frac{2^{\frac{2}{3}} \left(\cos \phi + \sqrt{3} \sin \phi \right)}{12} \chi^{\frac{1}{3}} \left[ 1- \frac{C}{\left(\frac{\chi}{2}\right)^{\frac{2}{3}}}  \right],
    \end{split}
\end{equation}
 where we find that formally two cases need to be distinguished:
\begin{align}
  \left\{\chi,\,\phi\right\} & =  \left\{A+\sqrt{|B|},\,\frac{7\pi}{6}\right\}  \hspace{2.7cm}   \text{if}   \, B\leq0,\label{eq:Blt0}\\
   \left\{\chi,\,\phi\right\} & = \left\{\sqrt{A^2 + B},\,\frac{1}{3}\left(\arctan \frac{-A}{\sqrt{B}} + 4 \pi\right)\right\}  \hspace{0.4cm} \text{if} \, B\geq0.
\end{align}
However, for $B\geq0$, we have $\chi = \sqrt{4 C^3}$ which readily yields $\sigma_{R,\text{DSI}}=0$ from (\ref{eq:gr_dsi}), as the expression inside the bracket vanishes. This implies that growth can only occur for $B<0$.

From now on we will restrict our attention to wavenumbers $k_{x}\gg k_{z}$ such that $\mu_{z}\to 0$ and $\mu_{x}\to 1$, which simplifies the calculation. Specifically, we wish to verify analytically that the region in $k_{z}$-space within which the DSI exists becomes increasingly small and is concentrated around $k_{z,\text{res}}$ [Eq.~(\ref{eq:kzrdi})], as indicated by Figure \ref{fig:dsi_vbuo}.

To proceed, we will first find the vertical wavenumbers $k_{z}$ for which $B=0$, as these delimit the region of positive growth rates. 
By setting $B=0$ one can derive from Eqs. (\ref{eq:A}) and (\ref{eq:B}) the quadratic equation
\begin{equation*}
    y^2 108 \left[  N_{z}^2 + \epsilon_{0} \right] + y \left[ 108 N_{z}^4 - \left(27 \epsilon_{0} + 18 N_{z}^2\right)^2\right] + 108 N_{z}^6 =0,
\end{equation*}
where we defined $y \equiv a_{d,z}^2$, with $a_{d,z}^2$ given by (\ref{eq:adv_dz}).
The solution of this equation is represented by the pair
\begin{equation}
    a_{d,z}^{\pm} = \left[\frac{27\epsilon_{0}^2 + 36 \epsilon_{0} N_{z}^2 + 8 N_{z}^4 \pm\sqrt{\epsilon_{0}\left(9 \epsilon_{0} + 8 N_{z}^2\right)^3}   }{8\left(\epsilon_{0} + N_{z}^2\right)}  \right]^{\frac{1}{2}}.
\end{equation}
In the limit $N_{z}^2 \gg \epsilon_{0}$ we find $a_{d,z}^{\pm}\to N_{z}$ such that indeed $k_{z} \to N_{z}/(\tau z_{0}) = k_{z,\text{res}}$. This means that with increasing vertical buoyancy $N_{z}^2$ the region of growing modes in $k_{z}$-space shrinks toward the resonant wavenumber (\ref{eq:kzrdi}).


Finally, we want to show that the maximal value of the growth rate is independent of the value of $N_{z}$. In the aforementioned limits we have $a_{d,z} \approx N_{z}$ and expressions (\ref{eq:A}), (\ref{eq:B}) read
\begin{equation}
    A  = 27 \epsilon_{0} N_{z} + 16 N_{z}^3,
\end{equation}
and
\begin{equation}
   B  = 256 N_{z}^6 \left[1 - \left(1+ \frac{27 \epsilon_{0}}{16 N_{z}^2}\right)^2\right] \approx - 864  \epsilon_{0} N_{z}^4,
\end{equation}
since $N_{z}^2 \gg \epsilon_{0}$,
such that $B<0$ as expected.
Hence we have
\begin{equation}
\begin{split}
\chi & =A + \sqrt{|B|}\\
\quad & = N_{z}\left(16 N_{z}^2 + 27 \epsilon_{0}\right) + 12\sqrt{6 \epsilon_{0}} N_{z}^2 \\
\quad & \approx 16 N_{z}^3  + 12\sqrt{6 \epsilon_{0}} N_{z}^2,
\end{split}
\end{equation}
since $N_{z} \gg \sqrt{\epsilon_{0}}$,
and thus
\begin{equation}\label{eq:chilim}
        \chi^{\pm \frac{1}{3}}  = 16^{\pm \frac{1}{3}} N_{z}^{\pm 1} \left[ 1 + \frac{3 \sqrt{6 \epsilon_{0}}}{4 N_{z}}\right]^{\pm \frac{1}{3}}  \approx 16^{\pm \frac{1}{3}} N_{z}^{\pm 1} \left[ 1 \pm \frac{\sqrt{6 \epsilon_{0}}}{4 N_{z}}\right].
\end{equation}
Using (\ref{eq:Blt0}) we find from (\ref{eq:gr_dsi})
\begin{equation*}
   \sigma_{R,\text{DSI}} = \frac{1}{2\sqrt{3}}\left[2^{\frac{7}{3}} N_{z}^2 \,\chi^{-\frac{1}{3}} -  2^{-\frac{1}{3}}\,\chi^{\frac{1}{3}} \right], 
\end{equation*}
which, using (\ref{eq:chilim}), readily yields the growth rate (\ref{eq:gr_dsi_lim}).

\section{Effect of Radial dust-gas drift on the VSI}\label{app:vsi_rad}

As outlined in \S \ref{sec:vsi_dust_p}, an equilibrium radial dust-gas drift has a damping effect on the VSI. In order to show this, we use Eqs. (\ref{eq:ug4p})---(\ref{eq:vd4p}) to derive an approximate dispersion relation for VSI modes in the presence of a background  dust-gas drift. Here we consider the isothermal limit without buoyancy and delete the gas entropy equation. Furthermore, as explained in \S \ref{sec:vsi_dust_mp}, we neglect terms involving dust density perturbations, $\delta \rho_d$. First, we rewrite the aforementioned equations, such that 
\begin{align}
\widehat{\sigma}_{g,x} \delta u_{g} & =  2\tau \mu_{z}^2 \delta v_{g} + \epsilon_{0} \left(\mu_{z}^2 \delta u_{d} + \mu_{x}^2 \frac{\delta u_{g}}{\widehat{\sigma}_{d,x}}\right)\label{eq:ug4},\\
\widehat{\sigma}_{g,x} \delta v_{g} & = \epsilon_{0}\delta v_{d} + \widetilde{q}_{z}\tau \delta u_{g}\label{eq:vg4},\\
\widehat{\sigma}_{d,x}\delta u_{d} & = \delta u_{g} + 2 \tau \delta v_{d}\label{eq:ud4},\\
\widehat{\sigma}_{d,x}\delta v_{d} & = \delta v_{g} -\frac{1}{2}\tau \delta u_{d} - q_{z}\tau \frac{k_{x}}{k_{z}}\frac{\delta u_{g}}{\widehat{\sigma}_{d,x}}\label{eq:vd4},
\end{align}
where we used (\ref{eq:linwd}) to replace the vertical dust velocity in (\ref{eq:ug4}) and (\ref{eq:vd4}),
and where we defined
\begin{align}
    \widehat{\sigma}_{d,x} & \equiv \sigma \tau + i a_{d,x}\tau  + 1,\label{eq:sighat_dx}\\
    \widehat{\sigma}_{g,x} & \equiv  \sigma \tau + i a_{g,x}\tau + \epsilon_{0}\label{eq:sighat_gx},
\end{align}
with $a_{d,x}$ given by (\ref{eq:adv_dx}), describing radial dust advection, and similarly
\begin{align}\label{eq:adv_gx}
    a_{g,x} = k_{x} u_{g0}, 
\end{align}
describing radial advection of the gas.
In what follows, we formally assume that the quantity 
\begin{equation}
   \widetilde{a}_{d,x} \equiv a_{d,x}\tau  \sim 1,
\end{equation}
 affording sufficiently large values of $k_{x}$. 
 On the other hand, since we assume $\epsilon_{0} \ll 1$ we neglect $|a_{g,x}| \sim \epsilon_{0} |a_{d,x}| \ll |a_{d,x}|$. Note that any common background drift of dust and gas can be absorbed into $\sigma$, and only leads to a frequency shift of the modes. It is only the \emph{relative} drift $a_{d,x}-a_{g,x}$ which can result in a possible damping of the VSI.

From Eqs. (\ref{eq:ug4})---(\ref{eq:vd4}) we find the dispersion relation
\begin{equation}
\begin{split}
    0 &= \left(\widehat{\sigma}_{g,x} \widehat{\sigma}_{d,x}-\epsilon_{0} \mu_{x}^2\right)\left(\epsilon_{0}\widehat{\sigma}_{d,x} - \widehat{\sigma}_{g,x}\left[\widehat{\sigma}_{d,x}^2 + \tau^2 \right] \right) \\
    \quad & + 2 \mu_{z}^2 \tau^2 \widetilde{q}_{z} \widehat{\sigma}_{d,x} \left( \widehat{\sigma}_{d,x}^2 + \tau^2 + \epsilon_{0} \right)\\
    \quad & - \mu_{z}^2 \epsilon_{0} \left(\epsilon_{0} \widehat{\sigma}_{d,x} - \widehat{\sigma}_{g,x} \left[ \widehat{\sigma}_{d,x}^2 + \tau^2 \right] - 2 \widetilde{q}_{z} \tau^2 \left[\widehat{\sigma}_{d,x}+\widehat{\sigma}_{g,x}\right]\right), 
    \end{split}
\end{equation}
which is of fifth-order and is formally written as
\begin{equation*}
    c_{0} + c_{1}\sigma + \ldots +c_{5} \sigma^5 =0.
\end{equation*}
In order to make analytical progress we first expand the coefficients $c_{0}\ldots c_{5}$ to second order in $\tau$.
We find 
\begin{align}
c_{0} & = \epsilon_{0}^2  \widetilde{a}_{d,x}^2  -2 \widetilde{A}_{d,x}^2 \mu_{z}^2  \widetilde{q}_{z}  \tau^2,\\
c_{1} & =  2\widetilde{A}_{d,x}\widetilde{a}_{d,x}\epsilon_{0} \tau,\\
c_{2} & = \widetilde{A}_{d,x}^2 \tau^2, 
\end{align}
where we also used $\epsilon_{0}\ll 1$ to further simplify the coefficients, and where we
temporarily define
\begin{equation}\label{eq:Ad}
     \widetilde{A}_{d,x} = -i + \widetilde{a}_{d,x},
\end{equation}
to shorten the notation.
The remaining coefficients $c_{3}, c_{4}$ and $c_{5}$ vanish, since these are of orders $\tau^3,\tau^4$ and $\tau^5$, respectively.
The corresponding quadratic dispersion relation can directly be solved. Its solutions are [replacing $\widetilde{A}_{d,x}$ using (\ref{eq:Ad})]
\begin{equation}\label{eq:sigg}
    \sigma =  \pm \mu_{z} \sqrt{2 \widetilde{q}_{z}} - \frac{\epsilon_{0} \widetilde{a}_{d,x}}{\left(-i +  \widetilde{a}_{d,x}\right) \tau}.
\end{equation}
The first term is the isothermal pure gas VSI growth rate, whereas the second term describes the effect of dust.
Splitting the second term into real and imaginary parts readily yields (\ref{eq:sig_vsi_1_p}).

\section{Effect of vertical dust settling on the VSI - the DSSI}\label{app:vsi_set}
 \subsection{Growth rate of the DSSI for general wavenumbers}\label{sec:gr_dssi}
 
Here we use Eqs. (\ref{eq:ug5p})---(\ref{eq:rhod5p}) to derive an analytical expression for the growth rate of the VSI in the presence of dust settling. As outlined in \S \ref{sec:vsi_dust_z}, here we ignore radial drift between dust and gas and consider the isothermal limit without buoyancy effects. First, we rewrite the above equations into 
\begin{align}
    \sigma \delta u_{g} & = 2 \mu_{z}^2 \delta v_{g} - \mu_{x} \mu_{z} \frac{w_{d0}}{\tau} \delta \rho_{d}\label{eq:ug5},\\
    \sigma \delta v_{g} & = \widetilde{q}_{z}  \delta u_{g}\label{eq:vg5},\\
    \widehat{\sigma}_{d,z} \delta u_{d} & = \delta u_{g} + 2 \tau \delta v_{d}\label{eq:ud5},\\
    \widehat{\sigma}_{d,z} \delta v_{d} & = \delta v_{g} - \frac{\tau}{2} \delta u_{d} + q_{z} \tau \frac{k_{x}}{k_{z}} \frac{\delta u_{g}}{\widehat{\sigma}_{d,z}}\label{eq:vd5},\\
   \left(\sigma + i a_{d,z}\right) \delta \rho_{d} & = -i \epsilon_{0} k_{x} \left(\delta u_{d} - \frac{\delta u_{g}}{\widehat{\sigma}_{d,z}}\right)\label{eq:rhod5},
\end{align}
where we again used (\ref{eq:linwd}) to eliminate $\delta w_{d}$ from the equations, as well as (\ref{eq:sighat_z}). 
From the above equations, we obtain the dispersion relation
\begin{equation}\label{eq:disprel_vsi}
\begin{split}
    & \left(\sigma + i a_{d,z}\right)\left(\sigma^2- 2 \mu_{z}^2 \widetilde{q}_{z} \right) \left(\widehat{\sigma}_{d,z}^3 +\tau^2 \widehat{\sigma}_{d,z}\right) \\
    \quad & - 2 \mu_{x}^2 a_{d,z} \,i \epsilon_{0}\widetilde{q}_{z} \left(\widehat{\sigma}_{d,z} + \sigma \tau\right)=0.
    \end{split}
\end{equation}
In the limit of vanishing vertical shear ($q_{z}\to 0$), Eq.~(\ref{eq:disprel_vsi}) is identical to (\ref{eq:disp5}) in the limit of vanishing vertical buoyancy $N_{z}^2\to 0$.  
If we assume $\left|\sigma \tau\right| \ll 1$ and $\left|a_{d,z} \tau\right| \ll 1$, then we obtain the cubic dispersion relation (\ref{eq:det_dssi}).

An analytical expression for the growth rate of the fastest-growing mode resulting from (\ref{eq:det_dssi}) is given by 
\begin{equation}\label{eq:sig_dssi}
   \sigma_{R,\text{DSSI}} = \frac{2^{\frac{2}{3}} \cos \theta}{6} \chi^{\frac{1}{3}}\left[1- \frac{C}{\left(\frac{\chi}{2}\right)^\frac{2}{3}}\right].
\end{equation}
In this expression: 
\begin{equation}
    \chi = \left|A+\sqrt{|B|}\right|,
\end{equation}
with
\begin{align}
    A & = -a_{d,z} \left( 54 \epsilon_{0} \mu_{x}^2 \widetilde{q}_{z} + 36 \mu_{z}^2 \widetilde{q}_{z} + 2 a_{d,z}^2 \right),\label{eq:A_dssi}\\
    B & = 4 C^3 -A^2,\label{eq:B_dssi}\\
    C & = a_{d,z}^2 - 6 \mu_{z}^2 \widetilde{q}_{z}\label{eq:C_dssi}.
\end{align}
For growing solutions, which require $B<0$ (see below), the angle $\theta$ in Eq. (\ref{eq:sig_dssi}) is given by 
\begin{align}
  \theta & =  -\frac{\pi}{6}  \hspace{0.45cm} \text{if}   \,A+\sqrt{|B|}>0,\\
\theta & = \frac{5\pi}{6}  \hspace{0.5cm} \text{if} \, A+\sqrt{|B|}<0. 
\end{align}

If $a_{d,z} \to 0$ (no dust settling), we have $A=0$, $B=4 C^3$ and hence $\chi=2 \sqrt{ |C^3|}$. Using $\theta = -\pi/6$  we recover
\begin{equation}
     \sigma_{R,\text{DSSI}} \to \mu_{z} \sqrt{2 \widetilde{q}_{z}},
\end{equation}
which is the isothermal pure gas VSI growth rate [cf. Eq.~(\ref{eq:sig_vsi_iso})].

Similar to the case of the DSI described in Appendix \ref{app:dsi}, one finds that $\sigma_{R,\text{DSSI}}=0$ if $B>0$, which for brevity will not be discussed here. The condition $B=0$ distinguishes between stable and unstable systems, but the explicit expression  is rather involved. However, from (\ref{eq:A_dssi})---(\ref{eq:C_dssi}) it can directly be shown that the term $(4 C^3 - A^2)\propto \widetilde{q}_{z}$ and therefore $\widetilde{q}_{z}=0$ is one solution, equivalent to the threshold for the pure gas VSI (cf. Figure \ref{fig:dssi}).

\subsection{Growth rate of the DSSI in the limit of small vertical wavenumbers}\label{app:vsi_kz0}

Our numerical results in \S\ref{sec:vsi_kz0} show that in the presence of both vertical shear and dust settling, instability persists for small $k_z$. Here, we formally consider the limit $k_{z}\to 0$, for which we can obtain simple analytical expressions for the growth rate and the frequency of the DSSI, unlike for the case of general wavenumbers, considered in \S \ref{sec:gr_dssi}. Note that for $k_z=0$, the VSI, DSI, and the SI all vanish. 

Compared to (\ref{eq:ug5})---(\ref{eq:rhod5}) we start here from a slightly more general set of equations, where all drag force terms are retained, and where we do not a priori neglect radial and azimuthal drift between dust and gas. 
We do, however, as in \S \ref{sec:gr_dssi}, set $\eta=0$ so that there is no contribution to dust-gas drift from the global radial pressure gradient.
Due to incompressibility (\ref{eq:dincomp}), we now have $\delta u_{g}=0$. As before, we assume that the gas is isothermal ($\beta \to 0$). 

Thus, we now consider the set of linearised equations 
 \begin{align}
  (\sigma \tau + i a_{d,x}\tau )\, \delta \rho_{d} & = - i \epsilon_{0} k_{x} \tau \delta u_{d}\label{eq:rhod_kz0}, \\
  \widehat{\sigma}_{d,x} \delta u_{d} & =  2 \tau \delta v_{d}\label{eq:ud_kz0}, \\
  \widehat{\sigma}_{d,x} \delta v_{d} & =   \delta v_{g}  -\frac{1}{2}  \tau \delta u_{d}  -q_{z} \tau \delta w_{d}\label{eq:vd_kz0},\\
  \widehat{\sigma}_{d,x} \delta w_{d} & = \delta w_{g},  \label{eq:wd}\\
  \widehat{\sigma}_{g,x} \delta v_{g} & =  -q_{z} \tau \delta w_{g} + \epsilon_{0}\delta v_{d}  +  \delta \rho_{d} \Delta v_{0},\\
  \widehat{\sigma}_{g,x} \delta w_{g}  & =  \epsilon_{0} \delta w_{d}+   w_{d} \delta \rho_{d}\label{eq:wg},
 \end{align}
 where $\widehat{\sigma}_{d,x}$ and $\widehat{\sigma}_{g,x}$ are given by (\ref{eq:sighat_dx}) and (\ref{eq:sighat_gx}), respectively. 
From (\ref{eq:wd}) and (\ref{eq:wg}) we find
 \begin{equation}
    \delta w_{g} =\frac{w_{d0}}{\left(\sigma\tau + i a_{d,x}\tau\right)\left(1+\widehat{\epsilon}_{0}\right)} \delta \rho_{d},
 \end{equation}
 where we defined
 \begin{equation}
     \widehat{\epsilon}_{0}=\frac{\epsilon_{0}}{\widehat{\sigma}_{d,x}}
 \end{equation}
 and where we assumed $a_{d,x}=a_{g,x}$, such that radial drift between dust and gas is ignored. This is justified since we assume $\eta=0$.
 We are now left with Eqs. (\ref{eq:rhod_kz0}), (\ref{eq:ud_kz0}), as well as
 \begin{align}
     \widehat{\sigma}_{d,x} \delta v_{d} & = \delta v_{g} -\frac{\tau}{2} \delta u_{d} -\frac{q_{z} \tau w_{d0}}{\left(\sigma \tau + i a_{d,x}\tau\right)\left(1+\widehat{\epsilon}_{0} \right)\widehat{\sigma}_{d,x}}\delta \rho_{d}\label{eq:vd},\\
     \begin{split}
     \widehat{\sigma}_{g,x} \delta v_{g} & = \epsilon_{0} \delta v_{d} +\Delta v_{0}\delta \rho_{d} -\frac{q_{z} \tau w_{d0}}{\left(\sigma \tau + i a_{d,x}\tau\right)\left(1+\widehat{\epsilon}_{0}\right)}\delta \rho_{d}\label{eq:vg}.
     \end{split}
 \end{align}
We now assume $|\sigma \tau| \ll 1$ and $|a_{d,x}\tau|\ll 1$. From (\ref{eq:ud_kz0}) we then obtain
\begin{equation}\label{eq:udvd}
    \delta u_{d} = 2 \tau \delta v_{d}.
\end{equation}
Next we combine (\ref{eq:vd}) and (\ref{eq:vg}) to get
\begin{equation}
\begin{split}
    \left(\sigma \tau + i a_{d,x}\tau\right)\left(\epsilon_{0} \delta v_{d} + \delta v_{g} \right)  = & -\epsilon_{0}\tau^2 \delta v_{d} \\ 
    \quad & +\left(\Delta v_{0} -\frac{ q_{z} \tau w_{d0} }{\sigma \tau + i a_{d,x}\tau}\right) \delta \rho_{d},
    \end{split}
\end{equation}
where we also used (\ref{eq:udvd}).
Finally, we assume that the perturbed azimuthal motions are tightly coupled (i.e. $\delta v_{d} \approx \delta v_{g}$) and use (\ref{eq:udvd}) and (\ref{eq:rhod_kz0}) to arrive at the cubic dispersion relation
\begin{equation}\label{eq:drel_vssi}
    \begin{split}
    \left(\widehat{\sigma}_{d,x}-1\right)^3 \left(\epsilon_{0}+1\right) & = -\left[\Delta v_{0}\left(\widehat{\sigma}_{d,x}-1 \right) - q_{z}w_{d0}\tau \right] 2 i k_{x} \epsilon_{0}\tau^2,
    \end{split}
\end{equation}
where we used (\ref{eq:sighat_dx}) to shorten the notation.
This equation possesses three roots. 
The (exact) eigenvalue of the fastest-growing mode reads
\begin{equation}\label{eq:sigred}
    \sigma  = - i k_{x} u_{d0} -\frac{2 i \epsilon_{0} k_{x} \Delta v_{0}}{\left(3 \epsilon_{0} k_{x} q_{z} w_{d0}\right)^{\frac{1}{3}} X^{\frac{1}{3}}}  + \frac{\left(\epsilon_{0} k_{x} q_{z} w_{d0}\right)^{\frac{1}{3}} X^{\frac{1}{3}}}{3^{\frac{2}{3}}\left(1+\epsilon_{0}\right)}
\end{equation}
where
\begin{equation}\label{eq:xx}
\begin{split}
    X &= -9 i \left(1+\epsilon_{0}\right)^2 \\ \quad & + \sqrt{3}(1+\epsilon_{0})^{(3/2)}\sqrt{-27(1+\epsilon_{0}) - i \frac{8 \epsilon_{0} k_{x} \Delta v_{0}^3}{\left(q_{z} w_{d0}\right)^2}}\\
    \quad & \approx -18 i \left(1+\epsilon_{0}\right)^2.
\end{split}
\end{equation}
In the last step, we neglected the second term in the square root as it is much smaller than the first term for all wavenumbers $k_{x}$ considered here. Physically this implies that the effect  of azimuthal dust-gas drift (expressed through $\Delta v_{0}$) is much smaller than the effect of vertical dust settling (expressed through $w_{d0}$) in the current situation.
We also assumed that $q_{z} w_{d0}<0$ ( positive (negative) vertical shear below (above) the midplane).
Using (\ref{eq:xx}), the growth rate and frequency of the growing mode can be obtained from (\ref{eq:sigred}). Care is required when selecting the correct solution of the cube root. We find 
\begin{equation}\label{eq:gr_red}
\begin{split}
    \sigma_{R} & = \frac{1}{3\left(1+\epsilon_{0}\right)} \Bigg[\left(\frac{\epsilon_{0} k_{x} q_{z} w_{d0}}{\sqrt{2}\left(1+\epsilon_{0}\right)}\right)^{\frac{2}{3}}\tau \\
    \quad & +\left(54\epsilon_{0}|k_{x} q_{z} w_{d0}|\left(1+\epsilon_{0}\right)^2 \right)^{\frac{1}{3}}\frac{\sqrt{3}}{2}\Bigg] \\
    \quad & \approx \sqrt{3}\left(\frac{1}{4}  k_{x}q_{z} w_{d0}\frac{\epsilon_{0}}{1+\epsilon_{0}}\right)^{\frac{1}{3}}
    \end{split}
\end{equation}
 where we neglected the first term $\propto \tau$ in the bracket. 
Similarly,
\begin{equation}
\begin{split}
    \sigma_{I} & \approx -2 k_{x} q_{z} w_{d0} \frac{\epsilon_{0}}{1+\epsilon_{0}} +\left(\frac{ k_{x}q_{z} w_{d0}}{4 } \frac{\epsilon_{0}}{1+\epsilon_{0}}\right)^{\frac{1}{3}},
    \end{split}
\end{equation}
where we used
\begin{equation}
     \Delta v_{0} \approx \frac{\tau q_{z} w_{d0}}{1+\epsilon_{0} }\label{eq:dv0}
\end{equation}
which follows from (\ref{eq:vd0}) and (\ref{eq:vg0}) with $\eta=0$ and $\tau \ll 1$.

It is worth noting that if we neglect the azimuthal drift $\Delta v_{0}$ in the dispersion relation (\ref{eq:drel_vssi}) we directly arrive at (\ref{eq:gr_red}). 
On the other hand, if we neglect only the dust-settling term $\propto w_{d0}$ in (\ref{eq:drel_vssi}), we find the growing mode
\begin{equation}\label{eq:gr_lh22}
    \sigma = -i k_{x} w_{d0} +\left(1-i\right)\sqrt{\frac{\epsilon_{0} k_{x} \Delta v_{0}}{1+\epsilon_{0}}},
\end{equation}
which corresponds exactly to Eq.~(C8) in \citet{lh2022}.
These authors found that an external torque acting on the gas alone results in an azimuthal drift between dust and gas, which in turn serves as an energy source for another dust-gas drag instability.
Indeed, the equilibrium relative azimuthal and radial velocities between dust and gas in the current situation are given by (\ref{eq:dv0}) and
\begin{equation}
    \Delta u_{0} \approx \frac{2\tau^2  q_{z} w_{d0}} {\left(1+\epsilon_{0}\right)^2}\label{eq:du0},
\end{equation}
which directly follows from (\ref{eq:ug0}) and (\ref{eq:ud0}).
 From these we infer that in the absence of a radial pressure gradient ($\eta=0$, for example in the vicinity of a pressure bump) and the presence of vertical shear, gas and dust experience a relative azimuthal drift that is larger by a factor $\sim 1/\tau$ than the relative radial drift. Under these conditions one may expect the manifestation of an azimuthal drift-induced SI, as reported by \citet{lh2022}. 
  The difference is that here $\Delta v_{0}\propto q_z z_0$ is due to vertical shear in combination with dust settling, rather than an external torque acting on the gas. That being said, in absence of dust-settling (\ref{eq:gr_lh22}) would vanish as well.
  However, in the present situation (away from the disc midplane) we find that the settling of dust results in a much more vigorous instability (the DSSI) than azimuthal drift would do.
  

\bsp	
\label{lastpage}

\begin{thebibliography}{}
\makeatletter
\relax
\def\mn@urlcharsother{\let\do\@makeother \do\$\do\&\do\#\do\^\do\_\do\%\do\~}
\def\mn@doi{\begingroup\mn@urlcharsother \@ifnextchar [ {\mn@doi@}
  {\mn@doi@[]}}
\def\mn@doi@[#1]#2{\def\@tempa{#1}\ifx\@tempa\@empty \href
  {http://dx.doi.org/#2} {doi:#2}\else \href {http://dx.doi.org/#2} {#1}\fi
  \endgroup}
\def\mn@eprint#1#2{\mn@eprint@#1:#2::\@nil}
\def\mn@eprint@arXiv#1{\href {http://arxiv.org/abs/#1} {{\tt arXiv:#1}}}
\def\mn@eprint@dblp#1{\href {http://dblp.uni-trier.de/rec/bibtex/#1.xml}
  {dblp:#1}}
\def\mn@eprint@#1:#2:#3:#4\@nil{\def\@tempa {#1}\def\@tempb {#2}\def\@tempc
  {#3}\ifx \@tempc \@empty \let \@tempc \@tempb \let \@tempb \@tempa \fi \ifx
  \@tempb \@empty \def\@tempb {arXiv}\fi \@ifundefined
  {mn@eprint@\@tempb}{\@tempb:\@tempc}{\expandafter \expandafter \csname
  mn@eprint@\@tempb\endcsname \expandafter{\@tempc}}}

\bibitem[\protect\citeauthoryear{{Andrews}, {Wilner}, {Hughes}, {Qi}  \&
  {Dullemond}}{{Andrews} et~al.}{2009}]{andrews2009}
{Andrews} S.~M.,  {Wilner} D.~J.,  {Hughes} A.~M.,  {Qi} C.,   {Dullemond}
  C.~P.,  2009, \mn@doi [\apj] {10.1088/0004-637X/700/2/1502}, \href
  {https://ui.adsabs.harvard.edu/abs/2009ApJ...700.1502A} {700, 1502}

\bibitem[\protect\citeauthoryear{{Bae}, {Isella}, {Zhu}, {Martin}, {Okuzumi}
  \& {Suriano}}{{Bae} et~al.}{2022}]{bae2022}
{Bae} J.,  {Isella} A.,  {Zhu} Z.,  {Martin} R.,  {Okuzumi} S.,   {Suriano} S.,
   2022, arXiv e-prints, \href
  {https://ui.adsabs.harvard.edu/abs/2022arXiv221013314B} {p. arXiv:2210.13314}

\bibitem[\protect\citeauthoryear{{Balbus} \& {Hawley}}{{Balbus} \&
  {Hawley}}{1991}]{balbus1991}
{Balbus} S.~A.,  {Hawley} J.~F.,  1991, \mn@doi [\apj] {10.1086/170270}, \href
  {https://ui.adsabs.harvard.edu/abs/1991ApJ...376..214B} {376, 214}

\bibitem[\protect\citeauthoryear{{Barker} \& {Latter}}{{Barker} \&
  {Latter}}{2015}]{barker2015}
{Barker} A.~J.,  {Latter} H.~N.,  2015, \mn@doi [\mnras]
  {10.1093/mnras/stv640}, \href
  {https://ui.adsabs.harvard.edu/abs/2015MNRAS.450...21B} {450, 21}

\bibitem[\protect\citeauthoryear{{Barranco}, {Pei}  \& {Marcus}}{{Barranco}
  et~al.}{2018}]{barranco2018}
{Barranco} J.~A.,  {Pei} S.,   {Marcus} P.~S.,  2018, \mn@doi [\apj]
  {10.3847/1538-4357/aaec80}, \href
  {https://ui.adsabs.harvard.edu/abs/2018ApJ...869..127B} {869, 127}

\bibitem[\protect\citeauthoryear{{Carrera} \& {Simon}}{{Carrera} \&
  {Simon}}{2022}]{carrera2022}
{Carrera} D.,  {Simon} J.~B.,  2022, \mn@doi [\apjl]
  {10.3847/2041-8213/ac6b3e}, \href
  {https://ui.adsabs.harvard.edu/abs/2022ApJ...933L..10C} {933, L10}

\bibitem[\protect\citeauthoryear{{Carrera}, {Simon}, {Li}, {Kretke}  \&
  {Klahr}}{{Carrera} et~al.}{021a}]{carrera2021a}
{Carrera} D.,  {Simon} J.~B.,  {Li} R.,  {Kretke} K.~A.,   {Klahr} H.,  2021a,
  \mn@doi [\aj] {10.3847/1538-3881/abd4d9}, \href
  {https://ui.adsabs.harvard.edu/abs/2021AJ....161...96C} {161, 96}

\bibitem[\protect\citeauthoryear{{Carrera}, {Thomas}, {Simon}, {Small},
  {Kretke}  \& {Klahr}}{{Carrera} et~al.}{021b}]{carrera2021b}
{Carrera} D.,  {Thomas} A.,  {Simon} J.~B.,  {Small} M.~A.,  {Kretke} K.~A.,
  {Klahr} H.,  2021b, arXiv e-prints, \href
  {https://ui.adsabs.harvard.edu/abs/2021arXiv210808315C} {p. arXiv:2108.08315}

\bibitem[\protect\citeauthoryear{{Chen} \& {Lin}}{{Chen} \&
  {Lin}}{2020}]{chen2020}
{Chen} K.,  {Lin} M.-K.,  2020, \mn@doi [\apj] {10.3847/1538-4357/ab76ca},
  \href {https://ui.adsabs.harvard.edu/abs/2020ApJ...891..132C} {891, 132}

\bibitem[\protect\citeauthoryear{{Chiang} \& {Youdin}}{{Chiang} \&
  {Youdin}}{2010}]{chiang2010}
{Chiang} E.,  {Youdin} A.~N.,  2010, \mn@doi [Annual Review of Earth and
  Planetary Sciences] {10.1146/annurev-earth-040809-152513}, \href
  {https://ui.adsabs.harvard.edu/abs/2010AREPS..38..493C} {38, 493}

\bibitem[\protect\citeauthoryear{{Drazkowska} et~al.,}{{Drazkowska}
  et~al.}{2022}]{drazkowska2022}
{Drazkowska} J.,  et~al., 2022, arXiv e-prints, \href
  {https://ui.adsabs.harvard.edu/abs/2022arXiv220309759D} {p. arXiv:2203.09759}

\bibitem[\protect\citeauthoryear{{Dubrulle}, {Morfill}  \&
  {Sterzik}}{{Dubrulle} et~al.}{1995}]{dubrulle1995}
{Dubrulle} B.,  {Morfill} G.,   {Sterzik} M.,  1995, \mn@doi [\icarus]
  {10.1006/icar.1995.1058}, \href
  {https://ui.adsabs.harvard.edu/abs/1995Icar..114..237D} {114, 237}

\bibitem[\protect\citeauthoryear{{Dullemond}, {Ziampras}, {Ostertag}  \&
  {Dominik}}{{Dullemond} et~al.}{2022}]{dullemond2022}
{Dullemond} C.~P.,  {Ziampras} A.,  {Ostertag} D.,   {Dominik} C.,  2022,
  \mn@doi [\aap] {10.1051/0004-6361/202244218}, \href
  {https://ui.adsabs.harvard.edu/abs/2022A&A...668A.105D} {668, A105}

\bibitem[\protect\citeauthoryear{{Flock}, {Nelson}, {Turner}, {Bertrang},
  {Carrasco-Gonz{\'a}lez}, {Henning}, {Lyra}  \& {Teague}}{{Flock}
  et~al.}{2017}]{flock2017}
{Flock} M.,  {Nelson} R.~P.,  {Turner} N.~J.,  {Bertrang} G. H.~M.,
  {Carrasco-Gonz{\'a}lez} C.,  {Henning} T.,  {Lyra} W.,   {Teague} R.,  2017,
  \mn@doi [\apj] {10.3847/1538-4357/aa943f}, \href
  {https://ui.adsabs.harvard.edu/abs/2017ApJ...850..131F} {850, 131}

\bibitem[\protect\citeauthoryear{{Flock}, {Turner}, {Nelson}, {Lyra}, {Manger}
  \& {Klahr}}{{Flock} et~al.}{2020}]{flock2020}
{Flock} M.,  {Turner} N.~J.,  {Nelson} R.~P.,  {Lyra} W.,  {Manger} N.,
  {Klahr} H.,  2020, \mn@doi [\apj] {10.3847/1538-4357/ab9641}, \href
  {https://ui.adsabs.harvard.edu/abs/2020ApJ...897..155F} {897, 155}

\bibitem[\protect\citeauthoryear{{Fukuhara}, {Okuzumi}  \& {Ono}}{{Fukuhara}
  et~al.}{2021}]{fukuhara2021}
{Fukuhara} Y.,  {Okuzumi} S.,   {Ono} T.,  2021, \mn@doi [\apj]
  {10.3847/1538-4357/abfe5c}, \href
  {https://ui.adsabs.harvard.edu/abs/2021ApJ...914..132F} {914, 132}

\bibitem[\protect\citeauthoryear{{Goldreich} \& {Lynden-Bell}}{{Goldreich} \&
  {Lynden-Bell}}{1965}]{goldreich1965}
{Goldreich} P.,  {Lynden-Bell} D.,  1965, \mn@doi [mnras]
  {10.1093/mnras/130.2.125}, \href
  {http://adsabs.harvard.edu/abs/1965MNRAS.130..125G} {130, 125}

\bibitem[\protect\citeauthoryear{{Haghighipour} \& {Boss}}{{Haghighipour} \&
  {Boss}}{003a}]{haghighipour2003a}
{Haghighipour} N.,  {Boss} A.~P.,  2003a, \mn@doi [\apj] {10.1086/378950},
  \href {https://ui.adsabs.harvard.edu/abs/2003ApJ...598.1301H} {598, 1301}

\bibitem[\protect\citeauthoryear{{Haghighipour} \& {Boss}}{{Haghighipour} \&
  {Boss}}{003b}]{haghighipour2003b}
{Haghighipour} N.,  {Boss} A.~P.,  2003b, \mn@doi [\apj] {10.1086/345472},
  \href {https://ui.adsabs.harvard.edu/abs/2003ApJ...583..996H} {583, 996}

\bibitem[\protect\citeauthoryear{{Huang}, {Li}, {Isella}, {Miranda}, {Li}  \&
  {Ji}}{{Huang} et~al.}{2020}]{huang2020}
{Huang} P.,  {Li} H.,  {Isella} A.,  {Miranda} R.,  {Li} S.,   {Ji} J.,  2020,
  \mn@doi [\apj] {10.3847/1538-4357/ab8199}, \href
  {https://ui.adsabs.harvard.edu/abs/2020ApJ...893...89H} {893, 89}

\bibitem[\protect\citeauthoryear{{Ishitsu}, {Inutsuka}  \& {Sekiya}}{{Ishitsu}
  et~al.}{2009}]{ishitsu09}
{Ishitsu} N.,  {Inutsuka} S.-i.,   {Sekiya} M.,  2009, arXiv e-prints, \href
  {https://ui.adsabs.harvard.edu/abs/2009arXiv0905.4404I} {p. arXiv:0905.4404}

\bibitem[\protect\citeauthoryear{{Jacquet}, {Balbus}  \& {Latter}}{{Jacquet}
  et~al.}{2011}]{jacquet2011}
{Jacquet} E.,  {Balbus} S.,   {Latter} H.,  2011, \mn@doi [\mnras]
  {10.1111/j.1365-2966.2011.18971.x}, \href
  {http://adsabs.harvard.edu/abs/2011MNRAS.415.3591J} {415, 3591}

\bibitem[\protect\citeauthoryear{{Johansen} \& {Youdin}}{{Johansen} \&
  {Youdin}}{2007}]{johansen2007}
{Johansen} A.,  {Youdin} A.,  2007, \mn@doi [\apj] {10.1086/516730}, \href
  {https://ui.adsabs.harvard.edu/abs/2007ApJ...662..627J} {662, 627}

\bibitem[\protect\citeauthoryear{{Johansen}, {Blum}, {Tanaka}, {Ormel},
  {Bizzarro}  \& {Rickman}}{{Johansen} et~al.}{2014}]{johansen2014}
{Johansen} A.,  {Blum} J.,  {Tanaka} H.,  {Ormel} C.,  {Bizzarro} M.,
  {Rickman} H.,  2014, in {Beuther} H.,  {Klessen} R.~S.,  {Dullemond} C.~P.,
  {Henning} T.,  eds, Protostars and Planets VI. p.~547 (\mn@eprint {arXiv}
  {1402.1344}), \mn@doi{10.2458/azu\_uapress\_9780816531240-ch024}

\bibitem[\protect\citeauthoryear{{Johansen}, {Youdin}  \& {Mac Low}}{{Johansen}
  et~al.}{009a}]{johansen2009a}
{Johansen} A.,  {Youdin} A.,   {Mac Low} M.-M.,  2009a, \mn@doi [\apjl]
  {10.1088/0004-637X/704/2/L75}, \href
  {https://ui.adsabs.harvard.edu/abs/2009ApJ...704L..75J} {704, L75}

\bibitem[\protect\citeauthoryear{{Klahr} \& {Hubbard}}{{Klahr} \&
  {Hubbard}}{2014}]{klahr2014}
{Klahr} H.,  {Hubbard} A.,  2014, \mn@doi [\apj] {10.1088/0004-637X/788/1/21},
  \href {https://ui.adsabs.harvard.edu/abs/2014ApJ...788...21K} {788, 21}

\bibitem[\protect\citeauthoryear{{Krapp}, {Youdin}, {Kratter}  \&
  {Ben{\'\i}tez-Llambay}}{{Krapp} et~al.}{2020}]{krapp2020}
{Krapp} L.,  {Youdin} A.~N.,  {Kratter} K.~M.,   {Ben{\'\i}tez-Llambay} P.,
  2020, \mn@doi [\mnras] {10.1093/mnras/staa1854}, \href
  {https://ui.adsabs.harvard.edu/abs/2020MNRAS.497.2715K} {497, 2715}

\bibitem[\protect\citeauthoryear{{Laibe} \& {Price}}{{Laibe} \&
  {Price}}{2014}]{laibe2014}
{Laibe} G.,  {Price} D.~J.,  2014, \mn@doi [\mnras] {10.1093/mnras/stu355},
  \href {https://ui.adsabs.harvard.edu/abs/2014MNRAS.440.2136L} {440, 2136}

\bibitem[\protect\citeauthoryear{{Latter}}{{Latter}}{2016}]{latter2016}
{Latter} H.~N.,  2016, \mn@doi [\mnras] {10.1093/mnras/stv2449}, \href
  {https://ui.adsabs.harvard.edu/abs/2016MNRAS.455.2608L} {455, 2608}

\bibitem[\protect\citeauthoryear{{Latter} \& {Papaloizou}}{{Latter} \&
  {Papaloizou}}{2017}]{lp2017}
{Latter} H.~N.,  {Papaloizou} J.,  2017, \mn@doi [\mnras]
  {10.1093/mnras/stx2038}, \href
  {https://ui.adsabs.harvard.edu/abs/2017MNRAS.472.1432L} {472, 1432}

\bibitem[\protect\citeauthoryear{{Latter} \& {Papaloizou}}{{Latter} \&
  {Papaloizou}}{2018}]{latter2018}
{Latter} H.~N.,  {Papaloizou} J.,  2018, \mn@doi [\mnras]
  {10.1093/mnras/stx3031}, \href
  {https://ui.adsabs.harvard.edu/abs/2018MNRAS.474.3110L} {474, 3110}

\bibitem[\protect\citeauthoryear{{Lehmann} \& {Lin}}{{Lehmann} \&
  {Lin}}{2022}]{lehmann2022}
{Lehmann} M.,  {Lin} M.~K.,  2022, \mn@doi [\aap]
  {10.1051/0004-6361/202142378}, \href
  {https://ui.adsabs.harvard.edu/abs/2022A&A...658A.156L} {658, A156}

\bibitem[\protect\citeauthoryear{{Lesur}}{{Lesur}}{2020}]{lesur2020}
{Lesur} G.,  2020, arXiv e-prints, \href
  {https://ui.adsabs.harvard.edu/abs/2020arXiv200715967L} {p. arXiv:2007.15967}

\bibitem[\protect\citeauthoryear{{Lesur} \& {Latter}}{{Lesur} \&
  {Latter}}{2016}]{lesur2016}
{Lesur} G. R.~J.,  {Latter} H.,  2016, \mn@doi [\mnras]
  {10.1093/mnras/stw2172}, \href
  {https://ui.adsabs.harvard.edu/abs/2016MNRAS.462.4549L} {462, 4549}

\bibitem[\protect\citeauthoryear{{Lesur} et~al.,}{{Lesur}
  et~al.}{2022}]{lesur2022}
{Lesur} G.,  et~al., 2022, arXiv e-prints, \href
  {https://ui.adsabs.harvard.edu/abs/2022arXiv220309821L} {p. arXiv:2203.09821}

\bibitem[\protect\citeauthoryear{{Li} \& {Youdin}}{{Li} \&
  {Youdin}}{2021}]{li2021}
{Li} R.,  {Youdin} A.~N.,  2021, \mn@doi [\apj] {10.3847/1538-4357/ac0e9f},
  \href {https://ui.adsabs.harvard.edu/abs/2021ApJ...919..107L} {919, 107}

\bibitem[\protect\citeauthoryear{{Lin}}{{Lin}}{2019}]{lin2019}
{Lin} M.-K.,  2019, \mn@doi [\mnras] {10.1093/mnras/stz701}, \href
  {https://ui.adsabs.harvard.edu/abs/2019MNRAS.485.5221L} {485, 5221}

\bibitem[\protect\citeauthoryear{{Lin}}{{Lin}}{2021}]{lin2021}
{Lin} M.-K.,  2021, \mn@doi [\apj] {10.3847/1538-4357/abcd9b}, \href
  {https://ui.adsabs.harvard.edu/abs/2021ApJ...907...64L} {907, 64}

\bibitem[\protect\citeauthoryear{{Lin} \& {Hsu}}{{Lin} \& {Hsu}}{2022}]{lh2022}
{Lin} M.-K.,  {Hsu} C.-Y.,  2022, \mn@doi [\apj] {10.3847/1538-4357/ac3bb9},
  \href {https://ui.adsabs.harvard.edu/abs/2022ApJ...926...14L} {926, 14}

\bibitem[\protect\citeauthoryear{{Lin} \& {Youdin}}{{Lin} \&
  {Youdin}}{2015}]{lin2015}
{Lin} M.-K.,  {Youdin} A.~N.,  2015, \mn@doi [\apj]
  {10.1088/0004-637X/811/1/17}, \href
  {https://ui.adsabs.harvard.edu/abs/2015ApJ...811...17L} {811, 17}

\bibitem[\protect\citeauthoryear{{Lin} \& {Youdin}}{{Lin} \&
  {Youdin}}{2017}]{lin2017}
{Lin} M.-K.,  {Youdin} A.~N.,  2017, \mn@doi [\apj] {10.3847/1538-4357/aa92cd},
  \href {https://ui.adsabs.harvard.edu/abs/2017ApJ...849..129L} {849, 129}

\bibitem[\protect\citeauthoryear{{Lovascio} \& {Paardekooper}}{{Lovascio} \&
  {Paardekooper}}{2019}]{lovascio2019}
{Lovascio} F.,  {Paardekooper} S.-J.,  2019, \mn@doi [\mnras]
  {10.1093/mnras/stz2035}, \href
  {https://ui.adsabs.harvard.edu/abs/2019MNRAS.488.5290L} {488, 5290}

\bibitem[\protect\citeauthoryear{{Lyra}}{{Lyra}}{2014}]{lyra2014}
{Lyra} W.,  2014, \mn@doi [\apj] {10.1088/0004-637X/789/1/77}, \href
  {https://ui.adsabs.harvard.edu/abs/2014ApJ...789...77L} {789, 77}

\bibitem[\protect\citeauthoryear{{Lyra}, {Raettig}  \& {Klahr}}{{Lyra}
  et~al.}{2018}]{lyra2018}
{Lyra} W.,  {Raettig} N.,   {Klahr} H.,  2018, \mn@doi [Research Notes of the
  American Astronomical Society] {10.3847/2515-5172/aaeac9}, \href
  {https://ui.adsabs.harvard.edu/abs/2018RNAAS...2..195L} {2, 195}

\bibitem[\protect\citeauthoryear{{Malygin}, {Klahr}, {Semenov}, {Henning}  \&
  {Dullemond}}{{Malygin} et~al.}{2017}]{malygin2017}
{Malygin} M.~G.,  {Klahr} H.,  {Semenov} D.,  {Henning} T.,   {Dullemond}
  C.~P.,  2017, \mn@doi [\aap] {10.1051/0004-6361/201629933}, \href
  {https://ui.adsabs.harvard.edu/abs/2017A&A...605A..30M} {605, A30}

\bibitem[\protect\citeauthoryear{{Marcus}, {Pei}, {Jiang}, {Barranco},
  {Hassanzadeh}  \& {Lecoanet}}{{Marcus} et~al.}{2015}]{marcus2015}
{Marcus} P.~S.,  {Pei} S.,  {Jiang} C.-H.,  {Barranco} J.~A.,  {Hassanzadeh}
  P.,   {Lecoanet} D.,  2015, \mn@doi [\apj] {10.1088/0004-637X/808/1/87},
  \href {https://ui.adsabs.harvard.edu/abs/2015ApJ...808...87M} {808, 87}

\bibitem[\protect\citeauthoryear{{Nakagawa}, {Sekiya}  \& {Hayashi}}{{Nakagawa}
  et~al.}{1986}]{nakagawa1986}
{Nakagawa} Y.,  {Sekiya} M.,   {Hayashi} C.,  1986, \mn@doi [\icarus]
  {10.1016/0019-1035(86)90121-1}, \href
  {https://ui.adsabs.harvard.edu/abs/1986Icar...67..375N} {67, 375}

\bibitem[\protect\citeauthoryear{{Nelson}, {Gressel}  \& {Umurhan}}{{Nelson}
  et~al.}{2013}]{nelson2013}
{Nelson} R.~P.,  {Gressel} O.,   {Umurhan} O.~M.,  2013, \mn@doi [\mnras]
  {10.1093/mnras/stt1475}, \href
  {https://ui.adsabs.harvard.edu/abs/2013MNRAS.435.2610N} {435, 2610}

\bibitem[\protect\citeauthoryear{{Onishi} \& {Sekiya}}{{Onishi} \&
  {Sekiya}}{2017}]{onishi2017}
{Onishi} I.~K.,  {Sekiya} M.,  2017, \mn@doi [Earth, Planets, and Space]
  {10.1186/s40623-017-0637-z}, \href
  {https://ui.adsabs.harvard.edu/abs/2017EP&S...69...50O} {69, 50}

\bibitem[\protect\citeauthoryear{{Pfeil} \& {Klahr}}{{Pfeil} \&
  {Klahr}}{2019}]{pfeil2019}
{Pfeil} T.,  {Klahr} H.,  2019, \mn@doi [\apj] {10.3847/1538-4357/aaf962},
  \href {https://ui.adsabs.harvard.edu/abs/2019ApJ...871..150P} {871, 150}

\bibitem[\protect\citeauthoryear{{Raettig}, {Klahr}  \& {Lyra}}{{Raettig}
  et~al.}{2015}]{raettig2015}
{Raettig} N.,  {Klahr} H.,   {Lyra} W.,  2015, \mn@doi [\apj]
  {10.1088/0004-637X/804/1/35}, \href
  {https://ui.adsabs.harvard.edu/abs/2015ApJ...804...35R} {804, 35}

\bibitem[\protect\citeauthoryear{{Raettig}, {Lyra}  \& {Klahr}}{{Raettig}
  et~al.}{2021}]{raettig2021}
{Raettig} N.,  {Lyra} W.,   {Klahr} H.,  2021, \mn@doi [\apj]
  {10.3847/1538-4357/abf739}, \href
  {https://ui.adsabs.harvard.edu/abs/2021ApJ...913...92R} {913, 92}

\bibitem[\protect\citeauthoryear{{Squire} \& {Hopkins}}{{Squire} \&
  {Hopkins}}{2018}]{squire2018}
{Squire} J.,  {Hopkins} P.~F.,  2018, \mn@doi [\mnras] {10.1093/mnras/sty854},
  \href {https://ui.adsabs.harvard.edu/abs/2018MNRAS.477.5011S} {477, 5011}

\bibitem[\protect\citeauthoryear{{Stoll} \& {Kley}}{{Stoll} \&
  {Kley}}{2014}]{stoll2014}
{Stoll} M. H.~R.,  {Kley} W.,  2014, \mn@doi [\aap]
  {10.1051/0004-6361/201424114}, \href
  {https://ui.adsabs.harvard.edu/abs/2014A&A...572A..77S} {572, A77}

\bibitem[\protect\citeauthoryear{{Stoll} \& {Kley}}{{Stoll} \&
  {Kley}}{2016}]{stoll2016}
{Stoll} M. H.~R.,  {Kley} W.,  2016, \mn@doi [\aap]
  {10.1051/0004-6361/201527716}, \href
  {https://ui.adsabs.harvard.edu/abs/2016A&A...594A..57S} {594, A57}

\bibitem[\protect\citeauthoryear{{Takeuchi} \& {Lin}}{{Takeuchi} \&
  {Lin}}{2002}]{takeuchi2002}
{Takeuchi} T.,  {Lin} D.~N.~C.,  2002, \mn@doi [\apj] {10.1086/344437}, \href
  {https://ui.adsabs.harvard.edu/abs/2002ApJ...581.1344T} {581, 1344}

\bibitem[\protect\citeauthoryear{{Taki}, {Fujimoto}  \& {Ida}}{{Taki}
  et~al.}{2016}]{taki2016}
{Taki} T.,  {Fujimoto} M.,   {Ida} S.,  2016, \mn@doi [\aap]
  {10.1051/0004-6361/201527732}, \href
  {https://ui.adsabs.harvard.edu/abs/2016A&A...591A..86T} {591, A86}

\bibitem[\protect\citeauthoryear{{Tassoul}}{{Tassoul}}{1978}]{tassoul1978}
{Tassoul} J.,  1978, {Theory of rotating stars}.
{Princeton: University Press}

\bibitem[\protect\citeauthoryear{{Urpin}}{{Urpin}}{2003}]{urpin2003}
{Urpin} V.,  2003, \mn@doi [\aap] {10.1051/0004-6361:20030513}, \href
  {https://ui.adsabs.harvard.edu/abs/2003A&A...404..397U} {404, 397}

\bibitem[\protect\citeauthoryear{{Urpin} \& {Brandenburg}}{{Urpin} \&
  {Brandenburg}}{1998}]{urpin1998}
{Urpin} V.,  {Brandenburg} A.,  1998, \mn@doi [\mnras]
  {10.1046/j.1365-8711.1998.01118.x}, \href
  {https://ui.adsabs.harvard.edu/abs/1998MNRAS.294..399U} {294, 399}

\bibitem[\protect\citeauthoryear{{Volponi}}{{Volponi}}{2016}]{volponi2016}
{Volponi} F.,  2016, \mn@doi [\mnras] {10.1093/mnras/stw954}, \href
  {https://ui.adsabs.harvard.edu/abs/2016MNRAS.460..560V} {460, 560}

\bibitem[\protect\citeauthoryear{{Youdin} \& {Goodman}}{{Youdin} \&
  {Goodman}}{2005}]{youdin2005}
{Youdin} A.~N.,  {Goodman} J.,  2005, \mn@doi [\apj] {10.1086/426895}, \href
  {https://ui.adsabs.harvard.edu/abs/2005ApJ...620..459Y} {620, 459}

\bibitem[\protect\citeauthoryear{{Youdin} \& {Johansen}}{{Youdin} \&
  {Johansen}}{2007}]{youdin2007}
{Youdin} A.,  {Johansen} A.,  2007, \mn@doi [\apj] {10.1086/516729}, \href
  {https://ui.adsabs.harvard.edu/abs/2007ApJ...662..613Y} {662, 613}

\bibitem[\protect\citeauthoryear{{van der Marel} et~al.,}{{van der Marel}
  et~al.}{2021}]{vdmarel2021}
{van der Marel} N.,  et~al., 2021, \mn@doi [\aj] {10.3847/1538-3881/abc3ba},
  \href {https://ui.adsabs.harvard.edu/abs/2021AJ....161...33V} {161, 33}

\makeatother
\end{thebibliography}
\end{document}